\documentclass{lmcs}
\pdfoutput=1 %uncomment to ensure pdflatex processing (mandatatory e.g. to submit to arXiv)

\usepackage{lipsum}
\usepackage[style=alphabetic,backend=bibtex]{biblatex}

\makeatletter
\def\paragraph{\@startsection{paragraph}{4}%
  \z@\z@{-\fontdimen2\font}%
  {\normalfont\itshape}}
\makeatother

\ExplSyntaxOn
\NewDocumentEnvironment{restatethm}{m+b}
{% #1 (mandatory): a label
    \AddToHookNext{para/begin}{
        \iow_shipout:cn { @auxout } {
            \RESTATETHEOREM{#1}{#2}
        }
    }
    \begin{thm}\label{#1}#2\end{thm}
}{}

\NewDocumentCommand{\RESTATETHEOREM}{mm}
 {%
  \cs_gset:cpn {RESTATETHIS@#1} {#2}%
 }
\NewDocumentCommand{\restate}{m}
 {
  \group_begin:
  \renewcommand{\thethm}{\ref{#1}}
  \begin{thm} \use:c {RESTATETHIS@#1} \end{thm}
  \addtocounter{thm}{-1}
  \group_end:
}
\ExplSyntaxOff

\usepackage{caption}
\newcommand{\details}[1]{\captionsetup{skip=1pt,font=footnotesize}\caption*{#1}}
\usepackage{subcaption}
\usepackage{amsmath,amssymb,bm}
\usepackage[dvipsnames]{xcolor}
\usepackage{tikz}
\usepackage{tikz-cd}
\usepackage{stmaryrd}
\usepackage{tabularx}
\usepackage{multirow}
\usepackage{booktabs}

\usetikzlibrary{arrows}
\usetikzlibrary{arrows.meta}
\usetikzlibrary{decorations.markings}
\usetikzlibrary{decorations.pathmorphing}
\usetikzlibrary{decorations.pathreplacing}
\usetikzlibrary{patterns}
\usetikzlibrary{patterns.meta}
\usetikzlibrary{calligraphy}
\usetikzlibrary{shapes}
\usetikzlibrary{calc}

\newcommand{\redsquare}{\textcolor{red!30}{\ensuremath{\blacksquare}}}

\newcommand{\defn}[1]{\emph{#1}}

\newcommand{\N}{\mathbb{N}}
\newcommand{\Z}[1][]{\smash{\mathbb{Z}^{#1}}}
\newcommand{\Q}{\mathbb{Q}}
\newcommand{\R}{\mathbb{R}}

\newcommand{\interval}[1]{\ensuremath{\llbracket #1 \rrbracket}}
\newcommand{\norm}[1]{\ensuremath{\left\lVert#1\right\rVert}}
\newcommand{\card}[1]{\ensuremath{\lvert #1 \rvert}}
\newcommand{\lift}[1]{\ensuremath{{#1^{\uparrow}}}}
\newcommand{\liftfree}[1]{\ensuremath{{#1^{\Uparrow}}}}
\newcommand{\restr}[2]{#1|_{#2}}
\newcommand{\mto}{\rightrightarrows}

\newcommand{\compclass}[1]{\relax\ifmmode\texttt{#1}\else\ensuremath{\mathtt{#1}}\fi}
\newcommand{\decproblem}[1]{\textsc{#1}}
\newcommand{\encode}[1]{\ensuremath{\left\langle #1 \right\rangle}}

\newcommand{\Dencode}[1]{\ensuremath{\left[ #1 \right]}}

\newcommand{\subpattern}{\ensuremath{\sqsubseteq}}
\newcommand{\shiftlang}[2][]{\ensuremath{\mathcal{L}_{#1}(#2)}}

\newcommand{\blank}{\ensuremath{\texttt{\textvisiblespace}}}
\newcommand{\heads}{\ensuremath{\mathsf{H}}}
\newcommand{\tails}{\ensuremath{\mathsf{T}}}
\newcommand{\freezero}{\ensuremath{\mathsf{0}}}

\colorlet{gridborder}{black!15!white}
\colorlet{TilingColorA}{gray!70!white}
\colorlet{TilingColorB}{black}
\definecolor{TilingHighlight}{HTML}{FFBC3F}
\colorlet{TilingGrid}{black!20!white}
\newcommand{\gcross}[1]{\begin{tikzpicture}[scale=0.33,baseline={(0,0.09)}]\draw[gridborder!30] (0,0) rectangle (1,1); \draw[#1,thick] (0,0.5) -- ++(1,0); \draw[#1,thick] (0.5,0) -- ++(0,1);\end{tikzpicture}}
\newcommand{\ghline}[1]{\begin{tikzpicture}[scale=0.33,baseline={(0,0.09)}]\draw[gridborder!30] (0,0) rectangle (1,1); \draw[#1,thick] (0,0.5) -- ++(1,0);\end{tikzpicture}}
\newcommand{\gvline}[1]{\begin{tikzpicture}[scale=0.33,baseline={(0,0.09)}]\draw[gridborder!30] (0,0) rectangle (1,1);\draw[#1,thick] (0.5,0) -- ++(0,1);\end{tikzpicture}}
\definecolor{RainbowA}{HTML}{ec111a}
\definecolor{RainbowB}{HTML}{fb6330}
\definecolor{RainbowC}{HTML}{eed42f}
\definecolor{RainbowD}{HTML}{3ea908}
\definecolor{RainbowE}{HTML}{009dd6}
\definecolor{RainbowF}{HTML}{7849b8}
\definecolor{RainbowG}{HTML}{da70d6}

\newcommand{\pisync}{\ensuremath{\pi_{\mathrm{sync}}}}
\newcommand{\entdim}{\ensuremath{D}}
\newcommand{\upentdim}{\ensuremath{\overline{\entdim}}}
\newcommand{\lowentdim}{\ensuremath{\underline{\entdim}}}

\newcommand{\grid}[1]{\mathfrak{#1}}
\newcommand{\rect}[1]{\mathfrak{#1}}
\NewDocumentCommand{\substep}{O{}m}{\ifstrempty{#1}
  {\xrightarrow{#2}}
  {\xrightarrow[{\raisebox{0.5ex}[0.5ex]{$\scriptstyle #1$}}]{#2}}}
\NewDocumentCommand{\rsubstep}{O{}m}{\ifstrempty{#1}
  {\xleftarrow{#2}}
  {\xleftarrow[{\raisebox{0.5ex}[0.5ex]{$\scriptstyle #1$}}]{#2}}}
\NewDocumentCommand{\limspace}{mm}{\smash{\overleftarrow{#1}_{\!#2}}}

\newenvironment{subproof}[1][\proofname]{%
  \begin{proof}[#1]%
  }{%
  \end{proof}%
}

\begin{document}

\title[Computability of extender sets in multidimensional subshifts\dots]{Computability of extender sets in multidimensional subshifts: asymptotic growths, dynamical constraints}

% \thanks{thanks, optional.}	%optional

% affiliations are numbered automatically with a, b, c (see below)
% use the optional argument to indicate the affiliation(s) of each author
% omit the argument if there is only one author, or only one affiliation
\author[A.~Callard]{Antonin Callard\lmcsorcid{0000-0002-4673-4881}}[a]
\author[L.~Paviet Salomon]{Léo Paviet Salomon\lmcsorcid{0009-0005-4498-3832}}[a]
\author[P.~Vanier]{Pascal Vanier\lmcsorcid{0000-0001-9207-9112}}[b]

% affiliation 1 (automatically numbered a)
\address{ENS de Lyon, UCBL, CNRS, Inria, LIP; F-69342 Lyon Cedex 07, France}	%optional
% write emails for all authors having that affiliation
\email{contact@acallard.net, mail@lpaviets.org}  %optional

% affiliation 2 (automatically numbered b)
\address{Université Caen Normandie, ENSICAEN, CNRS, Normandie Univ, GREYC; F-14000 Caen, France}	%optional
\email{pascal.vanier@unicaen.fr}  %optional

% \funding{This research was partially funded by ANR JCJC 2019 19-CE48-0007-01.}%optional, to capture a funding statement, which applies to all authors. Please enter author specific funding statements as fifth argument of the \author macro.

% \acknowledgements{We are thankful to the referees for their many helpful remarks and suggestions.}%optional

% \nolinenumbers %uncomment to disable line numbering

\begin{abstract}
  Subshifts are sets of colorings of $\Z[d]$ defined by families of forbidden patterns. In a given subshift, the extender set of a finite pattern is the set of all its admissible completions. Since soficity of $\Z$ subshifts is equivalent to having a finite number of extender sets, it had been conjectured that the number of extender sets could provide a way to separate the classes of sofic and effective subshifts in higher dimensions.

  We investigate some computational characterizations of extender sets in multidimensional subshifts, and in particular their growth, in terms of \emph{extender entropies} \cite{french_pavlov19_follow_predec_exten_entrop} and \emph{extender entropy dimensions}. We prove here that sofic and effective subshifts have the same possible extender entropies (exactly the $\Pi_3$-computable real numbers of $[0,+\infty)$) and extender entropy dimensions, and investigate the computational complexity of these growth-type quantities under various dynamical and combinatorial constraints.
\end{abstract}

\maketitle

\def\sectionautorefname{Section}
\let\subsectionautorefname\sectionautorefname
\let\subsubsectionautorefname\sectionautorefname

 % \vspace*{-\baselineskip}
\section{Introduction}

For $\mathcal{A}$ a finite set called \emph{alphabet}, and $d \in \N$ a dimension, a \emph{coloring} (or \emph{configuration}) $x \in \mathcal{A}^{\Z[d]}$ is a function that maps each position of $\Z[d]$ to a color from the alphabet $\mathcal{A}$. In this context, a \emph{subshift} is a set of colorings $X \subseteq \mathcal{A}^{\Z[d]}$ that is both translation-invariant and closed for Cantor's topology \cite[Chapter 6]{lind_marcus21_introd_symbol_dynam_codin}. This definition suggests a dynamical viewpoint on subshifts, by considering how the colorings of a subshift $X$ behave under the action of $\Z[d]$ on $X$ by iterated translations. Yet, subshifts are also combinatorial objects: if a \emph{pattern} is a coloring of a finite portion of $\Z[d]$, then a set $X \subseteq \mathcal{A}^{\Z[d]}$ is a subshift if and only if there exists forbidden patterns $\mathcal{F}$ such that $X$ is the set of all colorings in which none of the forbidden patterns of $\mathcal{F}$ appear \cite[Chapter 1]{lind_marcus21_introd_symbol_dynam_codin}.

Seeing the configurations of a subshift as more geometrically-complex words, the combinatorial definition suggests to consider subshifts as a variation of formal languages in an infinite and multidimensional setting. Similarly to classical languages of finite words, which can be separated into the classes of local, regular, context-free, and computable languages\dots\ subshifts can be classified by the complexity of their presentations: a subshift $X \subseteq \mathcal{A}^{\Z[d]}$ is
\enlargethispage{\baselineskip}
\begin{enumerate}
\item \emph{of finite type} (SFT) if it can be defined by a finite set of forbidden patterns;
\item \emph{effective} if it can be defined by a computably enumerable set of forbidden patterns;
\item \emph{sofic} if there exists a cell-by-cell projection $\pi \colon \mathcal{B} \to \mathcal{A}$ and a subshift of finite type $X' \subseteq \mathcal{B}^{\Z[d]}$ such that $X = \pi(X')$.
\end{enumerate}

Subshifts of finite type (SFTs) have been independently studied under the formalism of Wang tiles in \cite{wang_pattern_recognition_ii}, for which the classical Domino problem (given a finite family of forbidden patterns, is the resulting subshift non-empty?) was proved undecidable~\cite{berger66:undecidability-domino-problem}. Sofic subshifts were introduced in \cite{weiss_sft_sofic_systems} as a generalization of subshifts of finite type. They actually form a subset of the effective subshifts, which are considered as the class of ``explicitly constructible'' subshifts and contains all subshifts mentioned in this article:
\[ [\text{Subshifts of finite type}] \subsetneq [\text{Sofic subshifts}] \subsetneq [\text{Effective subshifts}].\]
These inclusions can be proved strict.

\subsection{Soficity}
In the case of $\Z$ subshifts, the separations between subshifts of finite type, sofic and effective subshifts are well understood. For any language of finite words
$L \subseteq \mathcal{A}^{\ast}$, the \emph{syntactic congruence} is the equivalence relation defined as:
\[\forall u, v \in L, u \sim_{L} v \iff (\forall x, y \in L, xuy \in L
  \iff xvy \in L);\]
and the \emph{syntactic monoid} of $L$ is the quotient monoid $M(L) = L/\sim_{L}$. The Myhill-Nerode theorem shows that a language $L \subseteq \mathcal{A}^*$ is regular if and only if its syntactic monoid $M(L)$ is finite; adapting the syntactic monoid from $\mathcal{A}^*$ to $\mathcal{A}^{\Z}$, one can also prove that a $\Z$~subshift~$X$ is sofic if and only if it defines a finite syntactic monoid $M(X)$.

Thus, sofic subshifts on $\Z$ are also labelings of bi-infinite paths on finite automata and are often algorithmically presented via an associated finite automaton. In particular, most decision problems about $\Z$ sofic subshifts are actually decidable.

\medskip
However, the characterization of \emph{multidimensional} sofic subshifts, and the separation between sofic and effective multidimensional subshifts, are important open questions in symbolic dynamics \cite{kass_madden13_suffic_condit_non_sofic_higher_dimen_subsh, pavlov13_class_nonsof_multid_shift_spaces, guillon_jeandel15_infinite_communication_complexity,destombes_romashchenko22_resour_bound_kolmog_compl_provid, barbieri_sablik_salo25_sofic_free_exten_effec_subsh}. In the multidimensional setting, many problems are already undecidable on subshifts of finite type (hence on sofic subshifts), including the already mentioned Domino problem~\cite{berger66:undecidability-domino-problem} and most other non-trivial properties; and many subshifts with geometrically or computationally-complex structures have actually turned out to be sofic: substitution-based subshifts \cite{mozes89_tilin_subst_system_dynam_system}, the odd shift by J.~Cassaigne (black and white colorings in which every finite connected component of black cells must have odd size), the ``seas of squares'' subshifts \cite{westrick17_seas_squar_with_sizes_from_pi_set} (colorings made of independent black squares over a white background) with lengths restricted to arbitrary $\Pi^{0}_{1}$-computable sets, effective subshifts on $\mathcal{A} = \{0,1\}$ with sublinear densities of symbols $1$ \cite{destombes23_algor_compl_sofic_shift_dimen_two}\dots\ It even turns out that sofic subshifts of dimension $d+1$ capture all the possible behaviors and dynamics of effective subshifts of dimension $d$ \cite{hochman_dynamics_recursive_prop_multidimensional_symbolic_systems,durand12_fixed_point_tile_sets_their_applic,aubrun_sablik13_simul_effec_subsh_two_dimen}.

Nevertheless, some multidimensional subshifts have been proved not to be sofic, including for example the mirror subshift \cite[Proposition~57]{aubrun_barbieri_jeandel18_domino_probl_subshifts_groups}, in which configurations containing a single mirror plane of a special color must reflect the upper half-space into the lower half (see~\autoref{sec:ext:elementary-constructions}). All methods known by the authors to prove that a subshift is not sofic revolve around a counting argument: in a subshift of finite type, the compatibility of a $\interval{0,n-1}^d$ pattern with a complementary partial configuration only depends on their shared border, which is of size $O(n^{d-1})$; thus, in subshift of finite type (and, as projections of SFTs, in sofic subshifts), only $O(n^{d-1})$ bits of information can ``cross the border'' of a square pattern of domain $\interval{0,n-1}^d$. This restriction can sometimes prove the non-soficity of a subshift that packs too much information in small domains. The most recent argument in this vein uses a resource-bounded version of Kolmogorov complexity \cite{destombes_romashchenko22_resour_bound_kolmog_compl_provid}.

\subsection{Extender sets}
In this article, we are interested in another formalization of this information-quantifying argument: the notion of \emph{extender sets}. The extender set of a pattern $p$ in a subshift $X \subseteq \mathcal{A}^{\Z[d]}$ is the set $E_X(p)$ of partial colorings of $\Z[d]$ with a $p$-shaped hole that can extend $p$ into fully valid configurations in $X$.

On $\Z$, this notion coincides with Myhill-Nerode's \emph{syntactic congruence}. Thus, it fully characterizes $\Z$ sofic subshifts as the $\Z$ subshifts with a uniformly bounded number of extender sets \cite[Lemma~3.4]{ormes_pavlov16_exten_sets_multid_subsh}.

The generalization of extender sets to multidimensional subshifts goes back to \cite{kass_madden13_suffic_condit_non_sofic_higher_dimen_subsh,ormes_pavlov16_exten_sets_multid_subsh}. In \cite{ormes_pavlov16_exten_sets_multid_subsh}, it was proved that a subshift for which patterns of size $\interval{0,n-1}^d$ generate less than $n$ extender sets is actually sofic. This is much lower than the expected bound, as the counting argument from above implies, in particular, that patterns of domain $\interval{0,n-1}^d$ only generate $2^{O(n^{d-1})}$ distinct extender sets in a subshift of finite type \cite[Section 2]{kass_madden13_suffic_condit_non_sofic_higher_dimen_subsh}; as opposed to effective shifts, such as the mirror subshift, which can already fully realize $2^{O(n^d)}$ distinct extender sets of domain $\interval{0,n-1}^d$.

\medskip
The characterization of $\Z$ soficity, and the $2^{O(n^{d-1})}$ bound in the case of subshifts of finite type, have inspired the community to look for characterizations of multidimensional soficity based on counting the number $\card{E_X(\interval{0,n-1}^d)}$ of extender sets on a given domain. \cite{french16_charac_follow_exten_set_sequen} introduced the \emph{extender set sequence} $(\card{E_X(\interval{0,n-1}^d)}_{n \in \N})$ in the case of $\Z$ subshifts.

As is the case for more classical measures of complexity like the \emph{pattern complexity} $(\card{P_X(\interval{0,n-1}^d)})_{n \in \N}$, which counts the number of valid patterns of domain $\interval{0,n-1}^d$ in a given subshift $X$, the precise values and behavior of the sequence $(\card{P_X(\interval{0,n-1}^d)})_{n \in \N}$ can be difficult to compute~\cite{Greenfeld-Moreira-Zelmanov_2024:complexity-subshifts-and-infinite-words}. This motivated the study of the \emph{asymptotic growth} of the pattern complexity, \textit{i.e.}~of the behavior of $\log (\card{P_X(\interval{0,n-1}^d)}) \in O(n^{d-1})$ in the form of several \emph{growth-type invariants}:
\begin{itemize}
\item The classical \emph{(topological) entropy} $h(X)$ \cite[Chapter~4]{lind_marcus21_introd_symbol_dynam_codin}, which is informally defined as the quantity $h \in \R_+$ such that $\log (\card{P_X(\interval{0,n-1}^d)}) \sim h \cdot n^d$;
\item And more recently the \emph{entropy dimensions} \cite{meyerovitch_growth_type_invariants}, which are informally defined\footnote{The actual definitions for upper and lower entropy dimensions respectively involve $\limsup$ and $\liminf$, see~\autoref{def:entropy-dimension}.} as the quantity $D$ such that $\log (\card{P_X(\interval{0,n-1}^d)}) = \Theta(n^{D})$.
\end{itemize}

In the case of extender sets, considerations on the asymptotic growth of the number of extender sets $\log (\card{E_X(\interval{0,n-1}^d)}) \in O(n^d)$ has not yielded the expected results. It was already mentioned in \cite[Section~2]{kass_madden13_suffic_condit_non_sofic_higher_dimen_subsh} that several examples of subshifts with extender growth $\Omega(n^{d-1})$ were being found, but the authors nevertheless conjectured that this growth would always satisfy $\log (\card{E_X(\interval{0,n-1}^d)}) \in o(n^{d})$ in the case of $\Z[d]$ sofic subshifts: this conjecture was disproved in \cite[Example 5'': semi-mirror]{destombes_romashchenko22_resour_bound_kolmog_compl_provid}, which exhibited a sofic variation of the mirror subshift with extender growth $\Theta(n^d)$ (see \autoref{sec:ext:elementary-constructions}).

In this article, we are particularly interested in two growth-type quantities defined in terms of extender sets: the \emph{extender entropy} -- introduced on $\Z$ subshifts in \cite{french_pavlov19_follow_predec_exten_entrop} --, and the \emph{extender entropy dimension} -- which we introduce here. Inspired by their non-extender definitions, these quantities represent two adjacent aspects of the asymptotic growth of the number of extender sets of increasing domain sizes, which are respectively the asymptotic growth rate, and the rate of intermediate growth. Despite not decreasing under factor map applications, the extender entropy is in particular a \emph{conjugacy invariant}.

\subsection{Computational aspects of subshift invariants}

Going back to multidimensional subshifts, computability theory was first considered as a major obstruction to their analysis due to the undecidability of the Domino problem and its variations~\cite{berger66:undecidability-domino-problem}, \cite[Proposition 2]{harel84_gener_result_infin_tree_appli}. Nevertheless, the successful realization of $\Z[2]$ subshifts of finite type embedding universal computations of Turing machines has enabled to explicitly build subshifts whose geometry and dynamics are controlled by the ``embedded programming'', resulting in, for example: the realization of effective systems as subdynamics of systems of finite type in a Higman-like embedding fashion \cite{hochman_dynamics_recursive_prop_multidimensional_symbolic_systems}, the characterization of dynamical properties \cite{Zinoviadis_2015:hierarchy-expansiveness-in-2D-sfts,jeandel15_hardn_conjug_embed_factor_multid_subsh}, the realization of computationally complex sofic subshifts \cite{westrick17_seas_squar_with_sizes_from_pi_set,destombes23_algor_compl_sofic_shift_dimen_two} or other algebraic invariants \cite{paviet_vanier23_realiz_finit_presen_group_projec}.

A seminal result that marked a renewed interest in the applications of computability theory in symbolic dynamics was the complete characterization in computational terms of the possible values of (topological) entropies of subshifts of finite type~\cite{hochman10_charac_entrop_multid_shift_finit_type}, and of other growth-type invariants related to pattern complexity~\cite{meyerovitch_growth_type_invariants}. More precisely, the entropies of $\Z[2]$ subshifts of finite type were characterized as the $\Pi_1$-computable non-negative real numbers: while proving that such entropies must be $\Pi_1$-computable is routine, \cite{hochman10_charac_entrop_multid_shift_finit_type} achieved the non-trivial task of realizing every $\Pi_1$-computable real number as the entropy of an explicitly constructed SFT. And, in fact, all characterizations and realizations mentioned in the previous paragraph are directly phrased in terms of computability, and involve similar explicit embeddings of universal computations in suitably built multidimensional subshifts of finite type.

\enlargethispage{\baselineskip}
\subsection{Results}

In this article, we study the growth of extender sets of multidimensional subshifts -- namely, their \emph{extender entropies} and the \emph{extender entropy dimensions} --, and the interactions between the class of the subshift (of finite type, sofic, computable, effective) and various dynamical properties (general case, minimal subshift, mixing subshift).

In a similar vein to \cite{hochman10_charac_entrop_multid_shift_finit_type,meyerovitch_growth_type_invariants}, we achieve the characterization of the possible extender
entropies and extender entropy dimensions in terms of computability. Our main results include:

\restate{thm:ext-entropy-effective}

\restate{thm:ext-entropy-sofic}

Sofic subshifts were conjectured to have growth ${\log(\card{E_X(\interval{0,n-1}^d)}) \in o(n^d)}$ in~\cite{kass_madden13_suffic_condit_non_sofic_higher_dimen_subsh}, and thus extender entropy zero: while this was already disproved in \cite{destombes_romashchenko22_resour_bound_kolmog_compl_provid}, our characterization shows that the set of possible values is in fact dense in $[0,+\infty)$, and that extender entropies do not allow to separate sofic from effective multidimensional subshifts.

We also extend these results under additional mixing constraints in Theorems~\ref{thm:ext-entropy-block-gluing-effective} and~\ref{thm:ext-entropy-block-gluing-sofic}, refuting the intuition that mixing properties could influence the values of extender entropies in sofic multidimensional subshifts.

\medskip
We also provide similar results for (upper, lower and) extender entropy dimensions:
\restate{thm:ext-entropy-dim-effective}

\restate{thm:ext-entropy-dim-sofic}

\medskip
What is known by the authors is entirely summarized in the two tables below (\textcolor{gray}{gray} references are folklore or routine, while \textcolor{gray!70!blue}{highlighted} references mark new results):

\bigskip
{
  \newcommand{\refeasy}[1]{\textcolor{gray}{\autoref{#1}}}
  \newcommand{\refnew}[1]{\textcolor{gray!70!blue}{\autoref{#1}}}

  \centering

  \begin{tabular}{>{\centering\arraybackslash}p{1.5cm} | >{\centering\arraybackslash}p{2.5cm} | >{\centering\arraybackslash}p{4.3cm} | >{\centering\arraybackslash}p{4.8cm}}
    (Type) & (Restriction) & $\Z$ & $\Z[d]$ ($d \geq 2$) \\
    \midrule
    SFT & & \multicolumn{2}{c}{$\{0\}$ (\refeasy{clm:ext-entropy-sft})} \\
    \midrule
    \multirow{4}{*}{Sofic}
           & (None) & $\{0\}$ (\refeasy{clm:ext-entropy-sofic-1d}) & $\R_+ \cap \Pi_3$ (\refnew{thm:ext-entropy-sofic}) \\
           & Computable & $\{0\}$ (\refeasy{clm:ext-entropy-sofic-1d}) & $\R_+ \cap \Pi_2$ (\refnew{thm:ext-entropy-computable-sofic})\\
           & Minimal & \multicolumn{2}{c}{$\{0\}$ (\refeasy{cor:ext-entropy-minimal-sofic})} \\
           & $1$-block-gluing & $\{0\}$ (\refeasy{clm:ext-entropy-sofic-1d}) & $\R_+ \cap \Pi_3$ (\refnew{thm:ext-entropy-block-gluing-sofic}) \\
    \midrule
    \multirow{4}{*}{Effective}
           & (None) & \multicolumn{2}{l}{\hspace*{2.7cm}$\R_+ \cap \Pi_3$ \hspace*{0.1cm}(\refnew{thm:ext-entropy-effective})} \\
           & Computable & \multicolumn{2}{l}{\hspace*{2.7cm}$\R_+ \cap \Pi_2$ \hspace*{0.03cm}(\refnew{thm:ext-entropy-computable-effective})} \\
           & Minimal & \multicolumn{2}{l}{\hspace*{2.7cm}$\R_+ \cap \Pi_1$ (\refeasy{cor:ext-entropy-minimal-effective})} \\
           & $1$-block-gluing & \multicolumn{2}{l}{\hspace*{2.7cm}$\R_+ \cap \Pi_3$ \hspace*{0.03cm}(\refnew{thm:ext-entropy-block-gluing-effective})}
  \end{tabular}

  \bigskip
  Sets of possible extender entropies for various classes of subshifts.

  \bigskip
  \begin{tabular}{>{\centering\arraybackslash}p{1.5cm} | >{\centering\arraybackslash}p{2.5cm} | >{\centering\arraybackslash}p{4.3cm} | >{\centering\arraybackslash}p{4.8cm}}
    (Type) & (Limit) & $\Z$ & $\Z[d]$ ($d \geq 2$) \\
    \midrule
    \multirow{3}{*}{Sofic}
           & Upper & $\{0\}$ (\refeasy{clm:ext-entropy-dim-sofic-1d}) & $[0,d] \cap \Pi_3$ (\refnew{thm:ext-entropy-dim-effective}) \\
           & Lower & $\{0\}$ (\refeasy{clm:ext-entropy-dim-sofic-1d}) & $[0,d] \cap \Sigma_4$ (\refnew{thm:ext-entropy-dim-effective}) \\
           & (Limit) & $\{0\}$ (\refeasy{clm:ext-entropy-dim-sofic-1d}) & $[0,d] \cap \Delta_3$ (\refnew{thm:ext-entropy-dim-effective}) \\
    \midrule
    \multirow{3}{*}{Effective}
           & Upper & \multicolumn{2}{c}{$[0,d] \cap \Pi_3$ (\refnew{thm:ext-entropy-dim-sofic})} \\
           & Lower & \multicolumn{2}{c}{$[0,d] \cap \Sigma_4$ (\refnew{thm:ext-entropy-dim-sofic})} \\
           & (Limit) & \multicolumn{2}{c}{$[0,d] \cap \Delta_3$ (\refnew{thm:ext-entropy-dim-sofic})}
  \end{tabular}

  \bigskip
  Sets of possible extender entropy dimensions for various classes of subshifts.

}

\section{Definitions}\label{sec:definitions}

\subsection{Subshifts}\label{sec:def:subshifts}
Let $\mathcal{A}$ denote a finite set of symbols called an \emph{alphabet}, and $d \in \N$ a \emph{dimension}.
A \defn{configuration} is a coloring $x \in \mathcal{A}^{\Z[d]}$, and the color of
$x$ at position $p \in \Z[d]$ is denoted by $x_p$. A ($d$-dimensional)
\defn{pattern} over $\mathcal{A}$ is a coloring $w \in \mathcal{A}^P$ for some
set $P \subseteq \Z[d]$ called its \defn{support} (or~\emph{domain})\footnote{It is sometimes
convenient to consider patterns up to the translation of their support.
Usually, context will make it clear whether patterns are truly equal, or
only up to a $\Z[d]$ translation.}. For any pattern $w$
over $\mathcal{A}$ of support $P$, we say that $w$ \defn{appears} in a
configuration $x$ (and we denote $w \subpattern x$) if there
exists a translation $p_0 \in \Z[d]$ such that $w_{p} = x_{p + p_0}$ for all $p \in P$.

The \defn{shift functions} $(\sigma^t)_{t \in \Z[d]}$ act on configurations as
$(\sigma^{t}(x))_{p} = x_{p+t}$.  For $t \in \Z[d]$, a configuration $x$ is
\defn{$t$-periodic} if $\sigma^{t}(x) = x$. We sometimes consider patterns or
configurations by their restrictions: for $S \subseteq \Z[d]$ either finite or
infinite, and $x \in \mathcal{A}^{\Z[d]}$ a configuration (resp. $w$ a pattern),
we denote by $x|_S \in \mathcal{A}^S$ (resp. $w|_S \in \mathcal{A}^S$) the coloring it induces on
$S$.
\begin{defi}[Subshift]
  For any family of finite patterns $\mathcal{F}$, we define
  \[X_{\mathcal{F}} = \left\{x \in \mathcal{A}^{\mathbb{Z}^{d}} \mid
      \forall w \in \mathcal{F},\, w \not\subpattern x \right\}\!.\]
  A set $X \subseteq \mathcal{A}^{\mathbb{Z}^{d}}$ is called a
  \defn{subshift} if it is equal to some $X_{\mathcal{F}}$.
\end{defi}

Given a subshift $X$ and a finite support $P \subseteq \Z[d]$, we define
$\shiftlang[P]{X}$ as the set of patterns $w$ of support $P$ that appear in the
configurations of $X$. Such patterns are said to be \defn{globally admissible} in $X$.
We define the \defn{language} of $X$ as $\smash{\shiftlang{X}
= \bigcup\limits_{P \subseteq \Z[d] \text{ finite}} \shiftlang[P]{X}}$.
Slightly abusing notations, we denote
$\shiftlang[n]{X} = \shiftlang[\interval{0,n-1}^d]{X}$ for $n \in \N$.

\paragraph*{Factor maps} The functions that preserve the structure of subshifts, as dynamical systems, are
the block maps: they are also known as \emph{morphisms}, and correspond to the continuous functions that commute with the dynamics of the system. Factor maps are the surjective morphisms:
\begin{defi}[Factor map]
  For $X \subseteq \mathcal{A}^{\Z[d]}$ and $Y \subseteq \mathcal{B}^{\Z[d]}$ two subshifts, a function $\varphi \colon X \to Y$ is a \defn{factor map} if there exists some finite $N \subseteq \Z[d]$ and $f \colon \mathcal{A}^{N} \to \mathcal{B}$ such that $\varphi(x)_p = f(x|_{p+N})$. In which case, $Y$ is a \defn{factor} of $X$.

  $X$ and $Y$ are \defn{conjugate} if there exists a bijective factor map $\varphi \colon X \to Y$ (also called \defn{conjugacy}); in which case, $\varphi^{-1}$ can also be shown to be a factor map \cite[Curtis-Hedlund-Lyndon theorem]{hedlund69_endom_autom_shift_dynam_system}. Any mathematical object associated with subshifts that is preserved by conjugacy is a \defn{conjugacy invariant}.
\end{defi}

In the previous definition, the set $N$ is called a \emph{neighborhood} of the factor map $\varphi$. The \emph{radius} of $\varphi$ is the smallest $r \in \N$ such that $[-r,r]^d$ is a neighborhood of $\varphi$. If $\varphi$ is bijective, the \emph{biradius} of $\varphi$ is the maximum of the radii of $\varphi$ and $\varphi^{-1}$.

\paragraph*{Classes of subshifts}

Subshifts are traditionally classified as follows, and theses classes are invariant by conjugacy:
\begin{itemize}
\item A subshift $X \subseteq \mathcal{A}^{\Z[d]}$ is a \defn{subshift of finite type} (SFT) if there exists a finite family of forbidden patterns $\mathcal{F}$ such that $X = X_{\mathcal{F}}$;
\item A subshift $X \subseteq \mathcal{A}^{\Z[d]}$ is \defn{effective} if there exists a computably enumerable family of forbidden patterns $\mathcal{F}$ such that $X = X_{\mathcal{F}}$;
\item A subshift $Y \subseteq \mathcal{A}^{\Z[d]}$ is \defn{sofic} if there exists another alphabet $\mathcal{B}$, an SFT $X \subseteq \mathcal{B}^{\Z[d]}$ and a factor map $f \colon X \to Y$ such that $f(X) = Y$.\footnote{Actually, the SFT $X$ can be assumed to be \emph{local} (\textit{i.e.}~defined by adjacency constraints), and the associated factor map to be of radius~$0$: in other words, a letter-by-letter projection $\pi \colon \mathcal{B} \to \mathcal{A}$.}
\end{itemize}
By definition, a subshift of finite type is also sofic, and a sofic subshift is also effective.

\paragraph*{Operations on subshifts} Several operations allow to combine subshifts. For example, our constructions will involve \defn{layers}: for
a subshift of a cartesian product $X \subseteq \prod_{i \in I} L_i$, the layers of $X$ are the projections of $X$ onto each of the $L_i$, which are often named for convenience. For $J \subseteq I$, we will denote by $\pi_{L_{j_1} \times L_{j_2} \times \dots} \colon \prod_{i \in I} L_i \mapsto \prod_{j \in J} L_j$ the cartesian projection.

Another operation that we will use on subshifts are the two following types of \defn{lifts}:
\begin{defi}[Lifts]
  Given a $\Z[d]$ subshift $X \subseteq \mathcal{A}^{\Z[d]}$, define
  \begin{itemize}
  \item \defn{the periodic lift} $\lift{X} = \{\lift{x} \in \mathcal{A}^{\Z[d+1]} \mid x \in X \}$,
    where $(\lift{x})|_{\Z[d] \times \{ i \}} = x$ for all $i \in \Z$;
  \item \defn{the free lift} $\liftfree{X} = \{y \in \mathcal{A}^{\Z[d+1]} \mid \forall i \in
    \mathbb{Z},\, y_{\Z[d] \times \{ i \}} \in X\}$.
  \end{itemize}
\end{defi}
If $X$ is sofic (resp. effective), then both $\lift{X}$ and $\liftfree{X}$ are also sofic (resp. effective) since they can be defined by the same forbidden patterns. On the other hand:

\begin{thmC}[\cite{hochman_dynamics_recursive_prop_multidimensional_symbolic_systems},~{\cite[Theorem 3.1]{aubrun_sablik13_simul_effec_subsh_two_dimen}},~{\cite[Theorem 10]{durand12_fixed_point_tile_sets_their_applic}}]\label{thmC:effective-lift-sofic}
  If $X$ is an effective $\mathbb{Z}^d$ subshift, then $\lift{X}$ is a
  sofic $\mathbb{Z}^{d+1}$ subshift.
\end{thmC}

\paragraph*{Dynamical properties} As a dynamical system (under the natural action of $\Z[d]$ on a $d$-dimensional subshift by translating its configurations), various dynamical properties can be studied on subshifts.

\begin{defi}[Minimality]
  A subshift is \defn{minimal} if it contains no nonempty proper subshift.
\end{defi}
\noindent Thus, if $X = X_{\mathcal{F}}$ is a minimal subshift and $u \in \shiftlang{X}$, then the subshift $X_{\mathcal{F} \sqcup \{u\}}$ is empty.

Mixingness is another dynamical notion. In the context of $\Z$~subshifts, mixingness intuitively implies that for any pair of admissible words, there exists a configuration containing both of them at arbitrary positions, provided they are sufficiently far apart: more formally, a $\Z$ subshift $X$ is \defn{mixing} if
  \[\forall n > 0, \exists N > 0, \forall u, v \in \shiftlang[n]{X},
    \forall k \geq N, \exists w \in \shiftlang[k]{X},\, uwv \in
    \shiftlang{X}.\]
  We say that $X$ is $f(n)$-mixing for some function $f$ if $N$ can be taken
equal to $f(n)$ in the previous definition. When $f$ is constant $f(n) = N$, we
simply write that $X$ is $N$-mixing.

There exists various mixing notions in higher dimension. The results of this paper are phrased in terms of \defn{block-gluing} subshifts \cite{boyle_pavlov_schraudner10_multid_sofic_shift_without_separ_their_factor}:

\begin{defi}[Block-gluingness]
  For $X \subseteq \mathcal{A}^{\Z[d]}$ a subshift and $g \in \N$ a \emph{gap}, an increasing function, we say that $X$ is $g$-\defn{block-gluing} if
  \[\forall p, q \in \shiftlang[n]{X}, \forall u \in \mathbb{Z}^{d}, \norm{u}_{\infty} \geq g+n \implies (p \cup \sigma^{u}(q) \in \shiftlang{X}).\]
\end{defi}
\noindent Said differently, $X$ is $g$-block-gluing if any two square patterns of size $n$
can appear at any position as long as they are placed with a gap of size at
least $g$ between them.

\paragraph*{Pattern complexity} The traditional notion of complexity is called \defn{pattern complexity} and is defined by $P_X(n) = \shiftlang[{\interval{0,n-1}^d}]{X}$. It counts the number of valid patterns of domain $\interval{0,n-1}^d$ that appear in the configurations of $X$. The exponential growth rate of $\card{P_X(n)}$ is the \defn{topological entropy} \cite[Chapter~4]{lind_marcus21_introd_symbol_dynam_codin}:
\begin{defi}
  Given a subshift $X \subseteq \mathcal{A}^{\Z[d]}$, its \emph{topological entropy} is defined as:
  \[ h(X) = \lim_{n \mapsto +\infty} \frac{\log\, \card{P_X(n)}}{n^d}. \]
\end{defi}

 By \cite[Theorem~1]{capobianco08:multivariate-fekete-lemma}, this limit always exists, is equal to the infimum of
$(\frac{\log\, \card{P_X(n)}}{n^d})_{n > 0}$, and could actually be computed along any sequence of hyperrectangles eventually covering $\Z[d]$. Intuitively, for a subshift whose complexity function grows as
$\card{P_{X}(n)} \sim C2^{\alpha n^{d}}$, we have $h(X) = \alpha$: thus, the topological entropy is strictly positive when $P_{X}$ has an exponential growth in~$n^{d}$.

When considering intermediate growth rates, such as
$\card{P_{X}(n)} \sim C2^{\alpha n^{\beta}}$ with $\beta < d$, one can define another
kind of growth quantity named the \defn{entropy dimension} \cite{meyerovitch_growth_type_invariants,Carvalho_1997:entropy-dimension-dynamical-systems}, which gives information on
$\beta$ when the topological entropy of such a subshift is zero:
\begin{defi}[Entropy Dimension]\label{def:entropy-dimension}
  Let $X$ be a $d$-dimensional subshift. Its \defn{upper entropy dimension} $\upentdim_{h}(X)$ and its \emph{lower entropy dimension} $\lowentdim_{h}(X)$ are defined as:
  \[\upentdim_{h}(X) = \limsup_{n} \frac{\log \log\, \card{P_{X}(n)}}{\log n}
    \qquad \text{and} \qquad
    \lowentdim_{h}(X) = \liminf_{n} \frac{\log \log\, \card{P_{X}(n)}}{\log n}.
  \]
  When $\upentdim_{h}(X) = \lowentdim_{h}(X)$, we write $\entdim_{h}(X)$ this value and
  refer to it as the \defn{entropy dimension}.
\end{defi}

Contrary to the usual topological entropy, we usually do not
have $\lowentdim_{h}(X) = \upentdim_{h}(X)$ for an arbitrary subshift $X$.

\subsection{Computability notions}
\label{sec:def:computability-notions}

\subsubsection{Arithmetical hierarchy}
\label{sec:def:arithmetical-hierarchy}

The \defn{arithmetical hierarchy}~\cite[Chapter 4]{soare_turing_computability}
stratifies formulas of first-order arithmetic over $\N$ by the number of their alternating
unbounded quantifiers: for $n \in \N$, define
\begin{align*}
  \Pi^{0}_{n} &= \{ \forall k_1, \exists k_2, \forall k_3, \dots\ \phi(k_1,\dots,k_n) \mid
      \phi \text{ only contains bounded quantifiers} \}\\
  \Sigma^{0}_{n} &= \{\exists k_1, \forall k_2, \exists k_3, \dots\ \phi(k_1,\dots,k_n) \mid
      \phi \text{ only contains bounded quantifiers}\}.
\end{align*}

A decision problem is said to be in $\Pi^{0}_{n}$ (resp. $\Sigma^{0}_{n}$) if
its set of solutions $S \subseteq \N$ is described by a $\Pi^{0}_{n}$ (resp. $\Sigma^{0}_{n}$)
formula: in other words, $\Pi^{0}_{0} = \Sigma^{0}_{0}$ corresponds to the set
of computable decision problems; $\Sigma^{0}_{1}$ is the set of computably
enumerable decision problems, etc\dots

\subsubsection{Arithmetical hierarchy of real numbers}
\label{sec:def:hierarchy-real-numbers}
The \defn{arithmetical hierarchy of real
  numbers}~\cite{zheng_weihrauch01_arith_hierar_real_number} stratifies real
numbers depending on the difficulty of computably approximating them: for
$n \geq 0$, define
\begin{align*}
  \Sigma_{n} & = \{ x \in \R \mid \{ r \in \Q \mid r \leq x \} \text{ is a } \Sigma^{0}_n \text{ set} \} \\
  \Pi_{n} & = \{ x \in \R \mid \{ r \in \Q \mid r \geq x \} \text{ is a } \Sigma^{0}_n \text{ set} \} \textcolor{black!70}{\;= \{ x \in \R \mid \{ r \in \Q \mid r \leq x \} \text{ is a } \Pi^{0}_{n} \text{ set}\}}.
\end{align*}
In particular, $\Sigma_{0} = \Pi_{0}$ is the set of computable real numbers,
\textit{i.e.}\ numbers that can be computably approximated up to arbitrary
precision; $\Pi_{1}$ real numbers are also called right-computable, since they
can be computably approximated from above; etc\dots

Alternatively, this hierarchy is also defined by the number of alternating limit
operations needed to obtain a real number from the computable ones~\cite{zheng_weihrauch01_arith_hierar_real_number}. In other
words, for $n \geq 1$:
\begin{align*}
  \Sigma_{n} &= \Big\{\sup_{k_1 \in \N} \inf_{k_2 \in \N} \sup_{k_3 \in \N} \dots\ \alpha_{k_1,\dots,k_n} \mid
               (\alpha_{k_1,\dots,k_n})_{k_1,\dots,k_n \in \N} \in \Q^{\N^n} \text{ is computable} \Big\} \\
  \Pi_{n} &= \Big\{\inf_{k_1 \in \N} \sup_{k_2 \in \N} \inf_{k_3 \in \N} \dots\ \alpha_{k_1,\dots,k_n} \mid
      (\alpha_{k_1,\dots,k_n})_{k_1,\dots,k_n \in \N} \in \Q^{\N^n}\text{ is computable} \Big\}
\end{align*}

We also define $\Delta_{n} = \Sigma_{n} \cap \Pi_{n}$.

\medskip
As needed in later proofs of this paper, we can make some additional assumptions on the sequences
$(\alpha_{k_1,\dots,k_n})$ defining real numbers in $\Pi_{n}$ and $\Sigma_{n}$:

\pagebreak
\begin{lemC}[{Monotonicity,~\cite[Lemma~3.1]{zheng_weihrauch01_arith_hierar_real_number}}]\label{lem:arith-hierarchy-monotonicity}
  Let $\alpha \in \Pi_{3}$ be a real number. There exists sequences $(\alpha_{i,j,k})_{i,j,k \in \mathbb{N}^{3}}$ such that:
  \begin{itemize}
  \item For every $i,j \in \mathbb{N}^{2}$,
    $(\alpha_{i,j,k})_{k \in \mathbb{N}}$ is decreasing and converges towards some $\alpha_{i,j} \in \Pi_{1}$.
  \item For every $i \in \mathbb{N}$, $(\alpha_{i,j})_{j \in \mathbb{N}}$ is increasing\footnote{To avoid ambiguities with ``non-increasing'' not being the negation of increasing, this article uses an ``increasing'' and ``strictly increasing'' terminology.} and converges towards some $\alpha_{i} \in \Sigma_{2}$.
  \item $(\alpha_{i})_{i \in \mathbb{N}}$ is decreasing, and $\inf_{i} \alpha_{i} = \alpha$.
  \end{itemize}
\end{lemC}

\begin{lem}[Slowdown lemma]\label{lem:arith-hierarchy-slowdown-lemma}
  Let $(\alpha_{i})_{i \in \mathbb{N}}$ be a bounded sequence of uniformly
  $\Sigma_{2}$ real numbers, and let $Q$ be any polynomial. There exists another
  sequence $(\alpha'_{i})_{i \in \mathbb{N}}$ of uniformly $\Sigma_{2}$ real numbers such
  that:
  \begin{itemize}
  \item $(\alpha_{i})$ is a subsequence of $(\alpha'_{i})$, \textit{i.e.} there
    exists a strictly increasing $\varphi \colon \mathbb{N} \to \mathbb{N}$ such that for
    all $i \geq 0$, we have $\alpha_{i} = \alpha'_{\varphi(i)}$
  \item For any $i \geq 0$ and $\varphi(i) \leq \ell \leq \varphi(i + 1)$, the value $\alpha'_{\ell}$ lies between $\alpha_{i}$ and $\alpha_{i+1}$;
  \item The sequence $(\alpha'_{i})_{i \in \N}$ varies very slowly: $\sum_{j=i}^{i+Q(i)}\big|\alpha'_{j+1} - \alpha'_{j} \big| = o_{i \to +\infty}(1)$.
  \end{itemize}
\end{lem}
\noindent Note that the sequences $(\alpha_{i})$ and $(\alpha'_{i})$ have
the same $\limsup$ (resp. $\liminf, \lim$).

\begin{proof}
  We give an explicit construction of the sequence $(\alpha'_{i})$. Let
  $(\alpha_{i,j,k})_{i,j,k \in \mathbb{N}}$ be a computable sequence of rational numbers
  such that: for all $i$, $\alpha_{i} = \sup_{j} \inf_{k} \alpha_{i,j,k}$; for all $i,j$, $(\alpha_{i,j,k})_{k}$ is decreasing towards some $\alpha_{i,j}$; $(\alpha_{i,j})_{j}$ is increasing towards $\alpha_{i}$ (see~\autoref{lem:arith-hierarchy-monotonicity}); and there exists $M \in \mathbb{N}$ such that for all $i,j,k \in \N$, we have $-M < \alpha_{i,j,k} < M$.

  Such a sequence exists, as $(\alpha_{i})$ is a uniform sequence of $\Sigma_{2}$ real numbers. Now, for $n \geq 1$, let $\delta_{n} = \frac{1}{n \log n}$. We first construct a sequence $(\alpha''_{i})$ so that $\big|\alpha''_{i+1} - \alpha''_{i} \big| \leq \delta_{i+1}$. Assume by induction that we have built $\varphi(0)$, \dots, $\varphi(i) \in \N$ and $\alpha''_{0},\dots,\alpha''_{\varphi(i)}$. We define $\varphi(i+1)$ and the next terms of $\alpha''$ as follows:
  \begin{itemize}
  \item The series $\sum \delta_{n}$ diverges; let $\varphi(i+1)$ be the
    smallest $k \in \mathbb{N}$ such that
    $\sum_{n=\varphi(i)+1}^{k}\delta_{n} \geq 2M$.
  \item For $\varphi(i) < \ell < \varphi(i+1)$, let
    $\lambda_{\ell} = \sum_{n=\varphi(i)+1}^{\ell}\delta_{n}$.
  \item Now for $\varphi(i) < \ell < \varphi(i+1)$, define
    $\alpha''_{\ell} = (1-\frac{\lambda_{\ell}}{2M})\alpha_{i} +
    \frac{\lambda_{\ell}}{2M}\alpha_{i+1}$, and
    $\alpha''_{\varphi(i+1)} = \alpha_{i+1}$.
  \end{itemize}

  By definition, $(\alpha_{i})$ is a subsequence of $(\alpha''_{i})$ along the
  extraction $\varphi$, and monotonicity properties are preserved as the other
  terms are linear combinations with coefficients in $[0, 1]$ of consecutive
  terms of the initial sequence. And by construction, we have
  $\big|\alpha''_{i+1} - \alpha''_{i}\big| \leq \delta_{i+1}$.

  For $i \geq 0$, define now
  $\alpha'_{i} = \alpha''_{\lfloor \log (i+1)\rfloor}$. Monotonicity properties
  of the initial sequence are still preserved, and clearly $(\alpha_{i})$ is a
  subsequence of $(\alpha'_{i})$. Finally, let $i \geq 0$:
  \begin{align*}
    \sum_{j=i}^{i+Q(i)}\big|\alpha'_{j+1} - \alpha'_{j} \big|
    &= \sum_{j=\log (i+1)}^{\log(i+Q(i)+1)} \big|\alpha''_{j+1} - \alpha''_{j} \big|\\
    &\leq \sum_{j=\log (i+1)}^{\log(i+Q(i)+1)} \delta_{j+1}\\
    &\leq \log(i + Q(i) + 1) \delta_{\log i} \\
    &\leq O_{i}(\log i) \frac{1}{\log i \log \log i} \\
    &\leq o_{i}(1) \qedhere
  \end{align*}
\end{proof}

\section{Extender sets and extender entropies}

\subsection{Extender sets}
\label{sec:ext:def-extender-sets}

In this article, we focus on another notion of complexity based on extender sets:
\begin{defi}[Extender set]\label{def:extender-set}
  For $X \subseteq \mathcal{A}^{\mathbb{Z}^{d}}$ a $d$-dimensional subshift, $P \subseteq \mathbb{Z}^{d}$ and $w \in \mathcal{A}^P$ a pattern of support $P$, the \defn{extender set} of $w$ is the set of partial configurations
  \[E_{X}(w) = \{ x \in \mathcal{A}^{\Z[d] \setminus P} \mid x \sqcup w \in X \}, \]
  where $(x \sqcup w)_{p} = w_{p}$ if $p \in P$ and $(x \sqcup w)_{p} = x_{p}$ otherwise.
\end{defi}

In other words, $E_{X}(w)$ is the set of all valid ``completions'' of the pattern $w$ in~$X$. For example, for two patterns with the same support $w, w'$, we have $E_{X}(w) \subseteq
E_{X}(w')$ if and only if the pattern $w$ can be replaced by $w'$ every time it appears in any configuration of $X$.

We denote $E_{X}(n_1,\dots,n_d) = \{ E_{X}(w) \mid w \in \shiftlang[n_1,\dots,n_d]{X} \}$ its set of extender sets of domain $\interval{0,n_1-1} \times \dots \interval{0,n_d-1}$. In the special case $n_1 = \dots = n_d = n$, we denote $E_{X}(n) = E_X(n,\dots,n)$; which in turns defines the \defn{extender set sequence} $(\card{E_{X}(n)})_{n \in \N}$, introduced in the context of $\Z$ subshifts in \cite{french16_charac_follow_exten_set_sequen}.

\medskip
In the case of $\Z$ subshifts, extender sets are similar to the more classical notions of \defn{follower} (resp.\ \defn{predecessor}) \defn{sets}, which are the sets of right-infinite (resp. left-infinite) words that complete a finite given pattern (see for example~\cite{french16_charac_follow_exten_set_sequen}). Extender sets can also be considered as a generalization of Nerode congruence classes in a multidimensional setting: as mentioned in the introduction, counting the number of extender sets is the basis of several (non-) soficity arguments \cite[Lemma~3.4]{ormes_pavlov16_exten_sets_multid_subsh}, \cite{kass_madden13_suffic_condit_non_sofic_higher_dimen_subsh}.

\enlargethispage{\baselineskip}
\paragraph*{Examples}

\begin{enumerate}
  \item Consider $X = \mathcal{A}^{\mathbb{Z}^{d}}$ some full-shift in dimension $d$. Then $X$ has maximal topological entropy, but defines only a single extender set per domain: indeed, for any two patterns $w,w' \in \shiftlang[n]{X}$, we have $E_{X}(w) = E_{X}(w') = \{ \mathcal{A}^{\Z[d] \setminus \interval{0,n-1}^d} \}$; which implies that $\card{E_{X}(n)} = 1$ for every $n \in \N$.
  \item Consider $X$ a strongly periodic subshift: there exist $p_1,\dots,p_d \in \N$ such that, for $x \in X$ and $i \leq d$, we have $\sigma^{p_i \cdot e_i}(x) = x$. Then, for $n \geq \max p_{i}$ and $w \in \shiftlang[n]{X}$, $w$ is the only pattern $w'$ such that $E_X(w')=E_X(w)$; so that $\card{E_{X}(n)} = \card{\shiftlang[n]{X}} \leq p\mathcal{A}^{p}$ for $p = \prod_i p_i$.
\end{enumerate}

\subsection{Extender entropies}

Since counting the number of extender sets $\card{E_X(n_1,\dots,n_d)}$ can be somewhat tedious, \cite[Definition 2.17]{french_pavlov19_follow_predec_exten_entrop} suggested to consider the asymptotic growth rate of the extender set sequence $(\card{E_X(n)})_{n \in \N}$, thus introducing the \emph{extender entropy}\footnote{\cite{french_pavlov19_follow_predec_exten_entrop} defines extender entropies in the case of $\Z$ subshifts, but the definition makes sense for
higher dimensional subshifts.}:
\begin{defi}[{Extender entropy}]
  For a $\Z[d]$ subshift $X$, its \defn{extender entropy} is \[h_{E}(X) = \lim\limits_{n \to +\infty} \frac{\log\, \card{E_{X}(n)}}{n^{d}}.\]
\end{defi}
\begin{clm}
  The extender entropy is actually well-defined.
\end{clm}
\begin{proof}
  \cite{french_pavlov19_follow_predec_exten_entrop} already proved this on $\Z$. In order to generalize to arbitrary $\Z^d$, we prove that the map $(n_1,\dots,n_d) \in \N^d \mapsto \card{E_X(n_1,\dots,n_d)}$ is subadditive in every variable, and thus verifies the multivariate subadditive lemma \cite[Theorem~1]{capobianco08:multivariate-fekete-lemma}. More formally, for fixed $1 \leq i \leq d$, $n_1,\dots,n_d \in \N^d$ and $m_i \in \N$, consider the map
  \[ f \colon \mathcal{A}^{\interval{1,n_1} \times \dots \times \interval{1,n_i+m_i} \times \dots \times \interval{1,n_d}} \to E_X(n_1,\dots,n_i,\dots,n_d) \times E_X(n_1,\dots,m_i,\dots,n_d)\]
  defined as follows: for an arbitrary pattern $w \in \mathcal{A}^{\interval{1,n_1} \times \dots \times \interval{1,n_i+m_i} \times \dots \times \interval{1,n_d}}$, we decompose $w$ into ${w = u \sqcup v}$ with $u \in \mathcal{A}^{\interval{1,n_1} \times \dots \times \interval{1,n_i} \times \dots \times \interval{1,n_d}}$ and $v \in \mathcal{A}^{\interval{1,n_1} \times \dots \times \interval{n_i+1,m_i} \times \dots \times \interval{1,n_d}}$; then, we define $f(w)$ to be $(E_X(u),E_X(v))$.

  \smallskip
  We prove that if two patterns ${w,w' \in \mathcal{A}^{\interval{1,n_1} \times \dots \times \interval{1,n_i+m_i} \times \dots \times \interval{1,n_d}}}$ verify ${f(w) = f(w')}$, then $E_X(w) = E_X(w')$. Indeed, denoting $w = u \sqcup v$ and $w' = u' \sqcup w'$ as above, we have by definition of $f$ that $E_X(u) = E_X(u')$ and $E_X(v) = E_X(v')$. Considering an arbitrary $x \in \mathcal{A}^{\Z[d] \setminus (\interval{1,n_1} \times \dots \times \interval{1,n_i+m_i} \times \dots \times \interval{1,n_d})}$, we do successive replacements and obtain:
  \begin{align*}
    x \sqcup w \in X \iff & x \sqcup u \sqcup v \in X & \\
    \iff & x \sqcup u' \sqcup v \in X & \text{since $E_X(u) = E_X(u')$;}\\
    \iff & x \sqcup u' \sqcup v' \in X & \text{since $E_X(v) = E_X(v')$;}\\
    \iff & x \sqcup w' \in X;
  \end{align*}
  \noindent so that $E_X(w) = E_X(w')$. Notice that, since $x$, $u$/$u'$ and $v$/$v'$ have disjoint domains, the notation $x \sqcup u \sqcup v$ is not ambiguous.

  \begin{figure}[ht]
    \centering
    \begin{tikzpicture}[scale=0.12,decoration={random steps,segment length=0.2em,amplitude=0.1em}]
      \begin{scope}
        \draw (20,-1.5) node {$x$};
        \draw [gridborder,decorate,path picture={
          \path[even odd rule,pattern={Dots[distance=4pt]}] (-1,-1) rectangle (41,41) (6,6) rectangle (34,22);
          \fill[pattern={Lines[angle=45,distance=3pt]},even odd rule] (6,6) rectangle (18,22) (10.7,12.5) rectangle (13.3,15.5);
          \draw[black] (12,14) node {$u\vphantom{u'}$};
          \fill[pattern={Lines[angle=0,distance=3pt,yshift=1pt]},even odd rule] (18,6) rectangle (34,22) (24.7,12.5) rectangle (27.3,15.5);
          \draw[black] (26,14) node {$v\vphantom{u'}$};
          \draw[black,thick] (6,6) rectangle (18,22) (18,6) rectangle (34,22);
        }] (0.5,0.5) rectangle (39.5,27.5);
      \end{scope}
      \draw[->] (44,13.5) -- (56,13.5) node[midway,above]{\footnotesize $E_X(u) = E_X(u')$};
      \begin{scope}[xshift=60cm]
        \draw (20,-1.5) node {$x$};
        \draw [gridborder,decorate,path picture={
          \path[even odd rule,pattern={Dots[distance=4pt]}] (-1,-1) rectangle (41,41) (6,6) rectangle (34,22);
          \fill[pattern={Lines[angle=-45,distance=3pt]},even odd rule] (6,6) rectangle (18,22) (10.7,12.5) rectangle (13.3,15.5);
          \draw[black] (12,14) node {$u'$};
          \fill[pattern={Lines[angle=0,distance=3pt,yshift=1pt]},even odd rule] (18,6) rectangle (34,22) (24.7,12.5) rectangle (27.3,15.5);
          \draw[black] (26,14) node {$v\vphantom{u'}$};
          \draw[black,thick] (6,6) rectangle (18,22) (18,6) rectangle (34,22);
        }] (0.5,0.5) rectangle (39.5,27.5);
      \end{scope}
    \end{tikzpicture}
    \caption{Replacing $u$ by $u'$ in the configuration $x \sqcup w = x \sqcup (u \sqcup v)$.}
  \end{figure}

  \noindent Since $f(w) = f(w')$ implies that $E_X(w) = E_X(w')$, we obtain $\card{E_X(n_1,\dots,n_i+m_i,\dots,n_d)} \leq \card{E_X(n_1,\dots,n_i,\dots,n_d)} \cdot \card{E_X(n_1,\dots,m_i,\dots,n_d)}$ and conclude the proof.
\end{proof}

As a consequence of \cite[Theorem~1]{capobianco08:multivariate-fekete-lemma}, the extender entropy is actually an infimum and could be computed along any sequence of hyperrectangles that eventually fills~$\Z[d]$. In other words,
\[ h_{E}(X) = \inf_{n_1,\dots,n_d \in \N^d} \frac{\log\, \card{E_{X}(n_1,\dots,n_d)}}{n_1 \cdots n_d} = \inf_{n \in \N} \frac{\log\, \card{E_X(n)}}{n^d}.\]

\subsection{Extender entropy dimensions}\label{sec:ext:extender-entropy-dimension}
In~\autoref{sec:def:subshifts}, we mentioned two asymptotic aspects of the pattern complexity $P_X$: the topological entropy (which measures the exponential growth of $P_X$); and the entropy dimension (which measures the intermediate growth rate).

For the extender set sequence $(\card{E_X(n)})_{n \in \N}$, we suggest a similar definition. In the same way, extender entropies correspond to the ``topological entropy'' of the extender set sequence, we introduce the \emph{extender entropy dimension} to measure its intermediate growth:

\begin{defi}[Extender entropy dimension]
  Let $X$ be a $\Z[d]$ subshift. We define its upper and lower
  \defn{extender entropy dimensions} respectively as
  \[\upentdim_{h_{E}}(X) = \limsup_{n \to +\infty} \frac{\log(\log \card{E_{X}(n)})}{\log
      n} \]
  and
  \[\lowentdim_{h_{E}}(X) = \liminf_{n \to +\infty} \frac{\log(\log \card{E_{X}(n)})}{\log
      n} \]

  When both are equal, they are called the \defn{extender entropy dimension} of
  $X$.
\end{defi}

Intuitively, if the extender set sequence has asymptotic growth $\card{E_X(n)} \sim 2^{n^{\beta}}$, then $\beta \leq d$ is the extender entropy dimension. As is the case for the classical entropy dimension, the main purpose of the extender entropy dimension is to provide a measure of the intermediate growth of the extender set sequence, which becomes especially relevant in the case of systems with extender entropy zero.

\subsection{Some properties}
\cite{french_pavlov19_follow_predec_exten_entrop} proved several properties of extender entropies in the case of $\Z$ subshifts. We prove that these properties still hold for all $\Z[d]$ subshifts.

\begin{prop}
  For a subshift $X \subseteq \mathcal{A}^{\Z[d]}$, we have $h_E(X) \leq h(X)$.
\end{prop}
\begin{proof}
  Since there are more patterns than extender sets, we have $\card{E_X(n_1,\dots,n_d)} \leq \card{P_X(n_1,\dots,n_d)}$, and thus $h_E(X) \leq h(X)$.
\end{proof}

\begin{prop}
  For two subshifts $X,X' \subseteq \mathcal{A}^{\Z[d]}$, we have $h_E(X \times X') = h_E(X) + h_E(X')$; and $\upentdim_{h_E}(X \times X') = \max(\upentdim_{h_E}(X),\upentdim_{h_E}(X'))$ (resp $\lowentdim_{h_E}$\dots).
\end{prop}
\begin{proof}This follows from $E_{X \times X'}(w,w') = E_{X}(w) \times E_X(w')$ for arbitrary $w,w'$ over the same domain.
\end{proof}

\begin{thm}\label{thm:ext-conjugacy-invariants} On $\Z[d]$ subshifts:
  \begin{itemize}
  \item The extender entropy $h_{E}$ is a conjugacy invariant.
  \item If they are larger than $d-1$, the upper (resp. lower) extender entropy dimensions $\upentdim_{h_E}$ (resp. $\lowentdim_{h_E}$) are conjugacy invariants.
  \end{itemize}
\end{thm}

The classical topological entropy can be proved to be a conjugacy invariant, since topological entropy is (weakly) decreasing under factor map. Unfortunately, extender entropy is not monotonic under factor maps, as noticed in~\cite[Theorem 3.7]{french_pavlov19_follow_predec_exten_entrop}. Thus, we proceed differently.
\begin{proof}
  (\cite{french_pavlov19_follow_predec_exten_entrop} already proved this on $\Z$). Let $X \subseteq \mathcal{A}^{\Z[d]}$ and $Y \subseteq \mathcal{B}^{\Z[d]}$ be two conjugated subshifts by a bijective factor map $\varphi : X \to Y$ of biradius $r$. We prove that $h_E(X) = h_E(Y)$.

  For $D \subseteq \Z[d]$ a finite domain, denote by $\mathcal{I}_{r}(D)$ (resp. $\partial_{r}(D)$) the \emph{interior} (resp. \emph{frontier}) of $D$, \textit{i.e.}~the set of points in $D$ at distance less than (resp. larger than) $r$ from $\Z^d \setminus D$. Since $\varphi$ (resp.~$\varphi^{-1}$) has radius $r$, we denote the image by $\varphi(w)$ (resp.~$\varphi^{-1}(w)$) the pattern of domain $\mathcal{I}_{r}(D)$ that is the image of a pattern $w$ by $\varphi$.

  \begin{clm}\label{claim:extender-sets-factor-map}
    Let $D \subseteq \Z[d]$ be a finite domain and $w,w' \in \mathcal{A}^{D}$ be two patterns such that $\restr{w}{\partial_{4r}(D)} = \restr{w'}{\partial_{4r}(D)}$ and $E_X(\restr{w}{\mathcal{I}_{2r}(D)}) = E_X(\restr{w'}{\mathcal{I}_{2r}(D)})$; then $E_Y(\varphi(w)) = E_Y(\varphi(w'))$.
  \end{clm}
  Indeed, let $y$ be a configuration in $Y$ such that $\restr{y}{\mathcal{I}_{r}(D)} = \varphi(w)$. Since $\varphi^{-1}$ is bijective, there exists some $x$ such that $\varphi(x) = y$: since $\varphi^{-1}$ has radius~$r$, we deduce that $\restr{x}{\mathcal{I}_{2r}(D)} = \restr{w}{\mathcal{I}_{2r}(D)}$. Since $E_X(\restr{w}{\mathcal{I}_{2r}(D)}) = E_X(\restr{w'}{\mathcal{I}_{2r}(D)})$, we know that the configuration $x'$ defined by $x_{i}\!\!' = x_i$ if $i \notin \mathcal{I}_{2r}(D)$ and $x_{i}\!\!' = w_{i}\!'$ otherwise is valid in $X$. Finally, let us consider $y' = \varphi(x') \in Y$:
  \begin{itemize}
  \item Since $\varphi$ has radius $r$ and that $\restr{x}{\Z[d] \setminus \mathcal{I}_{4r}(D)} = \restr{x'}{\Z[d] \setminus \mathcal{I}_{4r}(D)}$, we have $\restr{y'}{\Z[d] \setminus \mathcal{I}_{3r}(D)} = \restr{y}{\Z[d] \setminus \mathcal{I}_{3r}(D)}$. In particular, since $\varphi$ has radius $r$ and $\restr{w}{\partial_{4r}(D)} = \restr{w'}{\partial_{4r}(D)}$,
    \begin{align*}
      \restr{y'}{\mathcal{I}_{r}(D) \setminus \mathcal{I}_{3r}(D)} & = \restr{y}{\mathcal{I}_{r}(D) \setminus \mathcal{I}_{3r}(D)} = \restr{\varphi(w)}{\mathcal{I}_{r}(D) \setminus \mathcal{I}_{3r}(D)} = \restr{\varphi(w')}{\mathcal{I}_{r}(D) \setminus \mathcal{I}_{3r}(D)} \\
      \restr{y'}{\Z[d] \setminus \mathcal{I}_{r}(D)} & = \restr{y}{\Z[d] \setminus \mathcal{I}_{r}(D)}.
    \end{align*}
  \item And since $\restr{x'}{\mathcal{I}_{2r}(D)} = \restr{w'}{\mathcal{I}_{2r}(D)}$, we have $\restr{y'}{\mathcal{I}_{3r}(D)} = \restr{\varphi(w')}{\mathcal{I}_{3r}(D)}$.
  \end{itemize}

  \noindent Gluing all these domains together, we obtain $\restr{y'}{\Z[d] \setminus \mathcal{I}_{r}(D)} = \restr{y}{\Z[d] \setminus \mathcal{I}_{r}(D)}$ and $\restr{y'}{\mathcal{I}_{r}(D)} = \varphi(w')$; so that ${E_Y(\varphi(w)) \subseteq E_Y(\varphi(w'))}$ (and by symmetry $E_Y(\varphi(w)) = E_Y(\varphi(w'))$).

  \medskip
  \noindent \textbf{End of the proof.} From the previous claim, we deduce that:
  \[ \card{E_Y(n_1-2r,\dots,n_d-2r)}  \leq 2^{\sum_{i=1}^d 2 \cdot 4r \cdot \prod_{j \neq i} n_j} \cdot \card{E_X(n_1 - 4r,\dots,n_d-4r)};\]
  and the symmetric bound exists when exchanging $X$ and $Y$. Completing the computations, we conclude that $h_E(X) = h_E(Y)$; and if $\upentdim_{h_E}(X) \geq d-1$ (resp.~$\lowentdim_{h_E}(X) \geq d-1$), then $\upentdim_{h_E}(X) = \upentdim_{h_E}(Y)$ (resp.~$\lowentdim_{h_E}(X) = \lowentdim_{h_E}(Y)$).
\end{proof}

Although classical entropy dimension is a conjugacy invariant (see~\cite[Lemma~2.1]{meyerovitch_growth_type_invariants}), we actually do not know whether the (lower, upper\dots) extender entropy dimension is a conjugacy invariant for values $\upentdim_{h_E} \leq d-1$. This would appear to require a completely novel proof, that would significantly differ from~\autoref{claim:extender-sets-factor-map} and \cite[Theorem~3.3]{french_pavlov19_follow_predec_exten_entrop}.

\subsection{Elementary constructions on extender sets}
\label{sec:ext:elementary-constructions}

We consider two constructions: the free lift and the semi-mirror constructions. These constructions are known to preserve the class of a subshift (SFT, sofic, effective), and are the first non-trivial examples realizing a large class of extender entropies in arbitrary dimensions.

\paragraph*{The free and periodic lifts}

We use this construction to generalize results on $\Z$ or $\Z^2$ subshifts to higher dimensions:
\begin{clm}
  \label{clm:ext-entropy-free-lift}
  For a subshift $X \subseteq \mathcal{A}^{\Z[d]}$, its free lift $\liftfree{X} \subseteq \mathcal{A}^{\Z[d+1]}$ satisfies $h_{E}(\liftfree{X}) = h_{E}(X)$.
\end{clm}
\begin{proof}
  Consider $\liftfree{X} \subseteq \mathcal{A}^{\Z[d+1]}$. Since each $d$-dimensional hyperplane of
  $\Z[d+1]$ contains an independent configuration, we have $\card{E_{\liftfree{X}}(n)} =
  \card{E_{X}(n)}^{n}$ and $h_E(\liftfree{X}) = h_E(X)$.
\end{proof}

\pagebreak
\begin{clm}\label{clm:ext-entropy-dim-lifts}
  Let $\mathcal{D} \in \{\entdim, \upentdim, \lowentdim\}$ and let $X$ be a
  $d$-dimensional subshift. Then:
  \begin{itemize}
  \item $\mathcal{D}(\lift{X}) = \mathcal{D}(X)$;
  \item $\mathcal{D}(\liftfree{X}) = \mathcal{D}(X) + 1$.
  \end{itemize}
\end{clm}
\begin{proof}
  Since $\card{E_{\lift{X}}(n)} = \card{E_{X}(n)}$, we have
  $\mathcal{D}(\lift{X}) = \mathcal{D}(X)$. For the second claim, having
  $\card{E_{\liftfree{X}}(n)} = \card{E_{X}(n)}^{n}$ implies that
  $\frac{\log (\log \card{E_{\liftfree{X}}(n)})}{\log n} = 1 + \frac{\log \log
    \card{E_{X}(n)}}{\log n}$.
\end{proof}

 % \vspace{-0.5\baselineskip}
\paragraph*{The (semi)-mirror construction}

\begin{clm}\label{clm:mirror}
  Let $Y$ be any $\Z[2]$ sofic subshift over an alphabet $\mathcal{A}$. There exists a $\Z[2]$ sofic subshift $Y_{\mathrm{mirror}}$ such that $h_E(Y_\mathrm{mirror}) = h(Y)$ ($= h(Y_{\mathrm{mirror}})$).
\end{clm}

A first idea to create one extender set per pattern of $Y$ is the \defn{mirror
  construction}: add a line of some special symbol $\ast$ to separate two
half-planes; the upper half-plane contains a half-configuration of $Y$, while
the lower half-plane contains its reflection by the line of $\ast$. As any two
patterns of $Y$ have distinct reflections, they generate different extender
sets: this results in a subshift $Y'$ satisfying $h_E(Y') = h(Y)$. Unfortunately,
$Y'$ is not always sofic, see for example~\cite[Proposition
57]{aubrun_barbieri_jeandel18_domino_probl_subshifts_groups}.

\begin{figure}[ht]
  \pgfmathdeclarerandomlist{MyRandomColors}{{TilingColorA}{TilingColorB}}
  \begin{subfigure}[t]{0.3\textwidth}
    \centering
    \begin{tikzpicture}[scale=0.22, decoration={random steps,segment length=6pt,amplitude=2pt}]
      \pgfmathsetseed{4}
      \draw[decorate,gridborder]  [path picture={
        \foreach \i in {0,...,14}{
          \foreach \j in {0,...,14}{
            \pgfmathrandomitem{\RandomColor}{MyRandomColors}
            \draw[draw=none,fill=\RandomColor] ($(\i,\j)+(0,7)$) rectangle ++(1,1);
            \draw[draw=none,fill=\RandomColor] ($(\i,-\j)+(0,7)$) rectangle ++(1,1);
          }
        }
        \fill[TilingHighlight] (0,7) rectangle ++(15,1);
        \foreach \i in {0.5,...,14.5} {
          \draw (\i,7.5) node[font=\scriptsize] {\textcolor{black}{$\ast$}};
        }
        \draw[opacity=.6] (0,0) grid (15,15);
      }]
      (0.5,0.5) -- (14.5,0.5) -- (14.5,14.5) -- (0.5,14.5) -- cycle;
    \end{tikzpicture}
    \subcaption{The (classical) mirror shift}
  \end{subfigure}
  \begin{subfigure}[t]{0.3\textwidth}
    \centering
    \begin{tikzpicture}[scale=0.22, decoration={random steps,segment length=6pt,amplitude=2pt}]
      \pgfmathsetseed{4}
      \draw[decorate,gridborder]  [path picture={
        \foreach \i in {0,...,14}{
          \foreach \j in {0,...,14}{
            \pgfmathrandomitem{\RandomColor}{MyRandomColors}
            \draw[draw=none,fill=\RandomColor] ($(\i,\j)+(0,7)$) rectangle ++(1,1);
            \ifnum \i=6
              \ifnum \j=4
                \draw[draw=none,fill=\RandomColor] ($(\i,-\j)+(0,7)$) rectangle ++(1,1);
              \fi
            \fi
          }
        }
        \fill[TilingHighlight] (0,7) rectangle ++(15,1);
        \foreach \i in {0.5,...,14.5} {
          \draw (\i,7.5) node[font=\scriptsize] {\textcolor{black}{$\ast$}};
        }
        \draw[opacity=.6] (0,0) grid (15,15);
      }]
      (0.5,0.5) -- (14.5,0.5) -- (14.5,14.5) -- (0.5,14.5) -- cycle;
    \end{tikzpicture}
    \subcaption{The semi-mirror shift}
  \end{subfigure}
    \begin{subfigure}[t]{0.3\textwidth}
    \centering
    \begin{tikzpicture}[scale=0.22, decoration={random steps,segment length=6pt,amplitude=2pt}]
      \pgfmathsetseed{4}
      \draw[decorate,gridborder]  [path picture={
        \foreach \i in {0,...,14}{
          \foreach \j in {0,...,14}{
            \pgfmathrandomitem{\RandomColor}{MyRandomColors}
            \draw[draw=none,fill=\RandomColor] ($(\i,\j)+(0,7)$) rectangle ++(1,1);
            \ifnum \i=6
              \ifnum \j=4
                \draw[draw=none,fill=\RandomColor] ($(\i,-\j)+(0,7)$) rectangle ++(1,1);
              \fi
            \fi
          }
        }
        \fill[TilingHighlight] (0,7) rectangle ++(15,1);
        \foreach \i in {0,...,5} {
          \draw[black,-{Latex[length=2pt]}] (\i.07,7.5) -- ++(.86,0);
        }
        \foreach \i in {6.5,...,10.5} {
          \draw (\i,7.5) node[font=\scriptsize] {\textcolor{black}{$\ast$}};
        }
        \foreach \i in {11,...,14} {
          \draw[black,{Latex[length=2pt]}-] (\i.07,7.5) -- ++(.86,0);
        }
        \draw[very thick,OrangeRed] (6.5,3.5) -- ++(4,4) -- ++(-4,4) -- (6.5,3.5);
        \draw[opacity=.6] (0,0) grid (15,15);
      }]
      (0.5,0.5) -- (14.5,0.5) -- (14.5,14.5) -- (0.5,14.5) -- cycle;
    \end{tikzpicture}
    \subcaption{SFT cover for the semi-mirror shift}
    \label{fig:sft-cover-mirror}
  \end{subfigure}
  \caption{Example configurations of the mirror and semi-mirror subshifts.}
\end{figure}

\enlargethispage{1.5\baselineskip}
To solve this non-soficity issue, the \defn{semi-mirror with large discrepancy}
from~\cite[Example~$5''$]{destombes_romashchenko22_resour_bound_kolmog_compl_provid}
reflects a single symbol instead of the whole upper-plane:

\begin{proof}
  For $\mathcal{A}' = \mathcal{A} \cup \{\square, \ast\}$, define
  $Y_{\mathrm{mirror}}$ over the alphabet~$\mathcal{A}'$ as follows:
  \begin{itemize}
  \item Symbols $\ast$ must be aligned in a row, and there is at most one such
    row per configuration.
  \item If a row of $\ast$ appears in a configuration $x$, then the lower
  half-plane contains at most one non-$\square$ position; and the upper
  half-plane must appear in a configuration of $Y$.
  \item If $x_{i,j} = \ast$ and $x_{i,j-k} \in \mathcal{A}$ for some $i \in
    \Z, j \in \Z, k \in \N$, then $x_{i,j+k} = x_{i,j-k}$. In other words, the
    only symbol of $\mathcal{A}$ in the lower half-plane must be the mirror of
    the same symbol in the upper half-plane, as reflected by the horizontal row
    of $\ast$ symbols.
  \end{itemize}

  \noindent Then $Y_{\mathrm{mirror}}$ is sofic by implementing reflection lines in an SFT cover (see~\autoref{fig:sft-cover-mirror}), and
  $h_E(Y_{\mathrm{mirror}}) = h(Y)$. Indeed, any two distinct patterns of $Y$ must
  appear in $Y_{\mathrm{mirror}}$ and have distinct extender sets, since they can
  have different reflections. More precisely, for any two distinct patterns $u, v$ of $Y$ with the same support $\interval{0,n-1}^{2}$, differing in position $(i, j) \in \interval{0,n-1}^{2}$, consider the following extension $y \in (\mathcal{A} \cup \{\square, \ast\})^{\mathbb{Z}^{2} \setminus \interval{0,n-1}^{2}}$:
  \begin{itemize}
  \item For any $(k, \ell)$ in the upper half-plane, that is, $\ell \geq 0$, $y_{k, \ell}$ is arbitrary in $\mathcal{A}$;
  \item For every $k \in \mathbb{Z}$, define $y_{k, -1} = \ast$, so that the ``mirror line'' is $y_{\mathbb{Z} \times \{-1\}}$;
  \item The reflected cell $y_{i, -j-2} = u_{i,j} \neq v_{i,j}$;
  \item For any other position $(k, \ell)$ below the reflection line, \textit{i.e.} $\ell < -1$, we have $y_{k, \ell} = \square$.
  \end{itemize}
  Then $y$ is a valid extension for $u$ but not for $v$.
\end{proof}

This construction shows that there exist subshifts with arbitrarily large extender entropy; and since every $\Pi_1$ real number is the topological entropy of some SFT, and thus sofic subshift~\cite{hochman10_charac_entrop_multid_shift_finit_type}, they can all be realized as the extender entropy of some sofic subshift. In particular, this further disproves the conjecture
from~\cite{kass_madden13_suffic_condit_non_sofic_higher_dimen_subsh} mentioned in the introduction.

\subsection{Decision problems on extender sets}\label{sec:ext:decidability-extender}

\subsubsection{Inclusion of extender sets}

Let us consider the following decision problem:

\smallskip
\begin{tabular}{|ll}
  \decproblem{Extender-inclusion}&\\
  \textbf{Input:}& An effective subshift $X \subseteq \mathcal{A}^{\Z[d]}$, and patterns $u, v \in \shiftlang{X}$,\\
  \textbf{Output:}&Whether $E_{X}(u) \subseteq E_{X}(v)$.\\
  \hline
\end{tabular}

\begin{prop}\label{prop:inclusion-ext-completeness}
  \decproblem{Extender-inclusion} is a $\Pi_{2}^{0}$-complete
  problem.
\end{prop}

\begin{proof}[Proof of inclusion]
  As $E_X(u) \subseteq E_X(v)$ if and only if $\forall  B \in \mathcal{A}^{\ast}, u \subpattern B
  \implies
  (B \not\in \shiftlang{X}
  \,\vee\,
  ((B \setminus u) \sqcup v) \in \shiftlang{X})$,
  we obtain that \decproblem{Extender-inclusion} is a $\Pi_{2}^{0}$ problem: indeed, for $X$ effective, deciding whether a pattern $w$ belongs in $\shiftlang{X}$ is a $\Pi^{0}_{1}$ problem.
\end{proof}

\noindent The $\Pi^{0}_{2}$-hardness of \decproblem{Extender-inclusion} is actually not necessary for our purposes, but can be proved easily:
\begin{proof}[Proof of $\Pi^{0}_{2}$-hardness]
  We reduce to a standard $\Pi^{0}_{2}$ problem~\cite[Theorem~4.3.2]{soare_turing_computability}:

  \smallskip
  \begin{tabular}{|ll}
    \decproblem{Total}&\\
    \textbf{Input:}& A deterministic Turing Machine $M$,\\
    \textbf{Output:}& Does $M$ halt on every input $w \in \{0,1\}^{\ast}$?\\
    \hline
  \end{tabular}

  \medskip Let $M$ be a deterministic Turing machine. We construct an effective subshift $X$ over the alphabet $\{\texttt{0}, \texttt{1}, \square\}$ as follows:
  \begin{itemize}
  \item Symbols $\texttt{0}$ and $\texttt{1}$ cannot appear together in a
    configuration. The symbol $\texttt{1}$ can only appear at most once in
    a configuration.
  \item If two symbols $\texttt{0}$ appear in a configuration at distance,
    say, $n > 0$, then the whole configuration is $n$-periodic;
  \item If $M$ halts on the $n$\textsuperscript{th} word of $\{0,1\}^{\ast}$ (for some fixed enumeration of $\{0,1\}^{\ast}$, then two symbols $\texttt{0}$ cannot appear at distance $n$ from one another.
  \end{itemize}
  As the rules above forbid an enumerable set of patterns, $X$ is an
  effective subshift.

  Finally,
  $E_X(\texttt{0}) \subseteq E_X(\texttt{1})$ if and
  only if $M$ is total.
  Indeed, the symbol $\texttt{0}$ can be extended either by semi-infinite
  lines of symbols $\square$, which also extend the symbol
  $\texttt{1}$; or by configurations containing $n$-periodic symbols
  $\texttt{0}$, which do \emph{not} extend the symbol $\texttt{1}$
  because of the first rule.  However, by the third rule,
  this $n$-periodic configuration does not exist if $M$ halts on the
  $n$\textsuperscript{th} word of $\{0,1\}^{\ast}$.
\end{proof}

\subsubsection{Computing the number of extender sets}

Let us determine the computational complexity of the problem of deciding whether ``$k \leq
\card{E_{X}(n)}$'', when given a subshift $X$, some size $n$ and some $k$.

\begin{lem}\label{lem:ext-formula-upper-bound}
  For an effective subshift $X$, the problem:

  \smallskip
  \begin{tabular}{|ll}
    \decproblem{Lower-Bound-Extender}&\\
    \textbf{Input:}& An effective subshift $X \subseteq \mathcal{A}^{\Z[d]}$, and integers $k,n \in \N$;\\
    \textbf{Output:}& Whether $k \leq \card{E_X(n)}$.\\
    \hline
  \end{tabular}

  \smallskip
  \noindent is a $\Sigma_{2}^{0}$-complete problem.
\end{lem}
\begin{proof}
  Notice that $k \leq \card{E_{X}(n)}$ is equivalent to the following:
  \[
    \bigvee_{v_{1}, \dots, v_{k} \in \shiftlang[n]{X}}
    \bigwedge_{1 \leq i < j \leq k} E_{X}(v_{i}) \neq E_{X}(v_{j}).
  \]
  Since $v_i \in \shiftlang[n]{X}$ is a $\Pi^{0}_{1} \subseteq \Sigma^{0}_{2}$ problem and that the class of $\Sigma^{0}_{2}$ problems is stable by finite disjunctions and conjunctions, we conclude from \autoref{prop:inclusion-ext-completeness} that \decproblem{Lower-Bound-Extender} is a $\Sigma^{0}_{2}$-problem.  For $\Sigma^{0}_{2}$-hardness, consider the shift $X$ from the proof of~\autoref{prop:inclusion-ext-completeness}: indeed, $3 \leq E_X(1)$ if and only if the Turing machine is not total.
\end{proof}

\section{Characterizations of extender entropies}

The topological entropies of multidimensional subshifts of finite type were classified in computational terms in~\cite{hochman10_charac_entrop_multid_shift_finit_type}:
\begin{thmC}[\cite{hochman10_charac_entrop_multid_shift_finit_type}]
  For any $d \geq 2$, the class of topological entropies of $\Z[d]$ SFTs is exactly $[0,+\infty) \cap \Pi_{1}$.
  The same holds for sofic and effective subshifts instead of SFTs.
\end{thmC}

This theorem can be decomposed into two statements: on the one hand, the entropy $h(X)$ of any effective $\Z[d]$ subshift is a $\Pi_1$ non-negative real number; on the other hand, for every real number $\alpha \in [0,+\infty) \cap \Pi_1$, there exists an SFT $X$ such that $h(X) = \alpha$.

In a similar fashion, this section provides computational characterizations of the \emph{extender} entropies of sofic and effective subshifts in the arithmetical hierarchy.

\subsection{Subshifts of finite type}

\begin{clm}\label{clm:ext-entropy-sft}
  Let $X \subseteq \mathcal{A}^{\Z[d]}$ be a subshift of finite type. Then $h_E(X) = 0$.
\end{clm}
\begin{proof}Let $\mathcal{F}$ be a finite family of forbidden patterns defining $X$, and let $r \in \N$ be such that (translations of) the patterns $w \in \mathcal{F}$ all fit within the domain $\interval{0,r-1}^d$. Since $X$ is of finite type, the extender set of a pattern $p$ in $X$ only depends on the border of its domain (taken with thickness $r$), and the number of extender sets of domain $\interval{0,n-1}^d$ is bounded by the number of possible patterns of the border $\partial_{r}\interval{0,n-1}^d$. In particular,
  \[ \log\, \card{E_X(n)} \leq 2dr \cdot \log \card{\mathcal{A}} \cdot n^{d-1} = O(n^{d-1}).  \]
  Taking the limit, we obtain $h_E(X) = 0$.
\end{proof}

\subsection{Effective subshifts}
\label{sec:entropy:effective}

\begin{restatethm}{thm:ext-entropy-effective}
  For $d \geq 1$, the set of extender entropies of $\Z[d]$ effective subshifts is exactly $[0,+\infty) \cap \Pi_3$.
\end{restatethm}

Similarly to topological entropies, this theorem can be decomposed into two statements:

\begin{clm}\label{clm:ext-entropy-upper-bound}
  For $X$ an effective subshift, $h_{E}(X) \in \Pi_{3}$.
\end{clm}

\begin{proof} For $X$ a fixed effective subshift, the sets $\{k \leq \card{E_{X}(n)}\}$ are uniformly $\Sigma^{0}_{2}$-computable sets by~\autoref{lem:ext-formula-upper-bound}.
This implies that $\frac{\log\, \card{E_{X}(n)}}{n^{d}}$ are uniform $\Sigma_{2}$ real numbers; and since we have
$h_{E}(X) = \inf_{n} \frac{\log\, \card{E_{X}(n)}}{n^d}$, we obtain
$h_E(X) \in \Pi_{3}$ as the infimum of $\Sigma_{2}$ real numbers.
\end{proof}

The rest of the section will focus on the converse statement. By \autoref{clm:ext-entropy-free-lift}, we reduce from arbitrary $\Z[d]$ to the one-dimensional case and are left with proving:
\begin{lem}\label{lem:ext-entropy-effective-realization}
  For any real number $\alpha \in [0,+\infty) \cap \Pi_3$, there exists an effective $\Z$ subshift $Z_{\alpha}$ such that $h_E(Z_{\alpha}) = \alpha$.
\end{lem}

In order to explicitly realize such a subshift $Z_{\alpha}$, we would like to have $\card{E_{Z_\alpha}(n)} \simeq 2^{\alpha n}$. To do so, we could create one extender set per pattern, and $2^{\alpha n}$ patterns of size $n$ (as the semi-mirror in~\autoref{sec:ext:elementary-constructions}); however, since effective subshifts have $\Pi_{1}$ entropies, this would not realize the whole class of $\Pi_{3}$ numbers.

Yet, realizing the right number of patterns is the main idea behind the proof that follows: we just do not blindly create one extender set per pattern, but only separate extender sets when some conditions are met.

\subsubsection{Preliminary: encoding integers with configurations   \texorpdfstring{$\encode{i}_{k}$}{<i>}}
\label{sec:entropy:effective:encoding-integer-in-shifts}

Before we begin our construction, we fix a way to encode integers in configurations: to encode the integer $i \in \N$, we use configurations where a symbol $\ast$ is $i$-periodic, and the rest is blank.

More formally, consider the alphabet $\mathcal{A}_\ast = \{ \ast, \blank \}$. Denote by $\encode{i}_{k_1}$ the $i$-periodic configuration
$\encode{i}_{k_1} = \sigma^{k_1} (\dots \blank \underbrace{\ast\, \blank \dots \blank \,\ast}_{i+1 \text{ symbols}} \blank \dots \blank \ast \blank \dots)$
properly defined as $(\encode{i}_{k_1})_p = \ast$ if and only if $p = k_1 \bmod i$. A configuration $\encode{i}_{k_1}$ is said to \defn{encode} the integer $i \in \N$. Let $X_\ast$ be the subshift generated by all the configurations $\encode{i}_{k_1}$ for $i \in \N$ and $k_1 \leq i$:
\[ X_\ast = \bigcup_{i \in \N} \{ \encode{i}_{k_1} \in \mathcal{A}_{\ast}^{\Z} \mid k_1 \leq i \} \cup \encode{\infty} \]
where $\encode{\infty} = \{ x \in \mathcal{A}_\ast^{\Z} \mid |x|_{\ast} \leq 1 \}$ is the set of configurations having at most one symbol~$\ast$.  The configurations of $\encode{\infty}$ are said to be \defn{degenerate}, and they appear when taking the closure of all $\encode{i}_{k_1}$.

\subsubsection{Preliminary: Toeplitz density in periodic configurations}\label{sec:entropy:effective:toeplitz}

Our construction will also need to build configurations with a controlled density of symbols, \textit{i.e.}~configurations on $\{0,1\}$ where the number of symbols $1$ in large patterns converges to some value: for some fixed $\alpha$, we want to build configurations $x \in \{0,1\}^{\Z}$ such that $\lim_{n \to +\infty} \frac{1}{n} \cdot |x|_{\interval{0, n-1}}|_1 = \alpha$.
Several explicit constructions of such configurations and subshifts exist. We choose to work with \defn{Toeplitz sequences}.

\paragraph*{Toeplitz density words}

Consider the ruler sequence $T = 0 1 0 2 0 1 0 3 \dots$ defined by $T_n = \max \{m \in \N : 2^m \mid n+1 \}$ (see \cite[\href{https://oeis.org/A001511}{A001511}]{oeis}). For a given $u = (u_n)_{n \in \N} \in \{0,1\}^{\N}$, we consider its Toeplitzification $T(u) \in \{0,1\}^\N$ defined as $T(u)_n = u_{T_n}$ for $n \in \N$.

In particular, for $\beta \in [0,1]$ a real number and $(\beta_n)_{n \in \N}$ its proper binary expansion, we consider the word $T(\beta) = (\beta_{T_n})_{n \in \N} = \beta_{0} \, \beta_{1} \, \beta_{0} \, \beta_{2} \, \beta_{0} \, \beta_{1}\dots$. Denoting by $|w|_1$ the numbers of letters $1$ in a binary word $w \in \{0,1\}^*$ and by $|w|$ its length, we have:
\begin{clm}\label{clm:toeplitz-density}
  For $\beta \in [0,1]$ and $w \subpattern T(\beta)$ a factor of $T(\beta)$, we have $|w|_1 = \beta \cdot |w| + O(1)$.
\end{clm}

\paragraph*{Toeplitz density in periodic configurations}

For our specific construction, let $\alpha \in [0,1]$ and $i \in \N$, and consider the subshift $T_{\leq \alpha,i}$ composed of $i$-periodic configurations made of truncated Toeplitz words:
\[ T_{\leq \alpha,i} = \{ x \in \{0,1\}^{\Z} \mid \exists \beta \leq \alpha, \exists k_1 \in \interval{0,i-1}, \forall p \in \Z, \,x_p = T(\beta)_{(p+k_1 \bmod i)}\} \]
We denote $T(\beta,i)_{k_1} \in \{0,1\}^{\Z}$ the configuration defined by $(T(\beta,i)_{k_1})_p = T(\beta)_{(p+k_1 \bmod i)}$ for $p \in \Z$.
Notice that, for any $\alpha \in [0,1]$, $i \in \N$ and $n \in \N$, there are $|\shiftlang[n]{T_{\leq \alpha,i}}| = 2^{\log(\min(i,n)) + O(1)}\cdot O(\min(i,n))$ factors of length $n$ in $T_{\leq \alpha,i}$.

\begin{clm}\label{clm:toeplitz-effective}
  Let $\alpha \in [0,1] \cap \Pi_1$. Then $T_{\leq \alpha,i}$ is an effective subshift, and a family of forbidden patterns realizing $T_{\leq \alpha,i}$ can be computably enumerated from $\alpha$.
\end{clm}
\begin{proof}
  Consider $\alpha \in \Pi_1$: the set $\{ r \in Q \mid r > \alpha\}$ is computably enumerable. Thus, the following family $\mathcal{F}$ of forbidden patterns that realizes $T_{\leq \alpha,i}$ is recursively enumerable: forbid finite patterns that are either not $i$-periodic, or do not respect the structure of the ruler sequence in an $i$-period; and inside an $i$-period, forbid patterns $r_{T_{0}} \, r_{T_1} \, r_{T_0} \ldots \in \{0,1\}^i$ that encode the finite expansion of a rational $r = \sum_{k=0}^{\log i} r_k 2^{-(k+1)}$ if $r$ is such that $r > \alpha$.
\end{proof}

\subsubsection{Construction: the effective \texorpdfstring{$\Z$}{Z} subshift
  \texorpdfstring{$Z_{\alpha}$}{Z\_alpha}}
\label{sec:entropy:effective:final-construction}

Let us now begin the construction to prove
\autoref{thm:ext-entropy-effective}. Let $\alpha \in \Pi_{3}$ be a
positive real number, $\alpha = \inf_{i} \sup_{j} \alpha_{i,j}$ for some
sequence $(\alpha_{i,j})$ of uniformly $\Pi_1$ real numbers. We can assume
$\alpha \leq 1$ since extender entropy is additive under cartesian products, and
using \autoref{lem:arith-hierarchy-monotonicity}, we can assume that for all $i$,
$(\alpha_{i,j})_{j \in \N}$ is increasing towards some $\alpha_i$; and
the sequence $(\alpha_{i})_{i \in \N}$ is decreasing towards~$\alpha$.

\paragraph*{Auxiliary subshift {$Z_{\alpha}'$}}
We create an auxiliary subshift $Z_{\alpha}'$ on the following three layers:
\begin{enumerate}
\item \defn{First layer $L_{1}$:} We take $L_1= X_\ast$ to encode integers
  $i \in \N$. Intuitively, $i$ will denote which $\Sigma_2$ number $\alpha_i$ is
  approximated in the configuration.

\item \defn{Second layer $L_{2}$:} We also set $L_2 = X_\ast$ to encode integers
  $j \in \N$, $j \geq i$. Intuitively, $j$ will denote which $\Pi_1$ number
  $\alpha_{i,j}$ is approximated in the configuration.
\item \defn{Density layer $L_{d}$:} We define the density layer as $L_d = \{0,1\}^{\Z}$. Whenever the first two layers are non-degenerate, this layer will be restricted to densities $\lesssim \alpha_{i,j}$. Since the real numbers $\alpha_{i,j}$ are $\Pi_1$, the subshifts $T_{\leq \alpha_{i,j},i}$ are effective from the numbers $\alpha_{i,j}$.
\end{enumerate}

\noindent Define now $Z_{\alpha}'$ as:
\begin{align*}
  Z_{\alpha}' &= \Big\{ (z^{(1)}, z^{(2)}, z^{(d)}) \in L_1 \times L_2 \times L_d \mid  z^{(2)} \in \encode{\infty} \Big\} \\
              & \hspace{0.2cm} \cup \bigcup_{i \in \N} \bigcup_{j \geq i} \Big\{ (z^{(1)}, z^{(2)}, z^{(d)}) \in L_1 \times L_2 \times L_d \mid \exists k_1,k_2 \in \N,\\[-10pt]
              & \hspace{3cm} z^{(1)} = \encode{i}_{k_1},\, z^{(2)} = \encode{j}_{k_2}\,\text{and } \exists \beta \leq \alpha_{i,j},\, z^{(d)} = T(\beta,i)_{k_1} \Big\}
\end{align*}
Notice that, if $z^{(1)} = \encode{\infty}$, then $z^{(2)} = \encode{\infty}$.

\begin{figure}[ht]
  \centering
  \begin{tikzpicture}
    % origin
    \draw[very thick,opacity=.5,red] (3.8, 0.9) --++(0, -0.8);
    % <i>
    \draw (-0.4, 0.8) --++ (9.6, 0);
    \foreach \i in {0,3,...,10}{ % problem: only 14 cells when
      % 15 are needed (assuming 0.2 gap)
      \node at (\i, 0.7) {$\ast$};
    }
    % <j>
    \draw (-0.4, 0.6) --++ (9.6, 0);
    \foreach \i in {0.4, 4, 7.6}{
      \node at (\i, 0.5) {$\ast$};
    }
    % w Toeplitz, w = 101110101011101
    % encoding of 1010
    \draw (-0.4, 0.4) --++ (9.6, 0);
    \foreach \k in {0, 1, 2}{
      \foreach \i/\b in {0/1, 1/0, 2/1, 3/1, 4/1, 5/0, 6/1, 7/0,
        8/1, 9/0, 10/1, 11/1, 12/1, 13/0, 14/1}{
        \node at (\i*0.2 + \k*3.0, 0.3) {{\tiny \b}};
      }
    }
  \end{tikzpicture}
  \caption{A proper configuration $z = (\encode{15}_{11}, \encode{18}_{1}, T(\frac{5}{8},15)_{10})$: between pairs of adjacent $\ast$ symbols on the first layer, the density layer $L_{d}$ contains a Toeplitz encoding of $\overline{.1010}^{2} = \frac{5}{8}$. The vertical red line indicates the origin.}
  \label{fig:example-configurations-effective}
\end{figure}

\begin{clm}\label{clm:effective-effectivity}
  The $\Z$ subshift $Z_{\alpha}'$ is an effective subshift.
\end{clm}
\begin{proof}
  Since the subshift $X_\ast$ is effective, the conditions on the first two layers $L_1$ and $L_2$ are straightforward to enforce. Furthermore, since the $\alpha_{i,j}$ are $\Pi_1$ real numbers enumerated by a single machine, by \autoref{clm:toeplitz-effective} we can obtain $Z_{\alpha}'$ as follows: a pattern $w = (w^{(1)},w^{(2)},w^{(d)}) \in \shiftlang{X_\ast} \times \shiftlang{X_\ast} \times \{0,1\}^n$ is forbidden whenever both $w^{(1)}$ and $w^{(2)}$ contain at least two symbols $\ast$ (so that $w^{(1)}$ encodes an integer $i \in \N$, $w^{(2)}$ encodes an integer $j \geq i$) and $w^{(d)}$ contains a pattern forbidden in $T_{\leq \alpha_{i,j},i}$.
\end{proof}

A configuration $z = (\encode{i}_{k_1},\encode{j}_{k_2},T(\beta,i)_{k_1}) \in Z_{\alpha}'$ is said to be \defn{proper}. A configuration $z = (z^{(1)},z^{(2)},\cdot\,) \in Z_{\alpha}'$ with $z^{(2)} \in \encode{\infty}$ is said to be \defn{degenerate}. Thus, we separate patterns into two categories: whenever $w \in \shiftlang{Z_{\alpha}'}$ only appears in degenerate configurations, we call it a \defn{degenerate pattern}; if $w$ can appear in a proper configuration, we call it a \defn{proper pattern}.

\medskip
On the one hand, degenerate patterns of $Z_{\alpha}'$ do not contribute much to the number of extender sets, despite being exponentially many:

\begin{clm}\label{clm:effective-aux-counting-degenerate}
  Let $n \in \N$, and consider $D_E(n) = \{ E_{Z_\alpha'}(w) \mid w \in \shiftlang[n]{Z_{\alpha}'} \text{ degenerate}\}$, the set of extender sets of degenerate patterns of size $n$. Then $|D_E(n)| = O(n^3)$.
\end{clm}
\begin{proof}
  Let $u,v \in \shiftlang[n]{Z_{\alpha}'}$ be two degenerate patterns. Whenever $u^{(1)} = v^{(1)}$ and $u^{(2)} = v^{(2)}$, we have $E_{Z_\alpha'}(u) = E_{Z_\alpha'}(v)$ because the density layer of such patterns can be anything. Since at most a single symbol $\ast$ can appear on the second layer of degenerate patterns ($O(n)$~possibilities), by counting the number of admissible first layers ($O(n^2)$ possibilities) we deduce that $|D_E(n)| = O(n^3)$.
\end{proof}

On the other hand, all proper patterns of $Z_{\alpha}'$ have distinct extender sets:
\begin{clm}\label{clm:effective-aux-ext-proper}
  Let $u,v \in \shiftlang[n]{Z_{\alpha}'}$ be two distinct proper patterns. Then $E_{Z_\alpha'}(u) \neq E_{Z_\alpha'}(v)$.
\end{clm}
\begin{proof}
  Let $u,v \in \shiftlang[n]{Z_{\alpha}'}$ be two distinct proper patterns. Since $u$ is proper, there exists a proper configuration $z = (\encode{i}_{k_1},\encode{j}_{k_2},z^{(d)}) \in Z_{\alpha}'$ such that ${z|_{\interval{0,n-1}} = u}$. By definition, $z$ is periodic of period $ij$: thus, $u = z|_{\interval{0,n-1}}$ is equal to, say, $z|_{\interval{n i j, n i j + n - 1}}$; and $z|_{\Z \setminus \interval{0,n-1}}$ cannot extend the pattern $v$ since $v \neq u$.
\end{proof}

However, there are only polynomially many distinct proper patterns of a given size in~$Z_{\alpha}'$. The next section will nevertheless create a subshift $Z_{\alpha}$ with the correct (exponential) amount of proper patterns, thanks to the following remark:
\begin{clm}\label{clm:effective-aux-density}
  For $i \in \N$, let $z \in Z_{\alpha}'$ be a proper configuration such that $z^{(1)} = \encode{i}_{k_1}$: then for any $n \leq i$ and any subword $v$ of length $n$ appearing in $z$, the density layer $v^{(d)}$ of $v$ contains at most $\alpha_n \cdot n + O(1)$ symbols $1$.
\end{clm}
\begin{proof}
  This follows from \autoref{clm:toeplitz-density} and the monotonicities of the sequences $(\alpha_{i,j})_{i,j \in \N^2}$.
\end{proof}

\paragraph*{Free bits in the subshift {$Z_{\alpha}$}}

To create the desired exponential number of extender sets, we create the subshift $Z_\alpha$ by adding \defn{free bits} (\textit{i.e.}~symbols \emph{head} or \emph{tail}) on top of the symbols $1$ of the density layer. Informally, if there were $\beta \cdot i + O(1)$ symbols $1$ in an $i$-period of the density layer in $Z_{\alpha}'$, adding free bits on top of the symbols $1$ creates $2^{\beta \cdot i + O(1)}$ patterns in~$Z_{\alpha}$. Thus, we add a fourth layer to $Z_{\alpha}'$:

\begin{enumerate}
  \setcounter{enumi}{3}
\item \defn{Free layer $L_{f}$:} We define the free layer as $L_f = \{\freezero,\heads,\tails\}^{\Z}$. Given the synchronizing map $\pisync \colon \{\freezero,\heads,\tails\} \to \{0,1\}$ defined as $\pisync(\freezero) = 0$ and $\pisync(\heads) = \pisync(\tails) = 1$, we say that two configurations $z^{(d)} \in L_d$ and $z^{(f)} \in L_f$ are \defn{synchronized} if $\pisync(z^{(f)}) = z^{(d)}$.
\end{enumerate}

\noindent and we define $Z_\alpha$ as:
\begin{align*}
  Z_{\alpha} &= \Big\{ (z^{(1)}, z^{(2)}, z^{(d)},z^{(f)}) \in L_1 \times L_2 \times L_d \times L_f \mid z^{(2)} \in \encode{\infty} \text{ and } \pisync(z^{(f)}) = z^{(d)} \Big\} \\
             & \cup \bigcup_{i \in \N} \bigcup_{j \geq i} \Big\{ (z^{(1)}, z^{(2)}, z^{(d)}, z^{(f)}) \in L_1 \times L_2 \times L_d \times L_f \mid \exists k_1,k_2 \in \N,\\[-7pt]
             & \hspace{3cm} z^{(1)} = \encode{i}_{k_1},\, z^{(2)} = \encode{j}_{k_2}\, \pisync(z^{(f)}) = z^{(d)}, \\
  & \hspace{3cm} \exists \beta \leq \alpha_{i,j},\, z^{(d)} = T(\beta,i)_{k_1}\,\text{and } z^{(f)} \text{ is $i$-periodic }\Big\}.
\end{align*}

\begin{clm}
  The $\Z$ subshift $Z_{\alpha}$ is effective.
\end{clm}
\begin{proof}
  Inside the subshift $Z_{\alpha}'$, we additionally forbid the symbols $a = (a^{(1)},a^{(2)},a^{(d)},a^{(f)})$ such that $\pisync(a^{(f)}) \neq a^{(d)}$; and we forbid all the patterns $w = (w^{(1)},w^{(2)},w^{(d)},w^{(f)})$ for which $w^{(1)}$ and $w^{(2)}$ both contain at least two symbols $\ast$ (in which case, denote by $i$ the distance between two symbols $\ast$ in $w^{(1)}$), but $w^{(f)}$ is not $i$-periodic.
\end{proof}

We extend the terminology from $Z_\alpha'$ to $Z_\alpha$ and call \defn{proper} the configurations of $Z_\alpha$ that encode integers $i \in \N$ and $j \geq i$ on their first two layers, and \defn{degenerate} those who do not. Similarly, a pattern is \defn{proper} if it can be extended into a proper configuration, and \defn{degenerate} if it only extends into degenerate configurations.

Since the free layer is required to be $i$-periodic only in proper configurations, \autoref{clm:effective-aux-counting-degenerate} and \autoref{clm:effective-aux-ext-proper} both extend from $Z_\alpha'$ to $Z_\alpha$ by the very same arguments:
\begin{clm}\label{clm:effective-aux-to-final}\leavevmode
  \begin{itemize}
  \item For $n \in \N$, consider $D_E(n) = \{E_{Z_\alpha}(w) \mid w \in \shiftlang[n]{Z_\alpha} \text{ degenerate}\}$. Then $\card{D_E(n)} = O(n^3)$.
  \item Let $u,v \in \shiftlang[n]{Z_{\alpha}}$ be two distinct proper patterns. Then $E_{Z_\alpha}(u) \neq E_{Z_\alpha}(v)$.
  \end{itemize}
\end{clm}

\enlargethispage{1.4\baselineskip}
\begin{lem}\label{lem:effective-counting-ext}
  Let $P(n) = \{ w \in \shiftlang[n]{Z_{\alpha}} \mid w \text{ is proper}\}$. Then
  \[ 2^{n \cdot \alpha_n + O(1)} \leq \card{P(n)} \leq \mathrm{poly}(n) \cdot \sum_{i=1}^n 2^{\alpha_i \cdot i + O(1)}. \]
\end{lem}

\begin{proof}[Proof: lower bound.]
  Consider the patterns $w' = (\encode{n}_{0},\encode{j}_{0},T(\alpha_{n,j},n)_{0})|_{\interval{0,n-1}}$ in the auxiliary effective subshift $Z_\alpha'$ for some fixed $n \in \N$ and $j \geq n$: by \autoref{clm:toeplitz-density}, they admit $\alpha_{n,j} \cdot n + O(1)$ symbols $1$ on their density layer $w'^{(d)}$. Since $\lim_{j \to +\infty} \alpha_{n,j} = \alpha_n$, by taking $j \geq n$ large enough we obtain a proper pattern $w' \in \shiftlang[n]{Z_{\alpha}'}$ whose density layer $w'^{(d)}$ contains $\alpha_n \cdot n + O(1)$ symbols~$1$.

  When transposing these patterns to $Z_{\alpha}$, each symbol $1$ in $w'^{(d)}$ leads to two distinct free layers (respectively with $\heads$ or $\tails$ superimposed with this symbol $1$): we thus obtain $2^{\alpha_n \cdot n + O(1)}$ proper patterns by considering the patterns $w \in \shiftlang[n]{Z_\alpha}$ such that $\pi_{L_1 \times L_2 \times L_d}(w) = w'$.
\end{proof}
\begin{proof}[Proof: upper bound.]
  For $n \in \N$, we claim that the number of proper patterns $\card{P(n)}$ satisfies:
  \[ \card{P(n)} \leq \sum_{i=0}^{n} \sum_{k_1 = 0}^{i-1} \sum_{j=0}^{n} \sum_{k_2=0}^{j-1} O(i^2) \cdot 2^{\alpha_i \cdot i + O(1)} + \sum_{k_1 = 0}^{n} \sum_{k_2 = 0}^{n} O(n^2) \cdot 2^{\alpha_n \cdot n + O(1)}. \]
  Indeed, we first count the possible positions of symbols $\ast$ on Layers 1 and 2. For fixed integers $i \leq j$, $k_1$ and $k_2$, the amount of possible Toeplitz subwords of a given length is quadratic. And after fixing a configuration $z' = (\encode{i}_{k_1}, \encode{j}_{k_2}, z'^{(d)}) \in Z_{\alpha}'$ with some valid density layer $z'^{(d)}$, we obtain by \autoref{clm:effective-aux-density} that:
  \begin{itemize}
  \item If $i \leq n$, the number of symbols $1$ in an $i$-period of $z'^{(d)}$ is bounded by $\alpha_i \cdot i + O(1)$;
  \item If $i > n$, the number of symbols $1$ in an $n$-subpattern of $z'^{(d)}$ is bounded by $\alpha_n \cdot n + O(1)$.
  \end{itemize}
  Since every symbol $1$ on the density layer of an auxiliary configuration $z' \in Z'_{\alpha}$ results in two configurations in $Z_{\alpha}$, we obtain the claimed upper bound. In particular, we deduce that $\card{P(n)} \leq \mathrm{poly}(n) \cdot \sum_{i=1}^{n} 2^{\alpha_i \cdot i + O(1)}$.
\end{proof}

Combining \autoref{lem:effective-counting-ext} with \autoref{clm:effective-aux-to-final}, we deduce that the number of extender sets in $Z_{\alpha}$ satisfies $2^{n \cdot \alpha_n + O(1)} \leq \card{E_{Z_{\alpha}}(n)} \leq \mathrm{poly}(n) \cdot \sum_{i=1}^{n} 2^{\alpha_i \cdot i + O(1)}$. By taking the limit over $\alpha_n \to \alpha$ and going through some straightforward Cesàro-like $\varepsilon$-calculations, we deduce that $h_{E}(Z_{\alpha}) = \alpha$ and thus conclude the proof.

\subsection{Sofic subshifts}
\label{sec:entropy:sofic}

In the one-dimensional case $d=1$, sofic subshifts have a bounded number of extender sets \cite[Lemma~3.4]{ormes_pavlov16_exten_sets_multid_subsh}. Thus:
\begin{clm}\label{clm:ext-entropy-sofic-1d}
  Any sofic subshift $Y \subseteq \mathcal{A}^{\Z}$ satisfies $h_E(Y) = 0$.
\end{clm}

In dimension $d \geq 2$, we prove a very different picture:
\begin{restatethm}{thm:ext-entropy-sofic}
  For $d \geq 2$, the set of extender entropies of $\Z[d]$ sofic subshifts is exactly $[0,+\infty) \cap \Pi_3$.
\end{restatethm}

As sofic subshifts are effective, one inclusion follows from \autoref{clm:ext-entropy-upper-bound}. For the converse direction, we reduce from arbitrary $\Z[d]$ to $\Z[2]$ by \autoref{clm:ext-entropy-free-lift} and are left with realizing every non-negative $\Pi_3$ real number as the extender entropy of some $\Z[2]$ sofic subshift.

To do so, we will extend~\autoref{thm:ext-entropy-effective} to multidimensional sofic shifts. A first idea could be to replace $i$-periodic words on $\Z$ in the previous construction with $(i,i)$-periodic squares on~$\Z[2]$. Unfortunately, such a subshift cannot be sofic\footnote{The argument proving that the classical mirror subshift cannot be sofic still applies here: there would be $2^{O(i^2)}$ distinct $i \times i$ patterns, but only $2^{O(i)}$ borders in the SFT cover.}.
Yet, making configurations \emph{periodic} is not necessary to ensure that two proper patterns $u$ and $v$ have distinct extender sets: it is enough to have a configuration that \emph{witnesses} the difference between $u$ and $v$ (by extending one but not the other). This was already illustrated in the semi-mirror shift (see \autoref{sec:ext:elementary-constructions}): instead of mirroring the whole half-plane (which is not sofic), non-deterministically reflecting a single bit from the upper to the lower half-plane is actually enough, since each bit can be reflected individually in some configuration (see \autoref{fig:illustration-sofic-idea-vs-construction}).

\pagebreak
\begin{figure}[ht]
  \centering
  \pgfmathdeclarerandomlist{MyRandomColors}{{TilingColorA}{TilingColorB}}
  \begin{subfigure}{0.4\textwidth}
    \centering
    \begin{tikzpicture}[scale=0.20, decoration={random steps,segment length=6pt,amplitude=2pt}]
      \draw[decorate,gridborder]  [path picture={
        \pgfmathsetseed{6}
        % Draw configuration in rectangle
        \foreach \x in {0,...,3} {
          \foreach \y in {0,...,3}{
            \pgfmathrandomitem{\RandomColor}{MyRandomColors}
            \foreach \i in {0,4,...,18} {
              \foreach \j in {0,4,...,18} {
                \fill[\RandomColor] ($(\i,\j) + (\x,\y)$) rectangle ++(1,1);
              }}}}
        % Draw rectangles
        \foreach \i in {0.1,4.1,...,18.1} {
          \foreach \j in {0.1,4.1,...,18.1} {
            \fill[fill=TilingHighlight,opacity=0.7] (\i,\j) rectangle ++(3.8,3.8);
            \draw[pattern color=black,pattern=north west lines,opacity=0.7] (\i,\j) rectangle ++(3.8,3.8);     %\draw[black,thick,fill=TilingHighlight,opacity=0.8] (\i,\j) rectangle ++(3.9,3.9);
          }}
        % Draw columns of stars
        \foreach \i in {0.5,4.5,...,17.5} {
          \foreach \j in {0.5,...,17.5} {
            \draw (\i,\j) node[font=\tiny] {\textcolor{black}{$\ast$}};
          }}
        \draw[gridborder] (0,0) grid (18,18);
      }]
      (0.5,0.5) -- (17.5,0.5) -- (17.5,17.5) -- (0.5,17.5) -- cycle;
    \end{tikzpicture}
  \end{subfigure}
  \begin{subfigure}{0.4\textwidth}
    \centering
    \begin{tikzpicture}[scale=0.20, decoration={random steps,segment length=6pt,amplitude=2pt}]
      \draw[decorate,gridborder]  [path picture={
        \pgfmathsetseed{6}
        % Draw configuration in rectangle
        \foreach \x in {0,...,3} {
          \foreach \y in {0,...,3}{
            \foreach \i in {0,4,...,18} {
              \foreach \j in {0,4,...,18} {
                \pgfmathrandomitem{\RandomColorAux}{MyRandomColors}
                \pgfmathsetmacro{\RandomColor}{ifthenelse(\x==2,ifthenelse(\y==1,"TilingColorA","\RandomColorAux"),"\RandomColorAux")}
                \fill[\RandomColor,opacity=0.5] ($(\i,\j) + (\x,\y)$) rectangle ++(1,1);
              }}}}
        % Draw rectangles
        \foreach \i in {0.05,4.05,...,18.05} {
          \foreach \j in {0.05,4.05,...,18.05} {
            \draw[black,thick,opacity=0.8] (\i,\j) rectangle ++(3.9,3.9);
          }}
        % Draw periodic bit
        \foreach \i in {2.05,6.05,...,18.05} {
          \foreach \j in {1.05,5.05,...,18.05} {
            \fill[fill=TilingHighlight,opacity=0.9] (\i,\j) rectangle ++(0.9,0.9);
            \draw[pattern color=black,pattern=north west lines] (\i,\j) rectangle ++(0.9,0.9);
          }}
        % Draw columns of stars
        \foreach \i in {0.5,4.5,...,17.5} {
          \foreach \j in {0.5,...,17.5} {
            \draw (\i,\j) node[font=\tiny] {\textcolor{black}{$\ast$}};
          }}
        \draw[gridborder] (0,0) grid (18,18);
      }]
      (0.5,0.5) -- (17.5,0.5) -- (17.5,17.5) -- (0.5,17.5) -- cycle;
    \end{tikzpicture}
  \end{subfigure}

  \bigskip
  \begin{minipage}{0.8\textwidth}
    \centering
    \footnotesize The periodized area is highlighted in color $\textcolor{TilingHighlight!80!white}{\blacksquare}$ and hatched. To make the figure readable, symbols for free bits are $\{\textcolor{TilingColorA!50!white}{\blacksquare},\textcolor{TilingColorB!50!white}{\blacksquare}\}$ instead of $\{b,b'\}$.
  \end{minipage}

  \caption{Entirely periodic squares vs. one periodic free bit per $i \times i$~square.}
  \label{fig:illustration-sofic-idea-vs-construction}
\end{figure}

 % \vspace*{-2.5\baselineskip}
\subsubsection{Preliminary: marking offsets with configurations \texorpdfstring{$\Dencode{2i}_{m_1,m_2}$}{[2i]}}

In the construction, we will need to mark some positions $(m_1 + i\Z, m_2 + i\Z)$. To do so, let $A_m$ be the alphabet $A_m = \{\square,\redsquare\}$. Denote by $\Dencode{2i}_{m_1,m2}$ the $(2i,2i)$-periodic configuration formally defined as $(\Dencode{2i}_{m_1,m_2})_p = \redsquare$ if and only if $p = (m_1,m_2) \bmod (2i,2i)$. A symbol $\redsquare$ is called a \defn{marker}.

For a configuration $x = \Dencode{2i}_{m_1,m_2}$ with $(m_1,m_2) \in \interval{0,2i-1}^2$, we say that a position $p \in \Z[2]$ is \defn{marked} if $p \in (m_1 + i\Z, m_2 + i\Z)$. This lattice has unit cells of size $i \times i$ instead of $2i \times 2i$: this is voluntary. In particular, some marked positions $p \in \Z[2]$ satisfy $x_p = \square$. The reason for this will be apparent in \autoref{lem:sofic-counting-ext}: informally, this allows a $i \times i$ pattern without any $\redsquare$ to be extended in any configuration $\Dencode{2i}_{m_{1}, m_{2}}$ for $(m_{1}, m_{2}) \in \interval{0, 2i-1}^{2}$.

Considering the closure of all the configurations $\Dencode{2i}_{m_1,m_2}$, we define the grid subshift:
\[ \mathcal{G} = \bigcup_{i \in \N} \{ \Dencode{2i}_{m_1,m_2} \mid (m_1,m_2) \in \interval{0,2i-1}^2 \} \cup \Dencode{\infty}\]
where $\Dencode{\infty} = \{ x \in A_m^{\Z[2]} \mid |x|_{\redsquare} \leq 1\}$ is the set of configurations having at most one marker symbol $\redsquare$: these are the configurations that appear when taking the closure of all $\Dencode{2i}_{m_1,m_2}$.

\subsubsection{Construction: the sofic \texorpdfstring{$\Z[2]$}{Z\^{}2} subshift \texorpdfstring{$Y_{\alpha}$}{Y\_alpha}}\label{sec:entropy:sofic:final-construction}

We now prove \autoref{thm:ext-entropy-sofic}. With the notations introduced in the proof of~\autoref{thm:ext-entropy-effective}, we
fix $\alpha \in [0,1] \cap \Pi_3$ such that
${\alpha = \inf_i \sup_j \alpha_{i,j}}$ for $\alpha_{i,j}$ a computable sequence
of $\Pi_1$ real numbers (we assume the monotonicity properties given
by~\autoref{lem:arith-hierarchy-monotonicity}). We define a subshift $Y_{\alpha}$ on
the following five layers:
\begin{itemize}
\item \defn{Lifted layers:} We define the first three layers of $Y_{\alpha}$ as $\lift{L_{1}} \times \lift{L_{2}} \times \lift{L_d}$, where $L_{1}, L_{2}$ and $L_{d}$ are the three layers of the subshift $Z_{\alpha}'$ defined in the proof of \autoref{thm:ext-entropy-effective}.

\item \defn{Marker layer $L_{m}$:} We define $L_{m} = \mathcal{G}$ to mark positions $p \in (m_1 + i\Z,m_2 + i\Z)$.
\item \defn{Free layer $L_{f}$:} We also define the free layer by $L_f = \{\freezero,\heads,\tails\}^{\Z[2]}$.
\end{itemize}
and we define $Y_{\alpha}$ as (see~\autoref{fig:example-configurations-sofic} for an illustration):
\begin{align*}
  Y_{\alpha} &= \Big\{ (y^{(1)\uparrow}, y^{(2)\uparrow}, y^{(d)\uparrow},y^{(m)}, y^{(f)}) \in \lift{L_1} \times \lift{\encode{\infty}} \times \lift{L_d} \times L_m \times L_f \mid \pisync(y^{(f)}) = y^{(d)\uparrow}, \\[-4pt]
             & \hspace{2cm} \forall i \in \N, (\exists k_1 \in \N,\ y^{(1)} = \encode{i}_{k_1} \iff \exists m_1,m_2 \in \N,\ y^{(m)} = \Dencode{2i}_{m_1,m_2}) \Big\} \\
             & \hspace{0.2cm} \cup \bigcup_{i \in \N} \bigcup_{j \geq i} \Big\{ (y^{(1)\uparrow}, y^{(2)\uparrow}, y^{(d)\uparrow}, y^{(m)}, y^{(f)}) \in \lift{L_1} \times \lift{L_2} \times \lift{L_d} \times L_m \times L_f \mid \exists k_1,m_1,m_2,k_2\in \N, \\[-7pt]
             & \hspace{2cm} y^{(1)} = \encode{i}_{k_1},\, y^{(m)} = \Dencode{2i}_{m_1,m_2},\, y^{(2)} = \encode{j}_{k_2},\, \pisync(y^{(f)}) = y^{(d)\uparrow}, \\
             & \hspace{2cm} \exists \beta \leq \alpha_{i,j},\, y^{(d)} = T(\beta,i)_{k_1}\,\text{and } y^{(f)}|_{(m_1 + i\Z) \times (m_2 + i\Z)} \text{ is constant}\Big\}.
\end{align*}

\begin{figure}[ht]
  \centering
  \begin{tikzpicture}[decoration={random steps,segment length=6pt,amplitude=2pt}, scale=0.3]
    \draw[decorate,gridborder]  [path picture={%
      \foreach \i in {2,8,14} {
        \foreach \j in {1,7,13}
        \fill[draw=none, red!30] (\i,\j) rectangle ++(1,1);
      }
      \foreach \i in {2.5,5.5,8.5,11.5,14.5} {
        \foreach \j in {1.5,4.5,7.5,10.5,13.5} {
          \draw (\i,\j) node {\textcolor{black}{\scriptsize $b$}};
        }
      }
      \foreach \i in {0.5,3.5,6.5,9.5,12.5} {
        \foreach \j in {0.5,1.5,...,14.5} {
          \draw (\i,\j) node {\textcolor{black!60}{\scriptsize $\ast$}};
        }
      }
      \draw[gridborder!50] (0,0) grid (15,15);
      \draw (0,0) rectangle (1,1);
    }] (0.5,0.5) -- (14.5,0.5) -- (14.5,14.5) -- (0.5,14.5) -- cycle;
  \end{tikzpicture}
  \caption{Projection of a proper configuration $y$ on
    $\lift{L_{1}} \times L_{m}$, with $y^{(1)} = \encode{3}_{k_{1}}$ for some
    $k_{1} \geq 0$. The second layer $\lift{L_{2}}$ is not represented, but
      should be periodic of period $j \geq 3$ here. The symbols
      $\textcolor{black!60}{\ast}$ are on $\lift{L_{1}}$, the symbols
      $\redsquare$ on $L_{m}$. The symbols $b$ are the only represented bits in
      $L_{f}$: they are exactly those in marked positions, and hence should all
      be equal to some $b \in \{\freezero,\heads,\tails\}$. All the other bits of $L_{f}$
      (not drawn here) are free.}
  \label{fig:example-configurations-sofic}
\end{figure}

In the subshift $Y_{\alpha}$, the lifted layers play the same role as they did in the proof of \autoref{sec:entropy:effective}, and ensure that we have the correct density of symbols $1$ in some configurations, and the free layer still adds free bits exactly on top of these $1$-symbols. Then, we further impose that a configuration $y \in Y_{\alpha}$ has a proper first layer, that is, an $i$-periodic configuration of period $\ast \blank^{i-1}$, if and only if it has a marker layer $y^{(m)}$ of period $2i$. If moreover the second layer $y^{(2)}$ is also periodic (of period $\ast \blank^{j}$ for some $j \geq i$), then we further impose that all the free bits in marked positions (according to the marker layer $y^{(m)}$) are equal.

Extending the terminology from $Z_{\alpha}$ to $Y_{\alpha}$, we call \defn{proper} the configurations of $Y_{\alpha}$ that encode integers $i \in \N$ and $j \geq i$ on their first two layers, and \defn{degenerate} those which do not. In particular, such a proper configuration has a periodic marker layer, and so a sublattice of mesh $(i, i)$ of constant free bits. Additionally we say that a pattern is \defn{proper} if it can be extended into a proper configuration, and \defn{degenerate} otherwise. We say that two proper patterns $u,v \in \shiftlang[n]{Y_\alpha}$ are \defn{similar} if they are equal on their first four layers (\textit{i.e.}~$\pi_{\lift{L_1} \times \lift{L_2} \times \lift{L_d} \times L_m}(u) = \pi_{\lift{L_1} \times \lift{L_2} \times \lift{L_d} \times L_m}(v)$).
\begin{clm}\label{clm:sofic-ext-proper}
  Two similar proper patterns $u,v \in \shiftlang[n]{Y_\alpha}$ have distinct extender sets if and only if there exists a proper configuration $y$ that extends $u$ and that \emph{marks} a position $p \in \interval{0,n-1}^2$ such that $u^{(f)}_{p} \neq v^{(f)}_{p}$.
\end{clm}

We would very much like an analog of \autoref{clm:effective-aux-to-final}: unfortunately,
not all proper patterns generate distinct extender sets. Indeed, by the previous
claim, similar proper patterns generate distinct extender sets only when the
positions at which they differ can be \emph{marked} by an extending
configuration (this depends on the relative position of an $n \times n$ window
covering the four quadrants of a $2i \times 2i$ square, etc\dots), as
illustrated in \autoref{fig:example-similar-patterns-sofic}.

\begin{figure}[ht]
  \centering
  \begin{subfigure}[b]{0.4\linewidth}
    \centering
    \begin{tikzpicture}[scale=0.4]
      \node at (-2, 3.5) {$u$};
      \node at (-2, -6.5) {$v$};
      \begin{scope}
        \draw[red, thick] (0, 0) rectangle (7, 7);
        \clip (0, 0) rectangle (7, 7);

        \foreach \j in {0.5,...,6.5}{
          \node at (2.5,\j) {\textcolor{black!80}{\scriptsize $\ast$}};
          \node at (6.5,\j) {\textcolor{black!80}{\scriptsize $\ast$}};
        }

        \foreach \p in {(1, 1), (1,5), (5, 1), (5, 5)}{
          \fill[pattern={Lines[angle=-45,distance={3pt}, line width=1pt]}, pattern color=blue!50]
          \p rectangle ++(1, 1);
        }

        \foreach \p in {(1, 3), (5, 3)}{
          \fill[pattern={Lines[angle=-45,distance={3pt}, line width=1pt]},
          pattern color=yellow]
          \p rectangle ++(1, 1);
        }

        \fill[pattern={Lines[angle=-45,distance={3pt}, line width=1pt]},
        pattern color=purple]
        (3, 3) rectangle ++(1, 1);

        % Teal
        \node at (1.5, 1.5) {\scriptsize $\heads$};
        \node at (5.5, 5.5) {\scriptsize $\tails$};

        % Yellow
        \node at (1.5, 3.5) {\scriptsize $\heads$};
        \node at (5.5, 3.5) {\scriptsize $\heads$};

        % Lavander
        \node at (3.5, 3.5) {\scriptsize $\heads$};

        \draw[gridborder] (0, 0) grid (7, 7);
        \draw[black] (-2, -2) grid[step=4, shift={(2, 2)}] (7, 7);
      \end{scope}

      \begin{scope}[shift={(0, -10)}]
        \draw[blue!50, thick] (0, 0) rectangle (7, 7);
        \clip (0, 0) rectangle (7, 7);
        \foreach \j in {0.5,...,6.5}{
          \node at (2.5,\j) {\textcolor{black!80}{\scriptsize $\ast$}};
          \node at (6.5,\j) {\textcolor{black!80}{\scriptsize $\ast$}};
        }

        \foreach \p in {(1, 1), (1,5), (5, 1), (5, 5)}{
          \fill[pattern={Lines[angle=-45,distance={3pt}, line width=1pt]}, pattern color=blue!50]
          \p rectangle ++(1, 1);
        }

        \foreach \p in {(1, 3), (5, 3)}{
          \fill[pattern={Lines[angle=-45,distance={3pt}, line width=1pt]},
          pattern color=yellow]
          \p rectangle ++(1, 1);
        }

        \fill[pattern={Lines[angle=-45,distance={3pt}, line width=1pt]},
        pattern color=purple]
        (3, 3) rectangle ++(1, 1);

        % Teal
        \node at (5.5, 1.5) {\scriptsize $\heads$};
        \node at (5.5, 5.5) {\scriptsize $\heads$};

        % Yellow
        \node at (1.5, 3.5) {\scriptsize $\tails$};
        \node at (5.5, 3.5) {\scriptsize $\tails$};

        % Lavander
        \node at (3.5, 3.5) {\scriptsize $\tails$};

        \draw[gridborder] (0, 0) grid (7, 7);
        \draw[black] (-2, -2) grid[step=4, shift={(2, 2)}] (7, 7);
      \end{scope}
    \end{tikzpicture}
  \end{subfigure}
  \hfill\vrule\hfill
  \begin{subfigure}[b]{0.5\linewidth}
    \begin{tikzpicture}[scale=0.4, decoration={random steps,segment length=6pt,amplitude=2pt}]
      \draw[decorate,gridborder]  [path picture={%
        \foreach \i in {-1.5, 2.5,...,17}{%
          \foreach \j in {0.5,...,16.5}{%
            \node at (\i,\j) {\textcolor{black!80}{\scriptsize $\ast$}};%
          }
        }
        \foreach \i in {1, 9, 17}{%
          \foreach \j in {3, 11}%
          \fill[red!30] (\i, \j) rectangle ++(1, 1);%
        }
        % Yellow
        \foreach \i in {1,5,...,17}{%
          \foreach \j in {-1,3,...,17}{%
            \node at (\i+0.5, \j+0.5) {\textcolor{black!80}{\scriptsize $\heads$}};%
          }
        }
        \begin{scope}[shift={(4, 4)}]%
          % Teal
          \node at (1.5, 1.5) {\textcolor{black!80}{\scriptsize $\heads$}};%
          \node at (5.5, 5.5) {\textcolor{black!80}{\scriptsize $\tails$}};%
          % Lavander
          \node at (3.5, 3.5) {\textcolor{black!80}{\scriptsize $\heads$}};%
        \end{scope}
        \draw[gridborder] (0, 0) grid (17, 17);
        \draw[black] (-2, -2) grid[step=4, shift={(2, 2)}] (17, 17);
        \draw[red, thick] (4, 4) rectangle ++(7, 7);
      }] (0.5,0.5) -- (15.5,0.5) -- (15.5,15.5) -- (0.5,15.5) -- cycle;

    \end{tikzpicture}
  \end{subfigure}
  \caption{Example of similar patterns. Only some of the free bits are depicted.
    The dashed patterns are simply decorations: although $u, v$ differ on some
    blue position, those positions do not satisfy the hypothesis
    of~\autoref{clm:sofic-ext-proper} as they cannot be marked in neither $u$
    nor $v$. Yellow positions can be marked in both, and $u, v$ differ on them,
    so they satisfy the hypothesis of~\autoref{clm:sofic-ext-proper}. The red
    position obviously satisfies the hypothesis. On the right, an example of
    configuration extending $u$ but not $v$.}
  \label{fig:example-similar-patterns-sofic}
\end{figure}

Yet, we do not need precise considerations to count the number of extender sets,
and simply prove the following bounds:

\enlargethispage{\baselineskip}
\begin{lem}\label{lem:sofic-counting-ext}
  Let $P_E(n) = \{ E_{Y_\alpha}(w) \in \shiftlang[n]{Y_\alpha} \mid w \text{ is proper}\}$. Then
  \[ 2^{\alpha_n \cdot n^2 + O(n)} \leq \card{P_{E}(n)} \leq \mathrm{poly}(n) \cdot \sum_{i=0}^n 2^{\alpha_i \cdot i^2 + O(i)}.\]
\end{lem}
\begin{proof}[Proof: lower bound.]
  For $j \geq n$, consider the set $J_{j} = \{ y \in Y_{\alpha} \mid {y^{(1)} =
  \encode{n}_0}, {y^{(2)} = \encode{j}_0},\linebreak[3] {y^{(d)} = T(\alpha_{n,j},n)}\}$. The
  number of symbols $1$ in an $(n \times n)$-period in the density layer of such configurations is $\alpha_{n,j} \cdot n^2 + O(n)$ by \autoref{clm:toeplitz-density}. Since $\alpha_{n,j} \to \alpha_n$, by taking $j \geq n$ large enough we obtain a set $J = J_{j}$ of proper configurations $y$ whose density layer $y^{(d)}$ contains $\alpha_n \cdot n^2 + O(n)$ symbols $1$ in an $(n \times n)$-period.

  \pagebreak
  Considering the free layer of such patterns, there are at least $2^{\alpha_n \cdot n^2 + O(n)}$ distinct patterns in the finite set $W = \{ y|_{\interval{0,n-1}^2} \mid y \in J_j \text{ and } y^{(m)}|_{\interval{0,n-1}^2} = \square^{\interval{0,n-1}^2}\}$, and we claim that they all generate distinct extender sets. Indeed, for any two distinct patterns $u,v \in W$, there exists a position $p \in \interval{0,n-1}^2$ such that $u^{(f)}_{p} \neq v^{(f)}_{p}$; and there exists a configuration $y \in J_j$ that extends~$u$ with~$y^{(m)} = \Dencode{2n}_{p + (n,n)}$: in particular, $y$ marks the position $p$.\footnote{Markers were chosen to be $(2i,2i)$-periodic for this reason: we need to be able to mark a position $p \in \interval{0,i-1}^2$ in a configuration without seeing a marker in the square $\interval{0,i-1}^2$.} By \autoref{clm:sofic-ext-proper}, we obtain $E_{Y_\alpha}(u) \neq E_{Y_\alpha}(v)$. This proves that $\card{P_{E}(n)} \geq 2^{\alpha_n \cdot n^2 + O(n)}$.
\end{proof}
\begin{proof}[Proof: upper bound.]
  We proceed as with the $\Z$ effective subshift $Z_\alpha$: to bound the cardinality of $P_E(n)$, we consider the restrictions $w = y|_{\interval{0,n-1}^2}$ for $y$ ranging in the proper configurations of $Y_{\alpha}$ (for all values of $\encode{i}_{k_1}$, $\encode{j}_{k_2}$, $T(\beta,i)$ and $\Dencode{2i}_{m_1,m_2})$, and count free layers by \autoref{clm:effective-aux-density}:
  \begin{itemize}
  \item If $i \leq n$, an $i \times i$ square of the density layer $w^{(d)}$ contains less than $\alpha_i \cdot i^2 + O(i)$ symbols~$1$.
  \item If $i > n$, the density layer $w^{(d)}$ contains less than $\alpha_n \cdot n^2 + O(n)$ symbols $1$.
  \end{itemize}

  Finally, when summing over all these cases, we overestimate the number of extender sets generated by the free layer by assuming that each position $p \in \interval{0,i-1}^2$ containing a symbol $1$ on the density layer can be marked by a proper configuration $y$ extending the pattern (while only a subset of such positions can be marked):
\begin{align*}
    \card{P_{E}(n)} & \leq \sum_{i = 1}^{n} \sum_{k_1 = 0}^{i-1} \sum_{j=1}^{n} \sum_{k_2=0}^{j-1} O(i^4) \cdot 2^{\alpha_i \cdot i^2 + O(i)} + \sum_{k_1=0}^{n} \sum_{k_2=0}^{n} O(n^4) \cdot 2^{\alpha_n \cdot n^2 + O(n)} \\
    & \leq \mathrm{poly}(n) \cdot \sum_{i=1}^{n} 2^{\alpha_i \cdot i^2 + O(i)}.\qedhere
  \end{align*}
\end{proof}

By taking the limit $\alpha = \lim_{n} \alpha_n$, we obtain that $h_{E}(Y_{\alpha}) = \alpha$. Thus, we are left to prove:
\begin{clm}\label{clm:sofic-soficity}
  The subshift $Y_{\alpha}$ is a sofic subshift.
\end{clm}

This proof is very standard and unsurprising, yet sketched for the sake of exhaustiveness.

\begin{proof}
  First, introduce a grid subshift $Y_{\mathrm{grid}}$ on the alphabet $\{\gcross{black},\ghline{black},\gvline{black}\}$ defined as the closure of all the square grid configurations (see~\autoref{fig:sofic-grid}). It is a sofic subshift: by enforcing the continuity of black lines between adjacent positions, we obtain an irregular grid; to obtain a regular square grid, we make each cross $\gcross{black}$ send diagonals in the SFT cover (since diagonals can only go through a cross, the grid becomes regular).

\begin{figure}[ht]
  \begin{subfigure}[t]{0.32\textwidth}
  \centering
  \begin{tikzpicture}[decoration={random steps,segment length=6pt,amplitude=2pt}, scale=0.15]
    \draw[decorate,gridborder]  [path picture={%
      \draw[step=6, black,thick,xshift=1.5cm,yshift=1.5cm](-5,-5) grid (23,23);
      \draw[gridborder,step=1] (0,0) grid (23,23);
    }] (0.5,0.5) -- (22.5,0.5) -- (21.5,22.5) -- (0.5,22.5) -- cycle;
  \end{tikzpicture}
  \subcaption{A square grid configuration of mesh $i \times i$.}
  \label{fig:sofic-grid}
\end{subfigure}
\hfill
\begin{subfigure}[t]{0.32\textwidth}
  \centering
    \begin{tikzpicture}[decoration={random steps,segment length=6pt,amplitude=2pt}, scale=0.15]
      \draw[decorate,gridborder]  [path picture={%
        \foreach \i in {3,9,...,25} {
          \foreach \j in {0,...,23} {
            \draw[blue] (\i.5,\j.5) node[font=\tiny] {$\ast$};
          }
        }
      \draw[step=6, black,thick,xshift=1.5cm,yshift=1.5cm](-5,-5) grid (23,23);
      \draw[gridborder,step=1] (0,0) grid (23,23);
    }] (0.5,0.5) -- (22.5,0.5) -- (22.5,22.5) -- (0.5,22.5) -- cycle;
  \end{tikzpicture}
  \subcaption{Vertical blue columns of symbols $\ast$ are $i$-periodic, the square grid has mesh $i \times i$.}
  \label{fig:sofic-intertwined-grids}
\end{subfigure}
\hfill
\begin{subfigure}[t]{0.32\textwidth}
  \centering
  \begin{tikzpicture}[decoration={random steps,segment length=6pt,amplitude=2pt}, scale=0.15]
    \draw[decorate,gridborder]  [path picture={%
      \foreach \i in {-4.5,1.5,7.5,...,25}{
        \foreach \j in {-4.5,1.5,7.5,...,25}{
          \fill[opacity=0.5,yellow] (\i, \j) --++ (6, 0) --++ (0, 6) -- cycle;
          \foreach \k in {1,2,3}{
            \draw[opacity=0.5, thin, red, -{Latex[length=0.5mm 1]}] (\i+1.75+\k, \j+2) --++ (0.75, 0);
          }
          \draw[opacity=0.5, thin, red, -{Latex[length=0.5mm 1]}] (\i+7.25, \j+2) --++ (-0.75, 0);
          \foreach \k in {0,4}{
            \draw[opacity=0.5, thin, red, {Latex[length=0.5mm 1]}-{Latex[length=0.5mm 1]}] (\i+1.5+\k, \j+2) --++ (1, 0);
          }
        }
      }
      \foreach \i in {3,9,...,25} {
        \foreach \j in {0,...,23} {
          \pgfmathparse{Mod(\j,6)==3?1:0}
          \ifnum\pgfmathresult=0
            \draw[blue] (\i.5,\j.5) node[font=\tiny] {$\ast$};
          \fi
        }
      }
      \draw[step=6, black,thick,xshift=1.5cm,yshift=1.5cm](-5,-5) grid (23,23);
      \draw[gridborder,step=1] (0,0) grid (23,23);
    }] (0.5,0.5) -- (22.5,0.5) -- (22.5,22.5) -- (0.5,22.5) -- cycle;
  \end{tikzpicture}
  \subcaption{An SFT cover enforcing the same period in $\lift{L_{1}}$ and the other grid in $Y_{\mathrm{grid}}$.}
  \label{fig:sft-intertwined-grids}
\end{subfigure}

\caption{Configurations using grids.}
\end{figure}

Synchronizing $Y_{\mathrm{grid}}$ with $\lift{L_1}$, we define $Y_{\mathrm{grid}\ast} \subseteq \lift{L_1} \times Y_{\mathrm{grid}}$ the set of configurations $(x^{(1)\uparrow},x^{(g)})$ such that $x^{(g)}$ has mesh $i \times i$ if and only if $x^{(1)}$ encodes some $i \in \N$ (see~\autoref{fig:sofic-intertwined-grids}).

\begin{clm}\label{clm:sofic-grid-aux}
  $Y_{\mathrm{grid}\ast}$ is a $\Z[2]$ sofic subshift.
\end{clm}
\begin{subproof}
  Using areas of colors in the SFT cover, ensure that exactly one black vertical line in $Y_{\mathrm{grid}}$ can appear between two vertical lines of symbols~$\ast$ in $\lift{L_1}$, see~\autoref{fig:sft-intertwined-grids} for an illustration.
\end{subproof}

Let us now prove that $Y_{\alpha}$ is a $\Z[2]$ sofic subshift. Intuitively, it follows from \autoref{thmC:effective-lift-sofic}: $Y_{\alpha}$ is a ``decorated version'' of $\lift{Z_{\alpha}'}$. The most tricky step is in the periodicity condition: periodicity of a free bit in $L^{(f)}$ should only be enforced whenever both layers $y^{(1)}$ and $y^{(2)}$ do not belong to $\lift{\encode{\infty}}$, \textit{i.e.}~whenever they both actually encode some integers $i \in \N$ and $j \in \N$.

To proceed, we slightly alter the $\Z$ subshift $Z_{\alpha}'$ to define a new
subshift~$Z_{\alpha}''$: it contains an additional layer $L_p$ (the
\defn{proper} layer) that can take two values (either~$\mathsf{p}^{\Z}$
or~$\mathsf{d}^{\Z}$), and is forced to be~$\mathsf{p}^{\Z}$ whenever both the
first and second layer do encode integers:
\begin{align*}
  Z_{\alpha}'' &= \Big\{ (z^{(1)}, z^{(2)}, z^{(d)}, z^{(p)}) \in L_1 \times L_2 \times L_d \times \{\mathsf{p}^{\Z},\mathsf{d}^{\Z}\} \mid  z^{(2)} \in \encode{\infty} \Big\} \\
              & \hspace{0.2cm} \cup \bigcup_{i \in \N} \bigcup_{j \geq i} \Big\{ (z^{(1)}, z^{(2)}, z^{(d)},z^{(p)}) \in L_1 \times L_2 \times L_d \times \{ \mathsf{p}^{\Z} \} \mid \exists k_1,k_2 \in \N,\\[-10pt]
              & \hspace{3cm} z^{(1)} = \encode{i}_{k_1},\, z^{(2)} = \encode{j}_{k_2}\,\text{and } \exists \beta \leq \alpha_{i,j},\, z^{(d)} = T(\beta,i)_{k_1} \Big\}.
\end{align*}

By a slight alteration of \autoref{clm:effective-effectivity}, the subshift $Z_{\alpha}''$ is effective whenever $\alpha$ is a $\Pi_3$ real number. By \autoref{thmC:effective-lift-sofic}, the $\Z[2]$ subshift $\lift{Z_{\alpha}''}$ is thus sofic. Then, we use the proper layer to enforce periodicity of a free bit in $L^{(f)}$ only whenever $y^{(p)} = \mathsf{p}^{Z^2}$, and define $Y_{\alpha}'$ as:
\begin{align*}
  Y_{\alpha}' & = \Big\{ (y^{(1)\uparrow},y^{(2)\uparrow},y^{(d)\uparrow},y^{(p)\uparrow}, y^{(g)}, y^{(f)}) \in \lift{Z_{\alpha}''} \times Y_{\mathrm{grid}} \times \{\freezero,\heads,\tails\}^{\Z[2]} \mid \\
  & \hspace{1.5cm} (y^{(1)\uparrow},y^{(g)}) \in Y_{\mathrm{grid}\ast},\quad\pi_{\mathrm{sync}}(y^{(f)}) = y^{(d)\uparrow}, \\
              & \hspace{1.5cm} \exists b \in \{\freezero,\heads,\tails\}, \forall p \in \Z[2],\ y^{(p)} = \mathsf{p}^{\Z} \wedge \,y^{(g)}_p = \gcross{black} \implies y^{(f)}_p = b\Big\}
\end{align*}

\begin{clm}
  $Y_{\alpha}'$ is a $\Z[2]$ sofic subshift.
\end{clm}
\begin{subproof} By the previous paragraph, the first four layers are sofic; and
  by \autoref{clm:sofic-grid-aux}, the synchronization $Y_{\mathrm{grid}\ast}$
  of $\lift{L_1}$ and $Y_{\mathrm{grid}}$ is sofic. To make a free bit periodic,
  one can use the grid defined by $Y_{\mathrm{grid}}$: more precisely, we can
  additionally superimpose another layer carrying a unique symbol
  $b_{\mathrm{grid}} \in \{\freezero,\heads,\tails\}$ on top of the black lines of
  $Y_{\mathrm{grid}}$ in an SFT cover, which must be the same everywhere.
  Furthermore, we enforce the following condition: on positions at which a cross
  symbol $\gcross{black}$ appears on the grid layer $y^{(g)}$, and a symbol
  $\mathsf{p}$ appears on the proper layer $y^{(p)\uparrow}$, the free bit in
  $y^{(f)}$ is then made equal to the symbol $b_{\mathrm{grid}}$. As the proper
  layer is a constant configuration over $\mathbb{Z}^{2}$, every position of $y$
  containing a symbol $\gcross{black}$ in its grid layer $y^{(g)}$ will
  therefore carry the same bit $b_{\mathrm{grid}}$ in its free layer $y^{(f)}$.
\end{subproof}

We can now prove that $Y_{\alpha}$ is sofic. Indeed, fix an SFT cover of $Y_{\alpha}'$ in which we color cross symbols $\gcross{black}$ into two colors alternatively: let us say, red and blue. On each horizontal and vertical line of the grid layer $Y_{\mathrm{grid}}$, crosses are now alternating between red and blue.
  We claim that we obtain the subshift $Y_{\alpha}$ by projecting this SFT cover as follows:
  \begin{itemize}
  \item Erase the proper layer.
  \item Projection of the grid layer: red crosses become $\redsquare$, and all other symbols become $\square$.
  \end{itemize}
  Indeed, projecting the grid layer as mentioned creates the marker layer $L_m$,
  as the grid layer was forced to have a mesh $(i, i)$ whenever the first layer
  has period $i$. The only condition that remains to be checked is the
  $(i,i)$-periodicity condition on a free bit.

  Notice that in $Y_{\alpha}'$, both cases $y^{(p)\uparrow} = \mathsf{p}^{\Z[2]}$ and $y^{(p)\uparrow} = \mathsf{d}^{\Z[2]}$ are possible whenever $y^{(1)} \in \encode{\infty}$ or $y^{(2)} \in \encode{\infty}$, so that erasing the proper layer in the projection merges the two cases together and removes the periodicity enforced on the free bit of $y^{(f)}$; while whenever $y^{(1)} = \encode{i}_{k_1}$ and $y^{(2)} = \encode{j}_{k_2}$, only the case $y^{(p)} = \mathsf{p}^{\Z[2]}$ is allowed so that when projecting, the periodicity condition is still enforced.
\end{proof}

\subsection{Reformulation as growth rates of syntactic monoids}
\label{sec:reformulation-syntactic-monoid}

We briefly show a different point on view on the previous results, by relating
extender sets with the classical syntactic monoid from the study of formal
languages.

For any finite alphabet $\mathcal{A}$ and any language
$L \subseteq \mathcal{A}^{\ast}$, one can define an equivalence relation, called
the \defn{syntactic congruence}, as (see for example~\cite[Definition
3.6]{anderson_automata_theory_modern_applications}):
\[\forall u, v \in L, u \sim_{L} v \iff (\forall x, y \in L, xuy \in L
  \iff xvy \in L)\]

\begin{defi}[Syntactic monoid]
  Let $L$ be a language. Then $M(L) = L/\sim_{L}$ with the concatenation
  operation is a monoid, called the \defn{syntactic monoid} of $L$.
\end{defi}

For $L$ a language over $\mathcal{A}$ and $u \in L$, we call
\defn{reduced length} of $u$ the non-negative integer
$\norm{u}_{L} = \min_{v \sim_{L} u} |v|$. The \defn{growth rate} of
$M(L)$ is then
\[ h(M(L)) = \lim_{n \to +\infty} \frac{\log\, \card{\{[u] \in M(L) \mid u \in L,
      \norm{u}_{L} \leq n\}}}{n} \]

For a $\mathbb{Z}$ subshift $X$, we define its syntactic monoid $M(X)$
as $M(\shiftlang{X})$. In this setting, we can reformulate
\autoref{thm:ext-entropy-effective} as:

\begin{thm}
  The growth rates of syntactic monoids of effective
  $\Z$ subshift are exactly the non-negative $\Pi_{3}$ real
  numbers.
\end{thm}

\begin{proof}[Sketch of proof]Consider the subshift $Z_{\alpha}$ from
  the proof of \autoref{thm:ext-entropy-effective}. Then we
  claim that $M(Z_{\alpha})$ has growth rate $\alpha$.
  Indeed, for any two words $u,v \in \shiftlang[n]{Z_{\alpha}}$ such that
  $u \not \sim_{\shiftlang{Z_{\alpha}}} v$, we have
  $E_{Z_{\alpha}}(u) \neq E_{Z_{\alpha}}(v)$. So
  $\card{\{[u] \in M(X) \mid u \in \shiftlang[X]{Z_{\alpha}}, \norm{u}_{\shiftlang{Z_{\alpha}}} \leq n\}} \leq \card{E_{Z_{\alpha}}(n)}$.
  On the other hand, the argument computing a lower bound on the number of extender sets
  of $Z_{\alpha}$ in
  \autoref{lem:effective-counting-ext} exhibits a
  family of roughly $2^{n\alpha + o(n)}$ words of $\shiftlang[n]{Z_{\alpha}}$
  such that any two words of this family cannot be syntactically congruent.
  This concludes the proof.
\end{proof}

\section{Additional dynamical and computational restrictions}
\label{sec:addit-dynam-comp-restr}

In this section, we study how additional restrictions on the subshifts that we consider restrict the set of possible extender entropies. The main result of this section is \autoref{thm:ext-entropy-block-gluing-effective}, which shows that a subclass of $\mathbb{Z}^{d}$ sofic subshifts satisfying strong dynamical restrictions still exhibits the full range of possible extender entropies.

\subsection{Computable subshifts}
\label{sec:entropy:computable}

Recall that a subshift $X$ is \defn{computable} if its language $\shiftlang{X}$ is decidable.
Following the proofs from \autoref{sec:ext:decidability-extender}, one proves that
extender entropies of computable subshifts are $\Pi_{2}$ real numbers.
In the converse direction, we obtain:

\begin{restatethm}{thm:ext-entropy-computable-effective}
  For $d \geq 1$, the set of extender entropies of $\Z[d]$ computable subshifts is exactly $[0,+\infty) \cap \Pi_2$.
\end{restatethm}

\begin{restatethm}{thm:ext-entropy-computable-sofic}
  For $d \geq 2$, the set of extender entropies of $\Z[d]$ sofic computable subshifts is exactly $[0,+\infty) \cap \Pi_2$.
\end{restatethm}

\begin{proof}[Sketch of proof]
  The subshift $Z_{\alpha}'$ constructed in~\autoref{thm:ext-entropy-effective} might
  not be computable whenever $\alpha \in \Pi_3$, since, given some $i,j \in \N$
  and some factor of $T(\beta,i)$, it might be undecidable to know whether
  $\beta \leq \alpha_{i,j}$ when $\alpha_{i,j} \in \Pi_1$.

  Yet, when taking $\alpha = \inf_{i} \alpha_{i} = \inf_{i} \sup_{j} \alpha_{i, j}
  \in \Pi_{2}$ for $(\alpha_{i, j})$ a computable sequence of rationals, the language of $Z_{\alpha}'$ (and thus, of $Z_{\alpha}$ and $Y_{\alpha}$) becomes decidable.
\end{proof}

\enlargethispage{\baselineskip}
\subsection{Minimal subshifts}
\label{sec:entropy:minimal}

Extender sets are much easier in minimal subshifts and do not even depend on the
computability of the language:

\begin{prop}\label{prop:minimal-inclusion-ext-decidable}
  Let $X$ be a minimal subshift over $\mathbb{Z}^{d}$. Then for any $n > 0$ and any patterns $u, v \in \shiftlang[n]{X}$, $E_{X}(u) \subseteq E_{X}(v) \iff u = v$.
\end{prop}

\begin{proof} Let $u, v \in \shiftlang[n]{X}$ and suppose that $E_{X}(u)
\subseteq E_{X}(v)$. Then any appearance of $u$ in a configuration can be
replaced by $v$: by iterating the process while ordering patterns
lexicographically (see~\cite[Lemma 2.2]{quas-trow00:subshifts_multidim_sfts} for
the complete argument), we obtain by compactness a configuration of $X$ in which
$u$ does not appear, which contradicts minimality.
\end{proof}

This implies that $h_{E}(X) = h(X)$ if $X$ is minimal. It is folklore that minimal SFTs have zero entropy (see \cite[Proposition~25]{gangloff22_charac_entrop_dimen_minim_mutid} for a proof), and since entropy decreases under factor maps, so do minimal sofic subshifts. On the other hand, since minimal effective subshifts have arbitrary $\Pi_{1}$ entropy (consider the proof of~\cite[Theorem~4.77]{kurka03_topol_symbol_dynam} with a computable sequence of integers $(k_n)_{n \in \N}$), we obtain:
\begin{cor}\label{cor:ext-entropy-minimal-sofic}
  Let $Y \subseteq \mathcal{A}^{\Z[d]}$ be a minimal sofic subshift. Then $h_E(Y) = 0$.
\end{cor}

\begin{cor}\label{cor:ext-entropy-minimal-effective}
    For $d \geq 1$, the set of extender entropies of $\Z[d]$ minimal effective subshifts is exactly $[0,+\infty) \cap \Pi_1$.
\end{cor}

\subsection{Mixing subshifts}
\label{sec:entropy:block-gluing}

We naturally expected that strong mixing conditions restrict the possible behaviors of extender sets: indeed, all the subshifts considered in this article have either strong mixing properties (the full shift, $\Z$ SFTs\dots) and zero extender entropy, or positive extender entropy but structural rigidity (periodicity, reflected positions, \dots). Nevertheless, we prove that mixing properties do not imply anything on extender entropies. This is in stark contrast with the classical topological entropy, where subshifts satisfying mixing properties of various strengths tend have restrictions on their possible entropies.

Although \cite{hochman10_charac_entrop_multid_shift_finit_type} characterizes the entropies of multidimensional SFTs as the positive $\Pi_{1}$~real numbers, subclasses of subshifts subject to ``mixing'' conditions have been investigated in \cite{pavlov_schraudner15_class_sofic_projec_subdy_multid}, \cite{gangloff19_effec_quant_irred_comput_subsh_entrop} and \cite{gangloff21_quant_block_gluin_multid_subsh_finit_type}. In particular, the entropies of multidimensional block-gluing SFTs are a strict subset of computable real numbers (see \cite[Theorem 1.4]{pavlov_schraudner15_class_sofic_projec_subdy_multid}). We do not know of any such result about sofic subshifts, but informally, most tools and methods known to the authors to realize sofic shifts of prescribed entropies (``lifts'' \cite{aubrun_sablik13_simul_effec_subsh_two_dimen}, fixed-point constructions \cite{durand12_fixed_point_tile_sets_their_applic}, multidimensional Toeplitz or variants of substitutive subshifts) admit non-trivial quasi-periodic factors, an obstruction to any kind of strong mixing property.

\begin{restatethm}{thm:ext-entropy-block-gluing-effective}
  For $d \geq 1$, the set of extender entropies of $1$-block-gluing effective $\Z[d]$ subshifts is exactly $[0,+\infty) \cap \Pi_3$.
\end{restatethm}

\subsubsection{Adding a block-safe symbol \texorpdfstring{$\texttt{\#}$}{\#}}
We prove this theorem by adding a block-safe symbol~$\texttt{\#}$ to the alphabet $\mathcal{A}$, which  creates mixingness while preserving extender entropies. More precisely, let $\rect{R}$ denote the set of (potentially infinite) hyperrectangles $\rect{r} \subseteq \Z[d]$. Given a subshift $\smash{X \subseteq \mathcal{A}^{\Z[d]}}$, for $\mathcal{A}_{\texttt{\#}} = \mathcal{A} \cup \{\texttt{\#}\}$ we define the subshift $\smash{X_{\texttt{\#}} \subseteq \mathcal{A}_{\texttt{\#}}^{\,\Z[d]}}$ drawing (potentially infinite) non-contiguous rectangular patterns from $X$ over a background of symbols~$\texttt{\#}$:
\begin{align*}
  X_{\texttt{\#}} = \{ x \in \mathcal{A}_{\texttt{\#}}^{\,\Z[d]} \mid & \exists J, \exists (\rect{r}_j)_{j \in J} \in \rect{R}^J,\ d(\rect{r}_{j_1},\rect{r}_{j_2}) \geq 1 \ \text{if}\ j_1 \neq j_2, \\
                                                                      & \sqcup_{j \in J} \rect{r}_j = \{p \in \Z[d] \mid x_{p} \neq \texttt{\#}\} \ \text{and}\ \forall j \in J,\ \restr{x}{\rect{r}_j} \in \shiftlang{X} \}.
\end{align*}

\begin{lem}\label{lem:block-gluing-from-effective}
  Let $X \subseteq \mathcal{A}^{\Z[d]}$ be an effective subshift. The subshift $X_{\texttt{\#}} \subseteq \mathcal{A}_{\texttt{\#}}^{\,\Z[d]}$ defined above is $1$-block-gluing, effective, and satisfies $h_E(X_{\texttt{\#}}) = h_E(X)$.
\end{lem}

\begin{figure}[b]
  \centering
  \begin{tikzpicture}[scale=0.22,decoration={random steps,segment length=0.4em,amplitude=0.15em}]
    \pgfmathsetseed{3}
    \draw [gridborder,decorate,path picture={
      \foreach \i in {-1,...,34} {
        \foreach \j in {-1,...,20} {
          \draw ($(\i,\j) + (0.5,0.5)$) node[black!60,font=\tiny] {$\texttt{\#}$};
        }
      }
      \fill[gray] (2,1) rectangle ++(5,4);
      \fill[gray] (1,9) rectangle ++(7,9);
      \fill[gray] (7,19) rectangle ++(2,2);
      \fill[gray] (9,4) rectangle ++(10,6);
      \fill[gray] (22,-3) rectangle ++(13,13);
      \fill[gray] (8,-1) rectangle ++(1,3);
      \fill[gray] (13,-3) rectangle ++(7,5);
      \fill[gray] (14,17) rectangle ++(4,4);
      \fill[gray] (14,12) rectangle ++(1,1);
      \fill[gray] (20,14) rectangle ++(3,7);
      \fill[gray] (26,17) rectangle ++(7,7);
      \fill[gray] (28,12) rectangle ++(7,4);
      \draw[gridborder!50] (-1,-1) grid (35,21);
    }] (-0.5,-0.5) rectangle (34.5,20.5);
  \end{tikzpicture}
  \caption{Layout of a configuration in $X_{\texttt{\#}}$: independent rectangles float over a background of $\texttt{\#}$ symbols.}
\end{figure}

\begin{proof}
  Let $X \subseteq \mathcal{A}^{\Z[d]}$. The subshift $X_{\texttt{\#}}$ is effective since the set of finite patterns in $\shiftlang{X}$ is $\Pi^{0}_{1}$-computable. It is $1$-block-gluing, since for any two rectangles $\rect{r},\rect{r}' \in \rect{R}$ such that $d(\rect{r},\rect{r}') \geq 1$ and any two patterns $w \in \shiftlang[\rect{r}]{X_{\texttt{\#}}}$, $w' \in \shiftlang[\rect{r}']{X_{\texttt{\#}}}$, the configuration $\smash{x \in \mathcal{A}_{\texttt{\#}}^{\Z[d]}}$ defined as follows is valid in $X_{\texttt{\#}}$: $x_{p} = w_{p}$ if $p \in \rect{r}$, $x_{p} = w'_{p}$ if $p \in \rect{r}'$, or $x_{p} = \texttt{\#}$ otherwise. To complete this proof, we are left with computing the extender sets of $X_{\texttt{\#}}$.

  \medskip
  For a domain $D \subseteq \Z[d]$, assume that two patterns $w,w' \in \mathcal{A}^{D}$ satisfy $E_X(w) \neq E_X(w')$: by definition, there exists some partial configuration ${x \in \mathcal{A}^{\Z[d] \setminus D}}$ that extends either $w$ or $w'$ but not the other. Since the same configuration still exists in $X_{\texttt{\#}}$, and still extends only one pattern among $w$ and $w'$, we obtain $E_{X_{\texttt{\#}}}(w) \neq E_{X_{\texttt{\#}}}(w')$. In particular, $h_E(X) \leq h_E(X_{\texttt{\#}})$.

  Conversely, let us consider the extender sets of patterns in $X_{\texttt{\#}}$. Given a domain $\interval{0,n-1}^d$ and a pattern $w \in \shiftlang[n]{X_{\texttt{\#}}}$, we define the \emph{geometry} of $w$ (denoted $G(w)$) as the set of maximal non-contiguous hyperrectangles $\rect{r} \in \rect{R}$  (\textit{i.e.}~$d(\rect{r},\rect{r}') \geq 1$ for distinct $\rect{r},\rect{r'} \in G(w)$) that cover the non-$\texttt{\#}$ symbols of~$w$ (\textit{i.e.} $\sqcup_{\rect{r} \in G(w)} \rect{r} = \{p \in \interval{0,n-1}^d \mid w_{p} \neq \texttt{\#}\}$). Considering the \emph{border geometry} $\partial{G}(w)$ defined by
  \[ \partial{G}(w) = \{ \rect{r} \in G(w) \mid \rect{r} \cap \partial_{1}(\interval{0,n-1}^d) \neq \emptyset \},\]
  the extender set $E_{X_{\texttt{\#}}}(w)$ in $X_{\texttt{\#}}$ is entirely determined by the border geometry $\partial{G}(w)$ and the extender sets $\{E_X(\restr{w}{\rect{r}}) \mid \rect{r} \in \partial{G}(w)\}$ in $X$.

  From these considerations, we bound the number of extender sets in $X_{\texttt{\#}}$. By subadditivity \cite[Theorem~1]{capobianco08:multivariate-fekete-lemma}, for $\varepsilon > 0$, there exists some $N \in \N^d$ such that, for all $n \in \N $,
  \[ (n_1,\dots,n_d) \geq (N_1,\dots,N_d) \implies \log\, \card{E_X(n_1,\dots,n_d)} \leq \big(h_{E}(X)+\varepsilon\big)\, n_1 \!\cdots n_d. \]
  Thus, for any rectangle $\rect{r} \subseteq \interval{0,n-1}^d$ of size $r_1 \times \dots \times r_d$ (and denoting $\lVert \rect{r} \rVert = r_1 \cdots r_d$ its volume), two cases arise:
  \begin{itemize}
  \item Either $r_i \geq N_i$ for all $1 \leq i \leq d$ (we denote $\rect{R}_L$ this set of \emph{large rectangles}): in this case, we bound the number of extender sets as follows: $\log \card{E_X(\rect{r})} \leq (h_{E}(X)+\varepsilon)\, \lVert \rect{r} \rVert$.
  \item Or there exists some $1 \leq i \leq d$ such that $r_i < N_i$ (we denote $\rect{R}_{S}$ this set of \emph{small rectangles}): in this case, we bound $\log \card{E_X(\rect{r})}$ with $\log \card{E_X(\rect{r})} \leq \log \card{\mathcal{A}_{\texttt{\#}}} \cdot \lVert \rect{r} \rVert$.
  \end{itemize}

  For a fixed border geometry $\partial{G}$, we have:
  \begin{align*}
    \prod_{\rect{r} \in \partial{G} \cap \rect{R}_{S}} \hspace{-0.6em} \card{E_{X_\texttt{\#}}(\rect{r})}
    & \leq \card{\mathcal{A}_{\texttt{\#}}}^{\sum_{\rect{r} \in \partial{G} \cap \rect{R}_{L}} \lVert \rect{r} \rVert}
      \leq \card{\mathcal{A}_{\texttt{\#}}}^{\sum_{i=1}^d 2\cdot n^{d-1} \cdot N_i}\,; \\[5pt]
    \prod_{\rect{r} \in \partial{G} \cap \rect{R}_{L}} \hspace{-0.6em} \card{E_{X_\texttt{\#}}(\rect{r})}
    & \leq 2^{(h_{E}(X)+\varepsilon) \cdot \sum_{\rect{r} \in \partial{G} \cap \rect{R}_{L}} \lVert \rect{r} \rVert}
      \leq 2^{(h_{E}(X)+\varepsilon) \cdot\, n^d}.
  \end{align*}

  Summing over all geometries, we obtain:
  \begin{align*}
    \hspace{-0.3em}\card{E_X(n)}
    & \leq \sum_{\substack{\partial{G} = \partial{G}(w) \mid \\ w \in \shiftlang[n]{X_{\texttt{\#}}}}}
      \prod_{\rect{r} \in \partial{G} \cap \rect{R}_{S}} \hspace{-0.6em} \card{E_{X_\texttt{\#}}(\rect{r})}
      \cdot \prod_{\rect{r} \in \partial{G} \cap \rect{R}_{L}} \hspace{-0.6em} \card{E_{X_\texttt{\#}}(\rect{r})} \\
    & \leq \sum_{\substack{\partial{G} = \partial{G}(w) \mid \\ w \in \shiftlang[n]{X_{\texttt{\#}}}}}
      \card{\mathcal{A}_{\texttt{\#}}}^{\sum_{i=1}^d 2 \cdot n^{d-1} \cdot N_i}
      \cdot
      2^{(h_{E}(X)+\varepsilon) \cdot n^d}.
  \end{align*}
  For a given pattern $w \in \shiftlang[n]{X_{\texttt{\#}}}$, the number of rectangles in $\partial{G}(w)$ is bounded by the hypersurface of $\interval{0,n-1}^d$, which has cardinal $2d \cdot n^{d-1}$. Furthermore, a $d$-dimensional hyperrectangle in $\interval{0,n-1}^d$ is entirely determined by two points. Thus, the number of possible border geometries is bounded by:
  \[
    \card{\{ \partial{G}(w) \mid w \in \shiftlang[n]{X_{\texttt{\#}}} \}}
    \leq (O(n^{2d}))^{2d \cdot n^{d-1}}
    \leq 2^{O(n^{d-1} \cdot \log n)}.
  \]
  Plugging all these equations together leads to:
  \[
    \log \card{E_{X_{\texttt{\#}}}(n)}
    \leq \big(h_E(X) + \varepsilon) \cdot n^d + O(n^{d-1} \cdot \log n)
  \]
  Since $\varepsilon$ was arbitrary, we conclude $h_E(X_{\texttt{\#}}) \leq h_E(X)$.
\end{proof}

\subsubsection{The sofic case}
\label{sec:entropy:block-gluing:sofic}
With a slightly more involved construction, one can prove that the $1$-block-gluing $\Z[d]$ sofic subshifts realize the same extender entropies as the effective subshifts:

\begin{restatethm}{thm:ext-entropy-block-gluing-sofic}
  For $d \geq 2$, the set of extender entropies of $1$-block-gluing sofic $\Z[d]$ subshifts is exactly $[0,+\infty) \cap \Pi_3$.
\end{restatethm}

By \autoref{clm:ext-entropy-free-lift} and the fact that the free lift of a $1$-block-gluing subshift is still $1$-block-gluing, we write the proof on $\Z[2]$ only.

\begin{proof} Let us fix $\alpha \in [0,+\infty) \cap \Pi_3$, and denote by $Y_{\alpha} \subseteq \mathcal{A}^{\Z[2]}$ the subshift that was built in the proof of \autoref{thm:ext-entropy-sofic}, but in which configurations $\encode{i}$ and $\encode{j}$ only occur for $i$ and $j$ powers of $2$ (up to defining $\alpha_{i,j}' = \alpha_{\log i,\log j}$, the extender entropy of $Y_{\alpha}$ is preserved since it can be computed along any sequence of rectangles; but $Y_{\alpha}$ now has a reduced ``density'' of information). Denoting $\rect{S} \subseteq \rect{R}$ the set of (possibly infinite) squares $\rect{s} \subseteq \Z[2]$ and $\rect{S}_{\!f} \subseteq \rect{S}$ the set of \emph{finite} squares, define:
\begin{align*}
  Y_{\texttt{\#}} = & \{ x \in \mathcal{A}_{\texttt{\#}}^{\,\Z[2]} \mid \exists J, \exists (\rect{r}_j)_{j \in J} \in (\rect{R})^J,\ d(\rect{r}_{j_1},\rect{r}_{j_2}) \geq 1 \ \text{if}\ j_1 \neq j_2, \\
                                                                            & \quad \sqcup_{j \in J} \rect{r}_j = \{p \in \Z[2] \mid x_{p} \neq \texttt{\#}\} \ \text{and}\ \forall j \in J,\ \rect{r}_j \in \rect{S}_{\!f} \implies \restr{x}{\rect{r}_j} \in \shiftlang{Y_{\alpha}} \}.
\end{align*}
Intuitively, $\mathcal{A}$-rectangles in $Y_{\texttt{\#}}$ can be filled with any pattern on $\mathcal{A}$; but $\mathcal{A}$-squares in $Y_{\texttt{\#}}$ can only be filled with valid patterns from $Y_{\alpha}$.

\smallskip
As in \autoref{lem:block-gluing-from-effective}, $Y_{\texttt{\#}}$ is $1$-block-gluing and $h_E(Y_{\texttt{\#}}) = h_E(Y_{\alpha})$:
  \begin{itemize}
  \item Since rectangles $\rect{r} \in \rect{R} \setminus \rect{S}_{\!f}$ that are \emph{not} finite squares can now contain every pattern of $\mathcal{A}^{\rect{r}}$, all these new patterns belong in the same extender class and the upper bound $h_E(Y_{\texttt{\#}}) \leq h_E(Y_{\alpha})$ still holds.
  \item However, since $\mathcal{A}^{\Z[2]} \subseteq Y_{\texttt{\#}}$, the inequality $h_E(Y_{\alpha}) \leq h_E(Y_{\texttt{\#}})$ needs to be proved with finite square patterns of $\shiftlang{X}$ instead of complete configurations: if $y^{(0)} \in Y_{\alpha}$ distinguishes the extender classes of two patterns $w,w'$ in $Y$, then there exists some $n \in \N$ such that the configuration $y \in Y_{\texttt{\#}}$ defined as $y_{p} = \smash{y^{(0)}_{p}}$ if $p \in \interval{-n,n}^d$, and $y_{p} = \texttt{\#}$ otherwise, distinguishes $w$ from $w'$ in $Y_{\texttt{\#}}$.
  \end{itemize}

  The next section will prove that $Y_{\texttt{\#}}$ is a sofic subshift, completing the proof.
\end{proof}

\subsubsection{Proof of soficity}

Proving the soficity of $Y_{\texttt{\#}}$ is highly non-trivial: we no longer have uniform vertically lifted configurations to apply~\autoref{thmC:effective-lift-sofic}; instead, the configurations of $Y_{\texttt{\#}}$ are made of independent rectangular $\mathcal{A}$-patterns over a background of symbols~$\texttt{\#}$, each rectangle embedding various pieces of information (encodings of $i$, $j$, density of symbols~$1$\dots). We are thus required to perform infinite computation checks on finite independent objects, which is much more difficult. As a similar (but simpler) problem was solved in~\cite{westrick17_seas_squar_with_sizes_from_pi_set} using the so-called ``fixpoint construction'' \cite{durand12_fixed_point_tile_sets_their_applic}, we will apply the same method on $Y_{\texttt{\#}}$.

\smallskip
The soficity of $Y_{\texttt{\#}}$ actually motivated the development the main lemma in \cite{Callard-Guillon-PavietSalomon-Vanier_2022:pimp-my-fixpoint}. We copy a simplified version in~\autoref{thmC:pixpoint}. It is quite technical, and its application to $Y_{\texttt{\#}}$ is not immediate either. During their first reading, we invite the reader who is not already familiar with previous applications of the construction ~\cite[\dots]{westrick17_seas_squar_with_sizes_from_pi_set,destombes23_algor_compl_sofic_shift_dimen_two} or with $S$-adic systems to skip this proof entirely.

We first define \emph{ndill maps}, which are substitutions with a form of context-awereness:
\begin{defi}[Ndill maps]
  An \emph{ndill map} is a local multifunction $\smash{\tau \colon \mathcal{A}^{\interval{-r,r}^2} \mto \mathcal{B}^{\interval{0,N-1}^2}}$ where $\mathcal{A}$, $\mathcal{B}$ are finite alphabets, $r \in \N$ is the \emph{radius} of $\tau$, and $N \in \N$ is the \emph{size} of its images.
\end{defi}
\noindent Formally, we say that a \emph{grid} is a partition $\grid{g} = (\rect{s}_{p})_{p \in \Z[2]}$ of $\Z[2]$ into adjacent\footnote{That is, the squares $\rect{s}_{p}$ and $\rect{s}_{p + (0,1)}$ (resp.~$\rect{s}_{p + (1,0)}$) must be neighbors in the direction $(0,1)$ (resp.~$(1,0)$).} squares $\rect{s}_p \subseteq \Z[2]$ of the same size $N \times N$. Then $\tau$ extends to entire configurations as follows: for $x \in \mathcal{A}^{\Z[2]}$ and $y \in \mathcal{B}^{\Z[2]}$ two configurations and $\grid{g}$ a grid of $N \times N$-squares, we say that $\smash{x \substep[\grid{g}]{\tau} y}$ if
\[ \forall p \in \Z^2,\; y|_{\rect{s}_p} \in \tau(x|_{p + \interval{-r,r}^2}). \]

As with classical substitutions (which are exactly the ndill maps of radius $0$), we are interested in the \emph{limit configurations} induced by sequences of ndill maps. More precisely, given a sequence of ndill maps $\tau_{\ell} \colon \mathcal{A}_{\ell+1}^{\interval{-r,r}^2} \mto \mathcal{A}_{\ell}^{\interval{0,N-1}^2}$, a configuration $x \in \mathcal{A}_{0}^{\Z[2]}$ is a \emph{limit configuration} if there exists some configurations $x^0, x^1, \dots$ and grids $\grid{g}_{0}, \grid{g}_1,\dots$ such that:
\[ x = x^{0} \rsubstep[\grid{g}_0]{\tau_0} x^{1} \rsubstep[\grid{g}_1]{\tau_1} x^{2}  \rsubstep[\grid{g}_2]{\tau_2} x^{3} \dots \]
For fixed $\bm{\tau} = (\tau_{\ell})_{\ell \in \N}$, the associated set of limit configurations actually forms an \emph{N-adic} subshift~$\limspace{X}{\bm{\tau}}$. In \cite{Callard-Guillon-PavietSalomon-Vanier_2022:pimp-my-fixpoint}, we prove that some \emph{N-adic subshifts} are actually sofic:
\begin{thmC}[{Simplified version of \cite[Theorem~3.2]{Callard-Guillon-PavietSalomon-Vanier_2022:pimp-my-fixpoint}}]\label{thmC:pixpoint}
  Let $\alpha < 1$ be a real number, and let $\bm{\tau} = (\tau_{\ell})_{\ell \in \N}$ be a sequence of ndill maps $\tau_{\ell} \colon \mathcal{A}_{\ell+1}^{\{-1,0,1\}^2} \mto \mathcal{A}_{\ell}^{2 \times 2}$ such that:
  \begin{enumerate}
  \item The symbols in the alphabet $\mathcal{A}_{\ell+1}$ can be encoded with $O(2^{\alpha \cdot \ell})$ bits;
  \item The ndill maps $\tau_{\ell}$ are uniformly computable in time $O(2^{\alpha \cdot \ell})$ in the $\log$-RAM model.
  \end{enumerate}
  Then the resulting N-adic limit space $\limspace{X}{\bm{\tau}} \subseteq \mathcal{A}_{0}^{\Z[2]}$ is a sofic subshift.
\end{thmC}

No knowledge of the $\log$-RAM computation model is required for the following proofs. Intuitively, it is a RAM model with some restrictions on the bit lengths of its variables (see \cite{Angluin-Valiant_1979:fast-probabilistic-algorithms-hamiltonian-and-rac}); so that its time complexity is very close to real-world programming languages.

\autoref{thmC:pixpoint} is proved by applying the ``fixpoint construction'' (\cite[\dots]{durand12_fixed_point_tile_sets_their_applic,westrick17_seas_squar_with_sizes_from_pi_set}), but said proof is completely outside the scope of this article. In what follows, we will just apply it to prove the soficity of the subshift $Y_{\texttt{\#}}$.

\paragraph*{Proof: soficity of $Y_{\texttt{\#}}$}
We once again define an auxiliary subshift on the first three layers of the alphabet $\mathcal{A}$ (the \emph{first layer}, the \emph{second layer} and the \emph{density layer}), and denote $\mathcal{B}$ the corresponding alphabet. We consider the set of configurations over a $\texttt{\#}$ background in which $\mathcal{B}$-patterns actually form valid squares:
\begin{align*}
  Y = \{ y \in \mathcal{B}_{\texttt{\#}}^{\,\Z[d]} \mid & \exists J, \exists (\rect{s}_j)_{j \in J} \in \rect{S}^J,\ d(\rect{s}_{j_1},\rect{s}_{j_2}) \geq 1 \ \text{if}\ j_1 \neq j_2, \\
                                                                            & \sqcup_{j \in J} \rect{s}_j = \{p \in \Z[d] \mid y_{p} \neq \texttt{\#}\} \ \text{and}\ \forall j \in J,\ \restr{y}{\rect{s}_j} \in \pi_{\lift{L_1} \times \lift{L_2} \times \lift{L_d}}(\shiftlang{Y_{\alpha}}) \}.
\end{align*}
We claim that, if $Y$ is a sofic subshift, then so is $Y_{\texttt{\#}}$. Indeed, one can add a marker layer on top of the $\mathcal{B}$-squares in $Y$, a free layer and the synchronization as in~\autoref{clm:sofic-soficity}; and then add arbitrary $\mathcal{B}$-rectangular and \emph{non-square} patterns inside the configurations of~$Y$.

\paragraph*{Outline of the proof}
We apply \autoref{thmC:pixpoint} to prove that $Y$ is a sofic subshift. To proceed, we say that a $\mathcal{B}$-pattern $w \in \mathcal{B}^{n \times m}$ with the first, second and density layer is \emph{plausible} if it is vertically periodic, its first and second layer consist of evenly-spaced columns of $\ast$ symbols, and its density layer is a vertically repeated subword of a Toeplitz sequence\footnote{Equivalently, $w$ is plausible if it appears in the configuration of $\lift{Z_{\alpha}'}$ for some $\alpha \leq 1$.}.

\noindent We will first consider how much information is needed to represent a valid $\mathcal{B}$-pattern. We will then design a sequence of efficiently computable ndill maps $\bm{\tau} = (\tau_{\ell})_{\ell \in \N}$ of limit space $Y$.

\paragraph*{Representing plausible $\mathcal{B}$-patterns}
Let $w \in \mathcal{B}^{n \times m}$ be a rectangular pattern of width $n$ appearing in $Y$. Then $w = (w^{(1)}, w^{(2)}, w^{(d)})$ is actually vertically periodic, and can be represented by the following pieces of information:
\begin{itemize}[itemsep=4pt]
\item The \emph{first layer} can be represented by a pair of variables $(i,k_1)$ for $i \in \interval{1,n} \cup \{\texttt{LARGE}\}$ and $k_1 \in \interval{0,n-1} \cup \{\texttt{NONE}\}$. Intuitively, $i = \texttt{LARGE}$ and $k_1 = \texttt{NONE}$ if the first layer $w^{(1)}$ does not contain any column of $\ast$ symbols; otherwise, $k_1 \in \interval{0,n-1}$ is the index of the leftmost column of $\ast$ symbols, and $i$ represents the distance between two such adjacent columns, or \texttt{LARGE} if a single one appears;
\item The \emph{second layer} can be similarly represented by a pair $(j,k_2)$ for ${j \in \interval{i,n} \cup \{\texttt{LARGE}\}}$ and $k_1 \in \interval{0,n-1} \cup \{\texttt{NONE}\}$;
\item Finally, the \emph{density layer} of $w$ is the vertical lift of a Toeplitz subword $v \subpattern T(u)$ for some $u \in \{0,1\}^{\N}$. To describe it, we use (see~\autoref{fig:example-toeplitz}):
  \begin{itemize}[itemsep=1pt,topsep=2pt]
  \item A variable $\bm{u} \in \{0,1\}^{\lceil \log i \rceil + 1}$ representing the prefix of length $\lceil \log i \rceil + 1$ of $u$ (if $i$ is $\texttt{LARGE}$, the variable $\bm{u}$ is of size $\lceil \log (n) \rceil + 1$);
  \item A variable $\gamma \in \{0,1\}^{\lceil \log i \rceil + 1}$ representing the position of each letter\footnote{More precisely, recall that $T(u) = u_{0} u_{1} u_{0} u_{2} u_{0} u_{1}\dots$ (see~\autoref{sec:entropy:effective:toeplitz}). In the Toeplitz subword $v$, either $v_{i} = u_{0}$ for all even $i$ (in which case, the $v_{i}$'s for odd $i$ contain the remaining values of $u$), or $v_{i} = u_{0}$ for all odd $i$ (vice-versa): this alternative is encoded in the first bit $\gamma_0$. Iteratively, the bit $\gamma_i$ encodes which alternative holds for $u_{i}$.} of $\bm{u}$ in $w^{(d)}$.
  \end{itemize}
\end{itemize}

\begin{figure}[ht]
  \centering
  \begin{tikzpicture}[scale=0.4]
    \newcommand{\posmax}{19}

    \clip (-13.5, -3.5) rectangle ($(\posmax + 1.5, 1.5)$);

    \node[anchor=east] at (-9.5, 0) {Toeplitz:};
    \node[anchor=east] at (-9.5, 1) {Level:};
    %% Word: 0110101
    % First level: period 2^1
    \foreach \i in {-8,-6,...,\posmax}{
      \node at (\i, 0) {$0$};
      \node at (\i, 1) {$0$};
    }

    % Second level: period 2^2
    \foreach \i in {-5,-1,...,\posmax}{
      \node at (\i, 0) {$1$};
      \node at (\i, 1) {$1$};
    }

    % Third level: period 2^3
    \foreach \i in {-3,5,...,\posmax}{
      \node at (\i, 0) {$1$};
      \node at (\i, 1) {$2$};
    }

    % Fourth level: period 2^4
    \foreach \i in {-7,9,...,\posmax}{
      \node at (\i, 0) {$0$};
      \node at (\i, 1) {$3?$};
    }

    % Fifth level and above
    \node at (17, 0) {$1$};
    \node at (17, 1) {$4?$};

    \node at (1, 0) {$0$};
    \node at (1, 1) {$8?$};

    \node at (-9 , 0) {\dots};
    \node at (-9, 1) {\dots};
    \node at (\posmax + 1, 0) {\dots};
    \node at (\posmax + 1, 1) {\dots};

    \draw[decorate, decoration={brace, mirror}, very thick] (10.7, -1) -- (\posmax+1, -1)
    node[midway, below, text width=4cm,align=center] {A possible Toeplitz word $T(u)$ containing $v$};

    \path[insert path={{(-8.5,-3.5) rectangle ($(\posmax + 2, 1.5)$)}},draw=none,fill=white,opacity=0.5,overlay,even odd rule] (-5.5,-1) rectangle ++(16, 4);

    \draw[decorate, decoration={brace, mirror}, very thick] (-5.5, -1) -- ++(16, 0)
    node[midway, below] {A Toeplitz subword $v$};
  \end{tikzpicture}
\caption{A Toeplitz subword $v$ and the associated ${\bm{u} = 01100, \gamma=10010}$.}
\details{The Toeplitz subword $v$ of length $2^4$ must contain a bit of level $3$, even though we cannot decide its position (denoted $3?$); and a position of some higher level (denoted $8?$). We use the last two bits of $\bm{u}$ to encode these bits, and set the last two bits of $\gamma$ accordingly.}
  \label{fig:example-toeplitz}
\end{figure}

\noindent In particular, any valid $\mathcal{B}$-pattern $w$ of width $n$ can be represented by $O(\log n)$ bits.

\paragraph*{The sequence of ndill maps $(\tau_{\ell})_{\ell \in \N}$}
We now design a sequence of ndill maps $\bm{\tau} = (\tau_{\ell})_{\ell \in \N}$ of limit space $Y$. The main idea is that, in the intermediate configurations $(y^{\ell})_{\ell \in \N}$ of an $N$-adic sequence
\[ y = y^{0} \rsubstep{\tau_0} y^{1} \rsubstep{\tau_1} y^{2} \rsubstep{\tau_2} y^{3} \dots \]
each tile in the configuration $y^{\ell}$ will each represent the intersection of the $\mathcal{B}$-square patterns in $y$ with the domain of size $2^{\ell} \times 2^{\ell}$ it (= the tile of $y^{\ell}$) induces in $y$. Since this is actually too much information to apply~\autoref{thmC:pixpoint}, we will actually only encode in $y^{\ell}$ the $\mathcal{B}$-square patterns of size $n \geq 2^{\ell-1}$.

\medskip
More precisely, we define a sequence of alphabets $\mathcal{B}_{\ell}$. Each symbol $a \in \mathcal{B}_{\ell}$ will represent various plausible $\mathcal{B}$-patterns appearing in a \emph{simulated domain} $D_{\ell} = \interval{0,2^{\ell}-1}^2$. We thus define $\mathcal{B}_{\ell}$ as the set of possible records $a \in \mathcal{B}_{\ell}$ on the following two fields:
\begin{itemize}
\item A \textbf{density set} $D_{\alpha}$ of rationals $q_{i,j} \in \Q$ for $i,j \in \interval{1,2^{\ell}}$ powers of $2$. Intuitively, each $q_{i,j}$ is a rational number encoded on $\ell$ bits that will satisfy $q_{i,j} \leq \alpha_{i,j}$;
\item A \textbf{rectangle list} $R$ of entries $(\rect{r},w)$, for $\rect{r} \in \rect{R}$ and $w \in \mathcal{B}^{\rect{r}}$, such that:
  \begin{itemize}
  \item For every entry $(\rect{r},w) \in R$, the rectangle $\rect{r}$ is a subset of the simulated domain $D_{\ell}$; furthermore, there could exist a square $\rect{s} \in \rect{S}_{\!f}$ of size $n \geq 2^{\ell-1}$ such that $\rect{r} = \restr{\rect{s}}{D_{\ell}}$;
  \item Given two entries $(\rect{r},w)$ and $(\rect{r}',w')$, the rectangles $\rect{r}$ and $\rect{r}'$ are disjoint.
  \end{itemize}
  The \textbf{rectangle list} intuitively represents the $\mathcal{B}$-patterns of large size (larger than $2^{\ell-1}$) appearing in the simulated domain $D_{\ell} = \interval{0,2^{\ell}-1}^2$ of the symbol $a$.
\end{itemize}
The ndill map $\tau_{\ell}$ will then non-deterministically substitute the configurations $x \in \mathcal{B}_{\ell+1}^{\Z[2]}$ (which, informally, correspond to the configurations of $Y$ containing only the $\mathcal{B}$-squares of size $n \geq 2^{\ell}$) to introduce the $\mathcal{B}$-squares of size $2^{\ell-1} \leq n < 2^{\ell}$.

\begin{figure}
  \newcommand{\lrect}[1]{\textbf{\textcolor{Rainbow#1!85!black}{Rectangle~\lowercase{#1}}}}
  \centering
  \begin{tikzpicture}[scale=.115,decoration={random steps,segment length=0.4em,amplitude=0.15em}]
    \pgfmathtruncatemacro{\ngrid}{32}
    \pgfmathtruncatemacro{\hgrid}{\ngrid/2}
    \pgfmathtruncatemacro{\pad}{20}
    % Original picture
    \begin{scope}[shift={(0,80)}]
      \pgfmathsetseed{6}
      \draw[TilingGrid,decorate,path picture={
        \draw[TilingGrid,opacity=.2] ($(-\pad-4,-\pad-4)$) grid ($(\ngrid+\pad+4,\ngrid+\pad+4)$);
        \foreach[count=\c from 10] \x/\y/\s/\i/\k in {10/8/17/8/4, -15/-66/70/32/5/, -15/24/20/16/12, 40/36/18/8/3, -5/8/9/4/3, 29/11/11/8/5} {
          \pgfmathtruncatemacro{\yb}{max(\y,-\pad-4)}
          \pgfmathtruncatemacro{\lb}{max(\y,-\pad)}
          \pgfmathtruncatemacro{\yt}{min(\y+\s,\ngrid+\pad+4)}
          \pgfmathtruncatemacro{\xm}{\x+\k}
          \pgfmathtruncatemacro{\xn}{\x+\k+\i}
          \pgfmathtruncatemacro{\xt}{min(\x+\s,\ngrid+\pad+4)}
          \pgfmathsetmacro{\Hc}{Hex(\c)}
          \pgfmathsetmacro{\hc}{hex(\c)}
          \ifnum\s>15{
              \draw[thick,Rainbow\Hc,fill=Rainbow\Hc!20!white] (\x,\y) rectangle ++(\s,\s);
              \draw[Rainbow\Hc!80!black,font=\small] ($(\x,\lb) + (1.5,1.5)$) node {\hc};
              \foreach \xs in {\xm,\xn,...,\xt} {
                \ifnum \xs<\xt\relax{
                    \foreach \ys in {\yb,...,\yt} {
                      \ifnum\ys<\yt\relax{
                          \draw[Rainbow\Hc!80!black,font=\tiny] ($(\xs,\ys) + (.5,.5)$) node {$\ast$};
                        }\fi
                    }
                  }\fi
              }
            }\fi
        }
        \draw[opacity=.8,ultra thick,TilingGrid!30!black,step=32] ($(-\pad-4,-\pad-4)$) grid ($(\ngrid+\pad+4,\ngrid+\pad+4)$);
      }] ($(-\pad,-\pad)$) rectangle ($(\ngrid+\pad,\ngrid+\pad)$);
    \end{scope}
    \draw[<-] (16,54) -- ++(0,4) node[midway,right] {$\tau_{\ell}$};
    % Substituted picture
    \begin{scope}
      \pgfmathsetseed{6}
      \draw[TilingGrid,decorate,path picture={
        \draw[TilingGrid,opacity=.2] ($(-\pad-4,-\pad-4)$) grid ($(\ngrid+\pad+4,\ngrid+\pad+4)$);
        \foreach[count=\c from 10] \x/\y/\s/\i/\k in {10/8/17/8/4, -15/-66/70/32/5/, -15/24/20/16/12, 40/36/18/8/3, -5/8/9/4/3, 29/11/11/8/5} {
          \pgfmathtruncatemacro{\yb}{max(\y,-\pad-4)}
          \pgfmathtruncatemacro{\lb}{max(\y,-\pad)}
          \pgfmathtruncatemacro{\yt}{min(\y+\s,\ngrid+\pad+4)}
          \pgfmathtruncatemacro{\xm}{\x+\k}
          \pgfmathtruncatemacro{\xn}{\x+\k+\i}
          \pgfmathtruncatemacro{\xt}{min(\x+\s,\ngrid+\pad+4)}
          \pgfmathsetmacro{\Hc}{Hex(\c)}
          \pgfmathsetmacro{\hc}{hex(\c)}
          \ifnum\s>7{
              \draw[thick,Rainbow\Hc,fill=Rainbow\Hc!20!white] (\x,\y) rectangle ++(\s,\s);
              \draw[Rainbow\Hc!80!black,font=\small] ($(\x,\lb) + (1.5,1.5)$) node {\hc};
              \foreach \xs in {\xm,\xn,...,\xt} {
                \ifnum \xs<\xt\relax{
                    \foreach \ys in {\yb,...,\yt} {
                      \ifnum\ys<\yt\relax{
                          \draw[Rainbow\Hc!80!black,font=\tiny] ($(\xs,\ys) + (.5,.5)$) node {$\ast$};
                        }\fi
                    }
                  }\fi
              }
            }\fi
        }
        % Broken square
        15/29/13/16/16
        \draw[thick,RainbowG,fill=RainbowG!20!white] (17,28) rectangle ++(13,13);
        \draw[RainbowG!80!black,font=\small] ($(17,28) + (1.5,1.5)$) node {g};
        \foreach \ys in {28,...,40} {
          \ifnum\ys<32\relax%
            {
              \foreach \xs in {20,24,...,29} {
                \draw[RainbowG!80!black,font=\tiny] ($(\xs,\ys) + (.5,.5)$) node {$\ast$};
              }
            }%
          \else
            {
              \foreach \xs in {18,26,...,29} {
                \draw[RainbowG!80!black,font=\tiny] ($(\xs,\ys) + (.5,.5)$) node {$\ast$};
              }
            }%
          \fi
        }
        \draw[opacity=.8,ultra thick,TilingGrid!70!black,step=\hgrid] ($(-\pad-4,-\pad-4)$) grid ($(\ngrid+\pad+4,\ngrid+\pad+4)$);
        \draw[opacity=.7,ultra thick,TilingGrid!40!black,step=32] ($(-\pad-4,-\pad-4)$) grid ($(\ngrid+\pad+4,\ngrid+\pad+4)$);
      }] ($(-\pad,-\pad)$) rectangle ($(\ngrid+\pad,\ngrid+\pad)$);
    \end{scope}
  \end{tikzpicture}
  \caption{An application of the ndill map $\tau_{\ell}$ on a configuration.}
  \details{Each tile in the original configuration is substituted towards a pattern of size $2 \times 2$. In the substituted configuration, the ndill map $\tau_{\ell}$ has non-deterministically introduced \lrect{E}, \lrect{F} and \lrect{G} whose sizes are in range ${\interval{2^{\ell-1},2^{\ell}-1}}$.\\
  Since \lrect{G} appears across two adjacent tiles, there is no reason for the ndill map $\tau_{\ell}$ to actually align both halves of the square, or to define the same first layer on both halves of the pattern. However, the next ndill map $\tau_{\ell-1}$ will check all blocks of $3 \times 3$ tiles, detect the error, and refuse to substitute this configuration any further: in particular, such errors cannot appear in $\limspace{X}{\bm{\tau}} \cap Y_{\rect{S}}$.}
\end{figure}

More precisely, given an input pattern $u \in \mathcal{B}_{\ell+1}^{\{-1,0,1\}^2}$, the ndill map $\tau_{\ell}$ is defined as the following algorithm:
\begin{enumerate}
\item We first check that all input symbols $u_{p}$ (for $p \in \{-1,0,1\}^2$) have the same \textbf{density set}. Furthermore, we check this density set against an enumeration of the $\Pi_{1}$ real numbers~$\alpha_{i,j}$. More precisely, for every $q_{i,j} \in \Q$ in the \textbf{density set} of $u_{(0,0)}$:
  \begin{itemize}
  \item Compute some $\ell$ steps of computations of a rational (upper) approximation of $\alpha_{i,j}$, and check that $q_{i,j}$ is indeed smaller.
  \end{itemize}
\item We then non-deterministically guess a global \textbf{rectangle list} $R'$ for the whole input pattern of size $3 \times 3$, \textit{i.e.}~for a simulated domain $D_{\ell+1}' = \interval{-2^{\ell+1},2 \cdot 2^{\ell+1}-1}$. More precisely, we guess a set of pairs $(\rect{s},w)$ of squares $\rect{s} \in \rect{S}_{\!f}$ and associated patterns $w \in \mathcal{B}^{\rect{r}}$ (using the representation defined above).
  \begin{itemize}
  \item We first check that all the squares $\rect{s}$ appearing in $R'$ are indeed disjoint squares.
  \item For each entry $(\rect{s},w)$ in $R'$, we consider the variables $i \in \N$, $j \in \N$ and ${\bm{u} \in \{0,1\}^{\lceil \log i \rceil + 1}}$ from the representation of $w$. Considering $\bm{u}$ as the binary expansion of some rational number~$q_{i,j}'$, we must have $q_{i,j}' \leq q_{i,j}$ (as seen in the \textbf{density set} of $u_{(0,0)}$);
  \end{itemize}
\item We then check the geometry of the input \textbf{rectangle lists} against $R'$: for any input symbol $u_p$ for $p \in \interval{-1,1}^2$ and for any entry $(\rect{r},w)$ in the \textbf{rectangle list} of $u_{p}$, there must exist some corresponding entry $(\rect{s},w')$ in $R'$ such that
  \begin{itemize}
  \item The square $\rect{s}$ has size $n \times n$ for some $n \geq 2^{\ell}$;
  \item The rectangle $\rect{r}$ coincides with the restriction of $\rect{s}$ on the corresponding quadrant;
  \item The pattern $w$ coincides with the restriction of $w'$ on the corresponding quadrant\footnote{We do not get into details here, but the representations of plausible patterns involve some tedious book-keeping: the values of $k_1$, $k_2$ and $\gamma$ may change due to the geometric translations of the quadrants, the values of $i$ (resp.~$j$) should coincide (unless the restriction of $w'$ on $\rect{r}$ now contains a single $\ast$-column), etc\dots};
  \end{itemize}
\item We then define the output pattern $v \in \mathcal{B}_{\ell}^{2 \times 2}$: for every quadrant index $p \in \interval{0,1}^2$, the symbol~$v_{p}$ is a tuple composed of:
  \begin{itemize}
  \item A \textbf{density set} of rationals $\{ q_{i,j}' \}$ for $i,j \in \interval{1,2^{\ell}}$ powers of $2$, where $q_{i,j}'$ is the truncation of the binary expansion of $q_{i,j}$ on $\ell$ bits;
  \item A \textbf{rectangle list} of entries $(\rect{r},w)$ coinciding with the restriction of the entries of the \textbf{rectangle list} $R'$ on the quadrant $2^{\ell} \cdot p + \interval{0,2^{\ell}-1}^2$.
  \end{itemize}
\end{enumerate}
At any point in this algorithm, if any check is not satisfied, then the associated computation crashes and returns nothing. Thus, the substitution of a whole configuration can only involve accepting checks and computations.

\medskip
Let us now consider the resulting limit space $\limspace{X}{\bm{\tau}}$. Denoting by $Y_{\rect{S}} \subseteq \mathcal{B}_{\texttt{\#}}^{\Z[2]}$ the set of configurations composed of disjoint $\mathcal{B}$-square patterns over a $\texttt{\#}$ background, and identifying $\mathcal{B}_0$ with $\mathcal{B}$ by associating empty \textbf{rectangle lists} with $\texttt{\#}$ and \textbf{rectangle lists} $\{ (\{(0,0)\},w) \}$ with~$w$, we claim that:
\begin{clm}
  $\limspace{X}{\bm{\tau}} \cap Y_{\rect{S}} = Y$.
\end{clm}
\begin{proof}
  \emph{We first prove that $Y \subseteq \limspace{X}{\bm{\tau}} \cap Y_{\rect{S}}$}. Fix $y \in Y$ and fix $(\rect{s}_{j})_{j \in J}$ the set of disjoint squares corresponding with the domains of $\mathcal{B}$-patterns in $y$. At every level $\ell \in \N$, we define a configuration $y^{\ell} \in \mathcal{B}_{\ell}^{\Z[2]}$ as follows: at position $p \in \Z[2]$, the symbol $y^{\ell}_{p}$ is the record composed~of
  \begin{itemize}
  \item A \textbf{density set} of rationals $\{q_{i,j}\}$ where: $i,j \in \interval{1,2^{\ell}}$ are powers of $2$, and $q_{i,j}$ is the truncation of $\alpha_{i,j}$ on $\ell$ bits;
  \item A \textbf{rectangle list} of entries $(\rect{r},w)$, where $\rect{r}$ and $w$ are the restrictions of the squares $\rect{s}_j$ and patterns $y|_{\rect{s}_j}$ with the domain $2^{\ell} \cdot p + \interval{0,2^{\ell}-1}^2$, and $\rect{s}_{j}$ is a square of size $n \geq 2^{\ell-1}$.
  \end{itemize}
  Then this sequence of configurations $(y^{\ell})_{\ell \in \N}$ satisfies $y^{\ell+1} \substep{\tau_{\ell}} y^{\ell}$, so that $y \in \limspace{X}{\bm{\tau}}$.

  \medskip
  \emph{For the converse direction, we prove that $\limspace{X}{\bm{\tau}} \cap Y_{\rect{S}} \subseteq Y$}. Let $y \in Y_{\rect{S}}$ a configuration that is also a limit configuration in $\limspace{X}{\bm{\tau}}$. Since $y \in Y_{\rect{S}}$, there exists some disjoint squares $(\rect{s}_j)_{j \in J}$ corresponding to the $\mathcal{B}$-patterns in $y$. And since $y$ is a limit configuration, there exists a sequence of configurations $y^{0},y^{1},y^{2},\dots$ such that $y^{\ell+1} \substep{\tau_{\ell}} y^{\ell}$ for every $\ell \in \N$.

  Inductively, we notice that the rationals $q_{i,j}$ in the \textbf{density sets} of the configurations $y^{\ell}$ increase with $\ell$, but that with increasing values of $\ell$ the ndill maps $\tau_{\ell}$ enumerate increasingly precise rational upper bounds on $\alpha_{i,j}$. Thus, in any \textbf{density set} in any $y^{\ell}$, the rational $q_{i,j}$ satisfies $q_{i,j} \leq \alpha_{i,j}$.

  Then, during any substitution step $y^{\ell+1} \substep{\tau_{\ell}} y^{\ell}$, the ndill map $\tau_{\ell}$ checks all blocks of $3 \times 3$ tiles in the configuration $y^{\ell+1}$. In particular, it entirely covers every square $\rect{s}_{j}$ of size $n \times n$ for $2^{\ell} \leq n \leq 2^{\ell+1}$ in the configuration $y$. Furthermore, its input checks on the \textbf{rectangle lists} ensure that the corresponding patterns $y|_{\rect{s}_j}$ are plausible; and if its first and second layers encode some integers $i,j \in \interval{1,n}$, then its density layer contains the Toeplitzification of the binary expansion of some rational number $q \in \Q$ such that $q \leq q_{i,j}$ (and, thus, $q \leq \alpha_{i,j})$. In particular, we deduce that $y \in Y$.
\end{proof}

We then conclude by considering the computability of the ndill maps $\tau_{\ell}$:
\begin{itemize}
\item Since the \textbf{rectangle list} in any symbol $a \in \mathcal{B}_{\ell+1}$ consists of restrictions of disjoint squares of size $n \geq 2^{\ell}$ intersecting the domain $\interval{0,2^{\ell+1}-1}^2$, all \textbf{rectangles lists} are of uniformly bounded lengths and all its entries can be encoded on $O(\ell)$ bits. Since for any rational $q_{i,j}$ in the \textbf{density set} of a symbol $a \in \mathcal{B}_{\ell+1}$ the integers $i$ and $j$ must be powers of $2$, there are at most $O(\ell^2)$ entries in such a \textbf{density set}, each being encoded on $O(\ell)$ bits.\\
  In particular, all symbols in the alphabet $\mathcal{B}_{\ell+1}$ can be encoded on $O(\ell^3)$ bits.
\item Since arithmetic operations on large integers can be performed in quasi-linear time in the $\log$-RAM model, we similarly deduce that all the ndill maps $\tau_{\ell}$ can be uniformly computed in time $\mathrm{poly}(\ell)$.
\end{itemize}

By~\autoref{thmC:pixpoint}, the subshift $Y$ is thus sofic. This concludes the proof.

\section{Characterization of extender entropy dimensions}

The classical entropy dimensions of multidimensional subshifts of finite type were classified in computational terms in~\cite{meyerovitch_growth_type_invariants}:
\begin{thmC}[\cite{meyerovitch_growth_type_invariants}]\label{thm:meyerovitch_entropy_dimensions}
  For any $d \geq 2$,
  \begin{enumerate}
  \item The class of upper entropy dimensions of $\mathbb{Z}^{d}$ SFTs is
    $[0,d] \cap \Pi_{3}$
  \item The class of lower entropy dimensions of $\mathbb{Z}^{d}$ SFTs is
    $[0,d] \cap \Sigma_{2}$
  \item The class of entropy dimensions of $\mathbb{Z}^{d}$ SFTs is
    $[0,d] \cap \Delta_{2}$
  \end{enumerate}
  The same holds for sofic and effective subshifts instead of SFTs.
\end{thmC}

In this section, we prove similar results in the case of \emph{extender} entropy dimensions (\textit{c.f.}~\autoref{sec:ext:extender-entropy-dimension}) of sofic and effective subshifts.

\subsection{Minimal subshifts}

We begin by briefly mentioning the case of minimal subshifts:
\begin{prop}\label{prop:extender-entropy-dim-minimal-sft}
  For $d \geq 2$, the set of extender entropy dimensions of minimal $\mathbb{Z}^{d}$ SFTs
  is exactly $[0,d-1] \cap \Delta_{2}$.
\end{prop}

\begin{proof}
  S.~Gangloff shows in \cite[Theorem
  2]{gangloff22_charac_entrop_dimen_minim_mutid} that for $d \geq 2$, the
  extender entropy dimensions of minimal SFTs are exactly the real numbers of
  $[0, d-1] \cap \Delta_{2}$. Then, \autoref{prop:minimal-inclusion-ext-decidable}
  shows that the complexity sequence is equal to the extender set sequence, and
  we immediately get that $\entdim_{h_{E}}(X) = \entdim(X)$ for a minimal SFT $X$.
\end{proof}

Besides this easy consequence of~\cite{gangloff22_charac_entrop_dimen_minim_mutid}, we do not investigate further
the extender entropy dimensions of minimal \textit{sofic} (or effective)
subshifts: in any case, \autoref{prop:minimal-inclusion-ext-decidable} shows that
the extender entropy dimension coincides with the usual entropy dimension.

\subsection{Effective subshifts}\label{sec:entropy-dim:effective}

\begin{restatethm}{thm:ext-entropy-dim-effective}
  For any $d \geq 1$,
  \begin{enumerate}
  \item The class of upper extender entropy dimensions of $\mathbb{Z}^d$ effective subshifts is $[0,d] \cap \Pi_{3}$;
  \item The class of lower extender entropy dimensions of $\mathbb{Z}^d$ effective subshifts is $[0,d] \cap \Sigma_{4}$;
  \item The class of extender entropy dimensions of $\mathbb{Z}^d$ effective subshifts is $[0,d] \cap \Delta_{3}$.
  \end{enumerate}
\end{restatethm}

\begin{clm}[Upper bounds]
  (Upper, lower) extender entropy dimensions of effective $\Z[d]$ subshifts are respectively $\Pi_3$, $\Sigma_4$ and $\Delta_3$ real numbers.
\end{clm}
\begin{proof}
  Recall that for $X$ a fixed effective subshift, the sets $\{k \leq \card{E_{X}(n)}\}$ are uniformly $\Sigma^{0}_{2}$-computable sets by~\autoref{lem:ext-formula-upper-bound}.
This implies that $\frac{\log\log\, \card{E_{X}(n)}}{\log n}$ are uniform $\Sigma_{2}$ real numbers. As $\limsup$, $\liminf$ and $\lim$ respectively of sequences of uniform $\Sigma_{2}$ real numbers, we obtain the claimed upper bounds.
\end{proof}

The realization of every $\Pi_{3}$ (resp. $\Sigma_{4}$, $\Delta_{3}$) real number of~$[0,d]$ as upper extender entropy dimensions (resp. lower extender entropy dimensions, extender entropy dimensions) reduces to the case of $\Z$ subshifts by \autoref{clm:ext-entropy-dim-lifts}. Then, we follow very closely the construction of \autoref{thm:ext-entropy-effective} (and keep the associated notations), as we will only replace the density layer with a counterpart suited for extender entropy \emph{dimensions}. Note that we actually prove something stronger than~\autoref{thm:ext-entropy-dim-effective}: we construct a single subshift $Z_{\alpha}$ having both a prescribed lower and upper extender entropy dimensions; we maintain this formulation for clarity's sake and for an easier parallel with~\autoref{thm:meyerovitch_entropy_dimensions}.

\paragraph*{Preliminaries}
Given $\alpha \in [0,1]$, our goal is to construct an effective $\Z$ subshift $Z_{\alpha}$ such that $\card{E_{Z_{\alpha}}(n)} \sim 2^{n^{\alpha}}$. Similarly to extender entropies, this is obtained in two steps: we build (``proper'') patterns with different extender sets; and we ensure that we have the correct number of such patterns. For the rest of the section, we fix some $\gamma > 0$ and a sequence $(\alpha_{i,j})_{i,j \in \N}$ of uniformly $\Pi_{1}$ real numbers in $[\gamma, 1-\gamma]$, and define:
\begin{align*}
  \alpha_{i} &= \sup_{j} \alpha_{i,j} \text{ for } i \geq 0 \\
  \overline{\alpha} &= \limsup_{i} \alpha_{i}\\
  \underline{\alpha} &= \liminf_{i} \alpha_{i}
\end{align*}

Applying~\autoref{lem:arith-hierarchy-slowdown-lemma}, we replace $(\alpha_{i,j})_{i,j}$ by a uniform sequence of $\Pi_{1}$ real numbers in $[\gamma,1-\gamma]$ that satisfies its conclusions for a polynomial
$Q(n) = O(n^{1/\gamma})$. In the following section, we will construct an effective subshift $Z_{\alpha}$ such that
$\upentdim_{h_{E}}(Z_{\alpha}) = \overline{\alpha}$ and
$\lowentdim_{h_{E}}(Z_{\alpha}) = \underline{\alpha}$. Note in particular that
this is a small abuse of notation: the final subshift depends on the sequence
$(\alpha_{i,j})$ and not only on $(\overline{\alpha}, \underline{\alpha})$.

\medskip
Recall from~\autoref{sec:entropy:effective:encoding-integer-in-shifts} that for $0 \leq k < i$,
$\encode{i}_{k}$ is an $i$-periodic configuration on
$\mathcal{A}_{\ast} = \{\ast, \blank\}$ with exactly one symbol $\ast$ per
period. We also define a new set of \defn{density words} on the alphabet
$\mathcal{A}_{d} = \{0, 1\}$ as follows: for $0 < i \leq j$, write $W_{i, j}$
the set of words $w$ of length $i$ satisfying (see~\autoref{fig:example-density-ext-entropy-dim} for an illustration):
\begin{enumerate}
\item The number of symbols $1$ in a density word $w$ satisfies $\card{w}_{1} \leq \lfloor i^{\alpha_{i,j}}\rfloor$.
\item These symbols ``1'' are evenly spaced, \textit{i.e.}~$w$ is a prefix of a word of
  $(10^{k})^{\ast}$ for some~$k$.
\end{enumerate}

\paragraph*{Auxiliary subshift $Z'_{\alpha}$}
Following~\autoref{sec:entropy:effective:final-construction}, we define $Z'_{\alpha}$ on the
following three layers:
\begin{enumerate}
\item \defn{First layer $L_{1}$:} We take $L_1= X_\ast$ to encode integers
  $i \in \N$. Intuitively, $i$ will denote which $\Sigma_2$ number $\alpha_i$ is
  approximated in the configuration.

\item \defn{Second layer $L_{2}$:} We also set $L_2 = X_\ast$ to encode integers
  $j \in \N$, $j \geq i$. Intuitively, $j$ will denote which $\Pi_1$ number
  $\alpha_{i,j}$ is approximated in the configuration.

\item \defn{Density layer $L_{d}$:} We define the density layer as
  $L_d = \{0,1\}^{\Z}$. Whenever the first two layers are non-degenerate, this
  layer will be restricted to configurations where patterns of length $i$
  contain $\lesssim i^{\alpha_{i,j}}$ symbols ``1''.
\end{enumerate}

Formally, we define an auxiliary subshift $Z_{\alpha}'$:
\begin{align*}
  Z_{\alpha}' &= \Big\{ (z^{(1)}, z^{(2)}, z^{(d)}) \in L_1 \times L_2 \times L_d \mid  z^{(2)} \in \encode{\infty} \Big\} \\
              & \hspace{0.2cm} \cup \bigcup_{i \in \N} \bigcup_{j \geq i} \Big\{ (z^{(1)}, z^{(2)}, z^{(d)}) \in L_1 \times L_2 \times L_d \mid \exists k_1,k_2 \in \N,\\[-10pt]
              & \hspace{3cm} z^{(1)} = \encode{i}_{k_1},\, z^{(2)} =
                \encode{j}_{k_2}\,\text{and } \exists w \in W_{i,j},\, \forall
                p \in \mathbb{Z},\, z^{(d)}_{p+k_{1}} = w_{p \bmod i} \Big\}
\end{align*}
where $W_{i,j}$ is the set of \emph{density words} defined above. For reasons similar to~\autoref{clm:effective-effectivity}, the subshift $Z_{\alpha}'$ is effective.

\begin{figure}[ht]
  \centering
  \begin{tikzpicture}[scale=0.45]
    \clip (-15.5, -3) rectangle (15.5, 2);
    % origin
    \draw[thin, red!30]  (0, -0.5) --++ (0, 1);

    \foreach \i in {-15,...,15}{
      \node at (\i, 0) {0};
    }

    \foreach \i in {-23.4,-13.4,...,10}{%
      \draw[decorate, decoration={brace, mirror}, thick] (\i, -1) --++ (9.8, 0);
    }

    \node[below] at (1.5, -1) {$w \in W_{10, j}$, a period of length 10};

    \foreach \i in {-15, -13, -9, -5, -3, 1, 5, 7, 11, 15}{
      \node[fill=white] at (\i, 0) {1};
    }

    \node (one) at (1, 1.5) {Evenly spaced ``1'' inside a period};

    % \draw[decorate, decoration={brace}, thick] (-3.4, 1) --++ (3.8, 0);
    % \draw[decorate, decoration={brace}, thick] (0.6, 1) --++ (3.8, 0);

    \foreach \i/\j in {-3/a, 1/b, 5/c}{
      \node[draw, rectangle, inner sep=0pt, minimum size=10pt] (one\j) at (\i, 0) {};
    }

    \draw (one.south) -| (onea.north);
    \draw (one.south) -- (oneb.north);
    \draw (one.south) -| (onec.north);
  \end{tikzpicture}
  \caption{Part of a valid configuration on the density layer, with $i = 10$,
    $j \geq i$, and $\alpha_{i,j} = 0.5$ so that
    $\lfloor i^{\alpha_{i,j}}\rfloor = \lfloor \sqrt{10} \rfloor = 3$. Hence,
    there are at most 3 symbols ``1'' per $i$-period. The vertical red bar
    indicates the origin.}
  \label{fig:example-density-ext-entropy-dim}
\end{figure}

\paragraph*{Adding free bits}

\begin{enumerate}[start=4]
\item \defn{Free layer $L_{f}$:} We define the free layer as $L_f = \{\freezero,\heads,\tails\}^{\Z}$. Given the synchronizing map $\pisync \colon \{\freezero,\heads,\tails\} \to \{0,1\}$ defined as $\pisync(\freezero) = 0$ and $\pisync(\heads) = \pisync(\tails) = 1$, we say that two configurations $z^{(d)} \in L_d$ and $z^{(f)} \in L_f$ are \defn{synchronized} if $\pisync(z^{(f)}) = z^{(d)}$.
\end{enumerate}

\noindent and we can now define the desired subshift $Z_\alpha$ as:
\begin{align*}
  Z_{\alpha} &= \Big\{ (z^{(1)}, z^{(2)}, z^{(d)},z^{(f)}) \in L_1 \times L_2 \times L_d \times L_f \mid  z^{(1)} \in \encode{\infty} \text{ or } z^{(2)} \encode{\infty} \Big\} \\
             & \cup \bigcup_{i \in \N} \bigcup_{j \geq i} \Big\{ (z^{(1)}, z^{(2)}, z^{(d)}, z^{(f)}) \in L_1 \times L_2 \times L_d \times L_f \mid \exists k_1,k_2 \in \N,\\[-7pt]
             & \hspace{3cm} z^{(1)} = \encode{i}_{k_1},\, z^{(2)} = \encode{j}_{k_2}\, \pisync(z^{(f)}) = z^{(d)}, \\
             & \hspace{3cm} \exists w \in W_{i,j},\, \forall
               p \in \mathbb{Z},\, z^{(d)}_{p+k_{1}} = w_{p \bmod i} \,\text{and } z^{(f)} \text{ is $i$-periodic }\Big\}.
\end{align*}

We then claim that $\overline{\alpha}$ and $\underline{\alpha}$ are realized as extender entropy dimensions:
\begin{lem}\label{lem:ext-entropy-dim-counting}
  $\upentdim_{h_{E}}(Z_{\alpha}) = \overline{\alpha}$ and $\lowentdim_{h_{E}}(Z_{\alpha}) = \underline{\alpha}$
\end{lem}

\begin{proof}
  The notion of proper and degenerate patterns and configurations can be defined as in~\autoref{sec:entropy:effective:final-construction}. The equivalent of~\autoref{clm:effective-aux-counting-degenerate} and~\autoref{clm:effective-aux-ext-proper} about the number of extender sets generated by degenerate and proper patterns still hold: in particular, we only have to compute the number $\card{P(n)}$ of proper patterns of a given size $n \in \N$.

  The lower bound $2^{n^{\alpha_{n}}-1} \leq \card{P(n)}$ is obtained exactly as in \autoref{lem:effective-counting-ext}. As for the upper bound, we also follow~\autoref{lem:effective-counting-ext} and overestimate the number of proper patterns $\card{P(n)}$ by considering the restrictions $w' = z'_{\interval{0,n-1}}$, where $z' = (\encode{i}_{k_{1}}, \encode{j}_{k_{2}}, z^{(d)})$ ranges over the proper configurations of $Z_{\alpha}'$, and bound the number of symbols $1$ of the density layer in each case:
  \begin{itemize}
  \item If $i \leq n$, an $i$-period of the density layer $w'^{(d)}$ contains less than $i^{\alpha_{i}}$ symbols~$1$ by definition of the density words of $W_{i,j}$;
  \item For $i > n$, $w'^{(d)}$, we cannot immediately bound the number of symbols ``1'' in the density layer by some function of $\alpha_n$ (as done in \autoref{clm:effective-aux-counting-degenerate}), as we cannot assume that the sequence $(\alpha_{i})_{i \in \mathbb{N}}$ is decreasing. \\
    Nevertheless, we can count the number of symbols $1$ in $w'^{(d)}$ as follows: by definition, for a word $u \in W_{i,j}$, the number of symbols~$1$ in $u$ satisfies $\card{u}_{1} \leq i^{\alpha_{i,j}} \leq i^{\alpha_{i}}$. Since $w'^{(d)}$ is a subword of size $n$, and that symbols~$1$ are evenly spaced, we deduce that $\card{w'^{(d)}}_{1} \leq \frac{n}{i}\cdot i^{\alpha_{i}} + O(1)$.
  \end{itemize}

    Recall that $\alpha_i$ is assumed to be bounded by $\gamma < \alpha_i < 1 - \gamma$: for $i > n^{1/\gamma}$, we obtain that the number of symbols~$1$ in $w'^{(d)}$ satisfies $\card{w}_{1} = O(1)$, since $\frac{n}{i}\cdot i^{\alpha_{i}} < 1$. Thus, all indices $i > n^{1/\gamma}$ generate together polynomially many (actually, $O(n^4)$) proper patterns.

  Since each symbol $1$ in an $i$-period of the density layer results in two distinct patterns in the free layer, and there are less than $O(i^2)$ possibilities for such periods, we obtain when summing individually for $i \leq n$, $n < i < n^{1/\gamma}$ and $i > n^{1/\gamma}$:
  \begin{align*}
    \card{P(n)} & \leq \sum_{i = 1}^{n} \sum_{k_1 = 0}^{i-1} \sum_{j=1}^{n}
           \sum_{k_2=0}^{j-1} O(i) \cdot 2^{i^{\alpha_{i}}} +
           \sum_{i=n+1}^{n^{1/\gamma}}\sum_{k_1 = 0}^{n} \sum_{k_2 = 0}^{n}
                  O(n^{2}) \cdot 2^{\frac{n}{i}\cdot i^{\alpha_{i}} + O(1)}
                  + O(n^4)
  \end{align*}

  In the earlier proof of~\autoref{lem:effective-counting-ext}, the sequence $(\alpha_{i})_{i \in \N}$ was decreasing. This allowed to conclude with a few lines of computations, but this assumption no longer holds here. Instead, we turn towards the slowdown lemma (\autoref{lem:arith-hierarchy-slowdown-lemma}): we can bound the variations between consecutive $\alpha_i$'s on ranges of polynomial size. More precisely, recall that we assume that the sequence $(\alpha_{i})_{i \in \N}$ satisfies the conclusions of said lemma for a polynomial $Q(n) = n^{\lceil 1/\gamma \rceil}$.

  \medskip
  Let us fix $0 < \varepsilon < \gamma$: by the previous paragraph, there exists $N$ such that for all $N < m$, we have $\smash{\sum_{i=m}^{m^{1/\gamma}} |\alpha_{i+1}-\alpha_{i}| < \varepsilon}$. Applying this lemma for $n^{\gamma} > N$, we obtain:
  \[ \sum_{i=n^{\gamma}}^{n} \big|\alpha_{i+1}-\alpha_{i}\big| < \varepsilon
    \qquad \text{and} \qquad
    \sum_{i=n}^{n^{1/\gamma}}\big|\alpha_{i+1}-\alpha_{i}\big| < \varepsilon.\]
  And in particular: $\forall i \in \interval{n^\gamma, n^{1/\gamma}},\; \big|\alpha_{i} - \alpha_{n}\big| < \varepsilon$.

  \medskip
  We now split the sums over $i$ in three parts: $1 \leq i \leq n^{\gamma}$, $n^{\gamma} < i \leq n$ and $n < i \leq n^{1/\gamma}$:

 %   \vspace{-1.6\baselineskip}
  \begin{align*}
    \intertext{For $1 \leq i \leq n^{\gamma}$, we bound crudely $\alpha_{i} \leq 1$ and obtain:}
    \sum_{i=1}^{n^{\gamma}} 2^{i^{\alpha_i}} & \leq \sum_{i=1}^{n^{\gamma}} 2^{i} \leq 2^{n^{\gamma}+1}; \\
    \intertext{For $n^{\gamma} < i \leq n$, we bound $|\alpha_{i} - \alpha_{n}| < \varepsilon$ and obtain:}
    \sum_{i=n^\gamma+1}^{n} 2^{i^{\alpha_i}} & \leq \sum_{i=n^{\gamma}+1}^{n} 2^{i^{\alpha_n+\varepsilon}} \leq 2^{n^{\alpha_n+\varepsilon}+1}; \\
    \intertext{For $n < i \leq n^{1/\gamma}$, we bound $\frac{n}{i}\cdot i^{\alpha_{i}} \leq \frac{n}{i}\cdot i^{\alpha_{n} + \varepsilon}$ since $|\alpha_{i} - \alpha_{n}| < \varepsilon$, and then $\frac{n}{i}\cdot i^{\alpha_{n} + \varepsilon} \leq n^{\alpha_{n} + \varepsilon}$ since $\alpha_{n} + \varepsilon - 1 < 0$; and obtain:}
    \sum_{i=n+1}^{n^{1/\gamma}} 2^{\frac{n}{i} \cdot i^{\alpha_{i}}} & \leq \sum_{i=n+1}^{n^{1/\gamma}} 2^{n^{\alpha_n+\varepsilon}} \leq \mathrm{poly}(n) \cdot 2^{n^{\alpha_n + \varepsilon}}
  \end{align*}

  And grouping these sums together, we conclude:
  \[ \card{P(n)} \leq \mathrm{poly}(n) \cdot 2^{n^{\alpha_n + \varepsilon}}. \]

  Since there are polynomially many extender sets of degenerate patterns of size
  $n$, and as proper patterns generate different extender sets, we get
  $\upentdim_{h_{E}}(Z_{\alpha}) = \limsup_{n} \alpha_{n} = \overline{\alpha}$,
  and
  $\lowentdim_{h_{E}}(Z_{\alpha}) = \liminf_{n} \alpha_{n} =
  \underline{\alpha}$.
\end{proof}

We now use this construction to prove the full characterization of extender entropy dimensions in~\autoref{thm:ext-entropy-dim-effective}:
\begin{proof}[Proof of~\autoref{thm:ext-entropy-dim-effective} on $\Z$]
  For lower extender entropy dimensions, let $\alpha \in \Sigma_{4} \cap [0,1]$:
  \begin{itemize}
  \item If $\alpha = 0$, let $X = \{0^{\mathbb{Z}}\}$. Then, $\lowentdim_{h_{E}}(X) = 0$.
  \item If $\alpha = 1$, the one-dimensional mirror shift on two letters $X = \{x \in \{0,1,\ast\}^{\Z} \mid {|x|_{\ast} \leq 1 \ \wedge\ \forall p_0,p \in \Z,\, (x_{p_0} = \ast) \implies (x_{p_0+p} = x_{p_0-p})} \}$ satisfies $\lowentdim_{h_{E}}(X) = 1$ (see the proof of~\autoref{clm:mirror}).
  \item Otherwise, by definition, there is a sequence of uniformly $\Pi_{1}$ real numbers $(\alpha_{i,j})$ such that $\alpha = \liminf_{i} \sup_{j} \alpha_{i,j}$. Without loss of generality, we can assume that there exists $\gamma > 0$ such that $\gamma < \alpha_{i,j} < 1 - \gamma$. Let $Z_{\alpha}$ be the subshift constructed above: by~\autoref{lem:ext-entropy-dim-counting}, $\lowentdim_{h_{E}}(Z_{\alpha}) = \underline{\alpha}$.
  \end{itemize}

  The same proof goes through for the lower extender entropy and extender entropy dimensions.
\end{proof}

\subsection{Sofic subshifts}\label{sec:entropy-dim:sofic}

In the one-dimensional case $d=1$, sofic subshifts have a bounded number of extender sets \cite[Lemma~3.4]{ormes_pavlov16_exten_sets_multid_subsh}. Thus:
\begin{clm}\label{clm:ext-entropy-dim-sofic-1d}
  For $d=1$, a sofic subshift $Y \subseteq \mathcal{A}^{\Z}$ satisfies $\upentdim(Y) = \lowentdim(Y) = \entdim(Y) = 0$.
\end{clm}

In dimension $d \geq 2$, we once again recover the same behaviors as effective subshifts:
\begin{restatethm}{thm:ext-entropy-dim-sofic}
  For any $d \geq 2$,
  \begin{enumerate}
  \item The class of upper extender entropy dimensions of $\mathbb{Z}^d$ sofic subshifts is $[0,d] \cap \Pi_{3}$;
  \item The class of lower extender entropy dimensions of $\mathbb{Z}^d$ sofic subshifts is $[0,d] \cap \Sigma_{4}$;
  \item The class of extender entropy dimensions of $\mathbb{Z}^d$ sofic subshifts is $[0,d] \cap \Delta_{3}$.
  \end{enumerate}
\end{restatethm}

As in the previous proofs, we reduce from arbitrary $\Z[d]$ ($d \geq 2$) to $\Z[2]$ using~\autoref{clm:ext-entropy-dim-lifts}.
\begin{proof}[Sketch of proof for~\autoref{thm:ext-entropy-dim-sofic} on {$\Z[2]$}]
  Similarly to previous proofs (\textit{c.f.}~\autoref{sec:entropy:sofic:final-construction}), we prove \autoref{thm:ext-entropy-dim-sofic} by generalizing the effective construction of extender entropy dimensions of~\autoref{sec:entropy-dim:effective} from $\Z$ to $\Z[2]$. Let us briefly sketch the required changes.

  \smallskip
  Using the notation from \autoref{sec:entropy-dim:effective}, we fix a sequence $(\alpha_{i,j})$ of uniformly $\Pi_{1}$ real numbers in $(\gamma,2-\gamma)$ and define the following three layers on $\Z^2$:
  \begin{itemize}
  \item We define the first two layers as $\lift{L_1} \times \lift{L_2}$, where $L_1$ and $L_2$ are the first two layers of the effective $\Z$ subshift $Z_{\alpha}'$;
  \item We define the \emph{density layer} as $L_d = \{0,1\}^{\Z^2}$.
  \end{itemize}
  To bound the number of symbols $1$ in the \emph{density layer}, we replace the density words $W_{i,j}$ of the effective $\Z$ subshift $Z_{\alpha}'$ with \emph{density squares} $S_{i,j} \subset \{0, 1\}^{\interval{0, i-1}^{2}}$, where: the number of symbols~$1$ in any pattern $w \in S_{i,j}$ is bounded by $\card{w}_{1} \leq \lfloor i^{\alpha_{i,j}}\rfloor$; and said symbols are evenly spaced in a grid-like fashion\footnote{That is: there exists $0 \leq m < i$ such that $w_{(p_{1},p_{2})} = 1$ if and only if $(p_{1}, p_{2}) = (0, 0) \mod (m, m)$.}.  We then denote by $Y_{\alpha}'$ the auxiliary $\Z[2]$ subshift on these three layers such that, in configuration $(\lift{\encode{i}_{k_1}}, \lift{\encode{j}_{k_2}},y^{(d)})$, the density layer $y^{(d)}$ is an $i \times i$-periodic configuration, whose period is a density square $w \in S_{i,j}$.

  \smallskip
  We claim that $Y_{\alpha}'$ is a sofic $\Z[2]$ subshift, as it can be built as follows:
  \begin{itemize}
  \item Denote $\alpha_{i,j}' = \alpha_{i,j} / 2$. The $(\alpha_{i,j}')$ are a sequence of uniformly $\Pi_1$-real numbers in $(\frac{\gamma}{2},1-\frac{\gamma}{2})$. Thus, we can redefine the effective subshift $\smash{Z'_{\alpha'}}$ from the $\Z$ effective construction of extender entropy dimensions (\autoref{thm:ext-entropy-dim-effective});
  \item Lifting $Z'_{\alpha'}$ to $\Z[2]$, we obtain a sofic subshift, where $i \times i$-periods in the density layer of non-degenerate configurations contain at most $\smash{\lfloor i^{\alpha_{i,j}'} \rfloor}$ evenly-spaced columns of symbols~$1$;
  \item Since the horizontal spacing between adjacent columns is already fixed, we can introduce a square-grid (as in~\autoref{clm:sofic-soficity}) to also mark $\smash{\lfloor i^{\alpha_{i,j}'} \rfloor}$ rows, so that $i \times i$-periods in the density layer of non-degenerate configurations contain $\simeq i^{2 \alpha_{i,j}} = i^{\alpha_{i,j}}$ marked symbols~$1$. We then use a projection mapping non-marked symbols to $0$, and marked symbols $1$ to $1$.
  \end{itemize}

  \smallskip
  To generate the desired number of extender sets from $Y_{\alpha}'$, we once again introduce free bits on another layer, and a marker layer (as in~\autoref{sec:entropy:sofic:final-construction}) to periodize them. Following the marking argument from~\autoref{clm:sofic-ext-proper}, the number of extender sets associated with proper patterns is directly related to the number of symbols $1$ in the $i \times i$-periods of non-degenerate configurations. Computing the upper and lower bounds on their number then becomes a routine adaptation of~\autoref{lem:ext-entropy-dim-counting}.
\end{proof}

\printbibliography
% \bibliography{biblio.bib}
\end{document}